Title: **Processing, evaluating and understanding FMRI data with afni_proc.py**

Richard C. Reynolds[1], Daniel R. Glen[1], Gang Chen[1], Ziad S. Saad[1], Robert W. Cox[1], Paul A. Taylor[1]

[1] Scientific and Statistical Computing Core, National Institute of Mental Health, NIH, Bethesda, USA

**ABSTRACT**

FMRI data are noisy, complicated to acquire, and typically go through many steps of processing before they are used in a study or clinical practice. Being able to visualize and understand the data from the start through the completion of processing, while being confident that each intermediate step was successful, is challenging. AFNI's *afni_proc.py* is a tool to create and run a processing pipeline for FMRI data. With its flexible features, afni_proc.py allows users to both control and evaluate their processing at a detailed level. It has been designed to keep users informed about all processing steps: it does not just process the data, but first outputs a fully commented processing script that the users can read, query, interpret and refer back to. Having this full provenance is important for being able to understand each step of processing; it also promotes transparency and reproducibility by keeping the record of individual-level processing and modeling specifics in a single, shareable place. Additionally, afni_proc.py creates pipelines that contain several automatic self-checks for potential problems during runtime. The output directory contains a dictionary of relevant quantities that can be programmatically queried for potential issues and a systematic, interactive quality control (QC) HTML. All of these features help users evaluate and understand their data and processing in detail. We describe these and other aspects of *afni_proc.py* here using a set of task-based and resting state FMRI example commands.

# INTRODUCTION

FMRI processing is complicated. It relies on many disparate types of computational procedures, including alignment, "data cleaning" (such as despiking and censoring), time series analysis and statistical modeling. Researchers perform many different types of studies, each with a particular acquisition and modeling design: task-based, resting state (Biswal et al., 1995) and naturalistic (Hasson et al., 2010) datasets have distinct considerations. For example, task-based paradigms might require consideration of response times or modulation, as well as careful choice of hemodynamic response modeling assumptions (Bellgowan et al., 2003; Lindquist et al., 2009; Prince et al., 2022; Chen et al., 2023). Furthermore, EPI data may be acquired with either a "traditional" single echo or with multiple echoes (Posse et al., 1999), with the latter becoming increasingly popular and requiring a choice of echo combination methods. Data can be acquired across human age ranges (infant, pediatric, adult, aging populations) and across different species (macaque, rat, mouse, fetal pig, etc.), with each scenario requiring special considerations and particular assumptions. Finally, analyses can take place in either volumetric or surface-based topologies, and include one or more runs to process simultaneously.

Here, we describe *afni_proc.py*, a program available within the open source, publicly available AFNI toolbox (Cox, 1996), to create full processing pipelines across this wide FMRI landscape. Briefly, *afni_proc.py* allows a researcher to set up a full (or partial) subject-level FMRI processing script, specifying a desired set of steps and options to manage reading in raw data through linear regression and quality control (QC). From its creation in 2006, the program was designed to balance several important aspects of processing by having these features:
1. Being readable and understandable, both by the researcher using it as well as those with whom it is shared.
2. Being flexible to accommodate the precise steps that a researcher wants for the given analysis.
3. Being easy to use, relative to the amount of provided control.
4. Facilitating reproducibility, just by sharing the command and code version number used.
5. Retaining the provenance (record) of all processing that has been performed in a commented script, so no steps are hidden or require guessing.
6. Also retaining the intermediate datasets from each processing step, to facilitate quality control and to expedite investigations when the final results appear "unreasonable".
7. Growing and adapting to new user needs.

These design choices have had added benefits that, as new acquisition methods have been developed (e.g., multi-echo EPI), *afni_proc.py* has been able to incorporate new processing steps within a consistent framework.

This text is organized as follows. In the Methods, we first describe the general usage and organization of *afni_proc.py*, a program for specifying single subject (ss) pipelines. While not all details and options can be discussed here, we highlight several important aspects. These include the convenience and rigor of afni_proc.py processing, such as directly providing alignment concatenation and integrated checks through regression modeling, as well as how it provides detailed analysis provenance and inherent reproducibility. We then discuss how it facilitates quality control (QC) with its automatically generated HTML report and summary dictionary. Then, we outline general pipeline management before describing the specific datasets and summary of processing examples presented here. In the Results section, we present practical examples of *afni_proc.py* commands for different analysis cases to guide the examination of major processing steps. We briefly describe different features and possible adjustments that might apply in various scenarios. Finally, in the Discussion we summarize afni_proc.py's design, development philosophy and various validations. We outline some important considerations for setting up pipelines, and present some notes on choosing a hemodynamic response function (HRF). Lastly we discuss future directions and integrations for the program.

**METHODS**

**Setting up the pipeline: modular processing blocks with applicable options**

AFNI's *afni_proc.py* is a program to generate a full (or partial, if desired) FMRI processing pipeline for a single subject (covering what some researchers refer to as "first and second level" processing). First, a researcher specifies input datasets (such as an anatomical volume, one or more EPI time series, and tissue segmentations) and any accompanying files (stimulus timing files, physiological regressors, pre-calculated warp datasets, etc.); then, the researcher specifies the necessary processing choices (Fig. 1A). In order to simplify pipeline specification and conceptualization, *afni_proc.py* has a hierarchical and modular organization for defining an analysis. There is a top-level list of the major "processing blocks" to be performed (e.g., EPI-to-anatomical alignment, blurring, etc.), and then a set of any desirable option flags and values can be provided for each block (e.g., a particular cost function and blur radius). This layered framework allows the code to be readable, since the sequence of block names summarizes the processing and associated options have related prefixes. It also allows for flexibility to fit appropriately with a study design, since users can set up the blocks and tailor any number of options for each block (with the possibility for adding more at a researcher's request, a frequent occurrence). Table 1 provides a list of the currently available processing blocks within *afni_proc.py*, with a brief description of each.

# Schematic example: afni_proc.py command usage and features

A) Components of *afni_proc.py* command: inputs, processing blocks and any desired options

```
afni_proc.py
  -subj_id                  sub-101                                            ⎫  Subject ID and
  -dsets                    sub-101/func/sub-101_run-*_bold.nii.gz             ⎬  primary input
  -copy_anat                sub-101/anat/sub-101_run-1_T1w.nii.gz              ⎪  datasets
  -anat_has_skull           yes                                                ⎭
  -blocks                   tshift align volreg mask blur scale regress        }  Main processing
                                                                                  stages, in order
  -radial_correlate_blocks  tcat volreg regress                                ⎫  Options for extra
  -tcat_remove_first_trs    0                                                  ⎬  QC and (automatic)
                                                                               ⎭  tcat block
  -tshift_opts_ts           -tpattern alt+z2
  -align_unifize_epi        local                                              ⎫  Options for each
  -align_opts_aea           -giant_move -cost lpc+ZZ -check_flip               ⎪  processing block,
  -volreg_align_to          MIN_OUTLIER                                        ⎪  to manage detailed
  -volreg_align_e2a                                                            ⎬  behavior (the start
  -volreg_warp_dxyz         3                                                  ⎪  of an option name
  -volreg_compute_tsnr      yes                                                ⎪  typically matches
  -mask_epi_anat            yes                                                ⎪  associated block)
  -blur_size                6                                                  ⎭
  ...
```

B) Example of detailed provenance within *afni_proc*.py's automatically commented "proc script"

```
...
# =================================== align ==================================
# for e2a: compute anat alignment transformation to EPI registration base
# (new anat will be intermediate, stripped, sub-101_run-1_T1w_ns+orig)
# run (localized) uniformity correction on EPI base
3dLocalUnifize -input vr_base_min_outlier+orig -prefix    \
    vr_base_min_outlier_unif

align_epi_anat.py -anat2epi -anat sub-101_run-1_T1w+orig  \
      -save_skullstrip -suffix _al_junk                   \
      -epi vr_base_min_outlier_unif+orig -epi_base 0      \
      -epi_strip 3dAutomask                               \
      -giant_move -cost lpc+ZZ -check_flip                \
      -volreg off -tshift off

# =================================== volreg =================================
# align each dset to base volume, to anat

# register and warp
foreach run ( $runs )
    # register each volume to the base image
    3dvolreg -verbose -zpad 1 -base vr_base_min_outlier+orig     \
             -1Dfile dfile.r$run.1D -prefix rm.epi.volreg.r$run  \
             -cubic                                              \
             -1Dmatrix_save mat.r$run.vr.aff12.1D                \
             pb01.$subj.r$run.tshift+orig

    # create an all-1 dataset to mask the extents of the warp
    3dcalc -overwrite -a pb01.$subj.r$run.tshift+orig -expr 1    \
           -prefix rm.epi.all1
...
```

C) Example single subject FMRI processing work flow, incorporating *afni_proc.py*

| **Preliminary** | **Set up analysis** | **Run afni_proc.py** | **Run proc script** | **Evaluate analysis** |
|---|---|---|---|---|
| sswarper2, physio_calc.py, FreeSurfer, ... | Specify inputs and options for afni_proc.py | Create the fully commented proc script | Perform analysis, plus make checks and QC features | Check processed data with the QC scripts and HTML |

***Figure 1.*** *Schematic features of afni_proc.py. A) Primary data inputs and descriptors are highlighted in green. The processing is managed hierarchically: first the user selects and orders the desired blocks (or major stages), and then for each can specify zero, one or more options. The array of hot colors highlights which options are associated with which block, by matching them: the "tshift" block label with*

*the "-tshift_opts_ts" option, etc. Note that the start of the option name typically matches the block, as well.  B) The afni_proc.py command creates a fully commented processing pipeline ("proc script"), so that the user has detailed understanding and provenance of all the steps of the analysis. C) An example workflow that uses afni_proc.py for a single subject analysis, utilizing some preliminary programs beforehand and incorporating automatically-generated data checks and quality control features at the end. This can simply be looped over all subjects in a data collection.*

```
--------------------------------------------------------------------
Automatic blocks (setup and initialization; not user specified, always performed)
   setup     : check args, set run list, make output directory, copy stim files
   tcat      : copy input datasets and remove unwanted initial TRs
Default blocks (standard steps; user may skip, modify or rearrange)
   tshift    : slice timing alignment on volumes
   volreg    : volume registration (for reduction of subject motion effects)
   blur      : blur each volume (default is 4mm FWHM)
   mask      : create a 'brain' mask from the EPI data
   scale     : scale each run mean to 100, for each voxel (max of 200)[1]
   regress   : regression analysis (stimulus model, filter, censor, etc.)
Optional blocks (the default is to _not_ apply these blocks, but the user can add)
   align     : align EPI and anatomy (linear affine)
   combine   : combine echoes into one
   despike   : truncate spikes in each voxel's time series
   empty     : placeholder for some other user command
   ricor     : RETROICOR[2] - removal of cardiac/respiratory regressors
   surf      : project volumetric data into the surface domain (via SUMA[3])
   tlrc      : warp anatomical volume to a standard space or template
Implicit blocks (not user-specified, but performed when appropriate)
   blip      : perform B0 distortion correction
   outcount  : temporal outlier detection
   QC_review : generate QC review scripts and HTML report (APQC HTML)
   anat_unif : anatomical uniformity correction
--------------------------------------------------------------------
```
**Table 1.** *Current processing blocks in afni_proc.py, with brief descriptions. The automatic and implicit blocks are not specified by the user in the "-blocks" option list; the former will always be performed, and the user can still manage their behavior with options, such as "-tcat_remove_first_trs" and more. The implicit blocks will be automatically added in an appropriate spot within the processing stream when the user adds a relevant option, such as "-blip_forward_dset" for the blip block. The users can see where and how these blocks are performed in detail within the created processing script.*
[1]Chen et al., 2017.  [2]Glover et al., 2000.  [3]Saad et al., 2004.

Within each processing block, there can be zero, one or more control parameters to specify. Several blocks and useful options are described in the code examples below. We note that the presence of the "empty" block allows researchers further flexibility to directly insert their own steps. Historically, however, many steps have been directly added to the program itself.

**Examples of not masking during processing: benefits for understanding and QC**

A) Unmasked TSNR shows notable ghosting (affecting correlation) that masking would hide

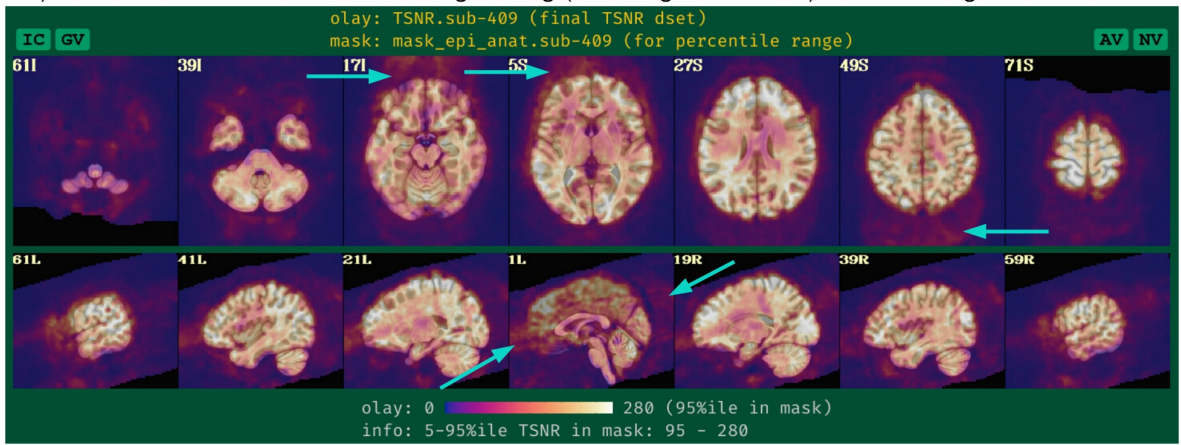

B) Unmasked TSNR map shows imperfect alignment that masking would hide

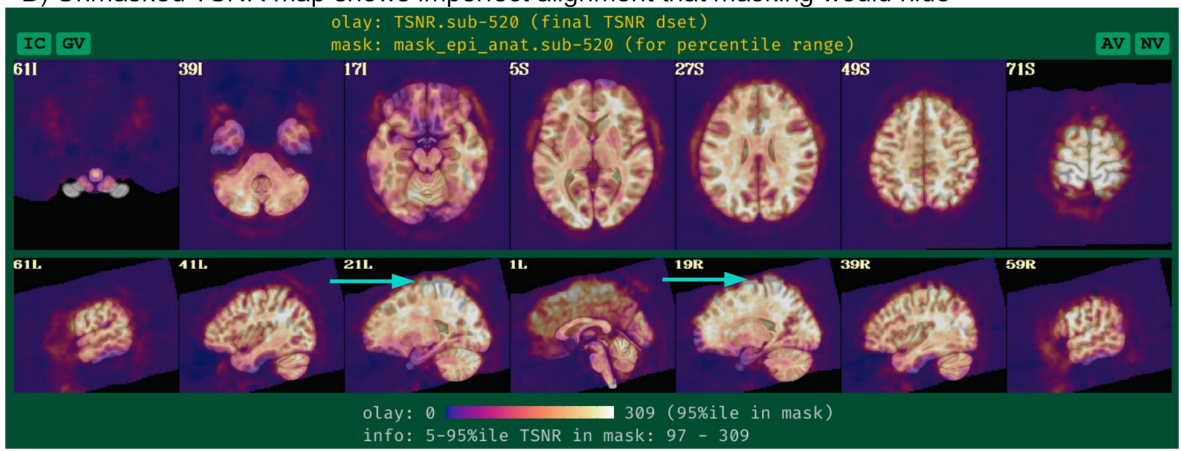

*Figure 2. While an EPI brain mask is estimated during afni_proc.py processing, it is not applied to the data, so that results throughout the whole FOV can be viewed. This facilitates understanding the data better, as well as improving QC evaluation. Two examples of this are shown from data processed within the FMRI Open QC project, showing TSNR in the final MNI space after regression modeling. In A, one can see strong ghosting outside the brain (cyan arrows), which helps explain some of the unexpected correlation patterns that are observed within the brain. In B, one can see from the TSNR pattern that part of the final EPI data is not well aligned in standard space; this is due to initially impressive skullstripping, which could then be fixed. In both cases, masking would have hidden the reality of what was happening and contributed to potentially biased results.*

An important note about the "mask" block is that it typically only involves *calculating* relevant masks from the EPI and anatomical volumes, which can later be used at the group level or for summary estimates. The masks are typically not *applied* during standard processing, except for defining regions within which to estimate summary statistics such as temporal signal-to-noise ratio (TSNR), so that modeling results from all voxels (even outside the brain) can be assessed. This allows one to more thoroughly check for the presence of artifacts (e.g., due to a bad coil), severe ghosting, misalignment or other features of the data, rather than to hide them away. Two examples of this are provided in Fig. 2, which show TSNR after processing (in MNI space) using data from the FMRI Open QC project (Taylor et al., 2023) from processing as part of Reynolds et al. (2023) contribution. In panel A, strong ghosting patterns are visible along the anterior-posterior axis, and understanding this helps explain some unexpected correlation patterns within the brain. In panel B, the full TSNR pattern reveals misalignment during initial processing that might otherwise be missed; this led to the discovery that anatomical skullstripping had been imprecise, which was missed due to low tissue contrast. By seeing data outside the brain, the problem was detected and reprocessing was applied to fix it—otherwise, misalignment would likely have gone unnoticed in masked data and simply skewed final results.

The above is one example of how, in general, afni_proc.py has been developed to allow the researcher to see and explore *more* in their data: more details about the processing code, more intermediate files to explore for verification, and more quality control images and quantitative warnings to help evaluate the processing. This facilitates understanding and improves confidence in both the data and the processing.

Processing convenience and rigor

Many underlying steps are managed within the *afni_proc.py* program itself when building the processing script, in ways to optimize mathematical benefits. For example, there are typically several volume registration or alignment steps within a given single subject pipeline, such as various combinations of: reverse-phase encoded EPI alignment for B0 distortion correction (implicit "blip" block); motion correction, possibly with multiple transformations ("volreg" block); EPI-to-anatomical alignment ("align" block); and subject anatomical-to-template alignment ("tlrc" block). Creating new datasets at each block would introduce unnecessary smoothing in the final data, as each regridding process involves interpolation. Instead, it is preferable first to concatenate all the estimated transforms and then to apply them in a single step (Jo et al., 2013); *afni_proc.py* performs this beneficial concatenation automatically, simplifying the procedure for the user. Additional conveniences include performing any bandpassing, censoring and regression as a single step, rather than as a mathematically inconsistent two-step process (Hallquist et al., 2013). This allows for a complete evaluation of the degrees of freedom used in processing and avoidance of mistakenly overusing them.

Having these programmatic conveniences occur automatically "under the hood" has several benefits. It simplifies the processing specification, reduces the chance of errors or subtle bugs occurring and increases the understandability of the processing itself. It also makes it easier to alter or update a processing stream, because one merely needs to add an option or change a parameter, rather than to reorganize potentially complicated logic within a script. Finally, it also means that more consistency checks can be performed automatically (e.g., keeping track of utilized degrees of freedom in the regression modeling appropriately).

Provenance and Reproducibility

Specifying a full FMRI pipeline with a set of options in an *afni_proc.py* command is convenient. However, it is also necessary to ensure that the researcher knows exactly what occurs during the processing—having the provenance of the results—and *afni_proc.py* also provides this. When executed, the *afni_proc.py* command first produces a full processing script (Fig. 1B), which is organized by processing block, automatically commented and even contains a copy of the generative *afni_proc.py* command. This "proc script" is then itself executed to process an individual's data and saved for later reference. This two stage approach—having the command generate a readable and commented script file, rather than simply carrying out the processing with no record—ensures that the researcher has all the details of the analysis at their fingertips. They are able to investigate the script to see the exact options used in each command at any time.

For example, a researcher might choose to smooth the EPI data during preprocessing by using the "blur" block, but not specify the blur radius explicitly and just use the program's default (which is not necessarily set as a recommendation for all types of data). Even so, it is still possible to directly see the actual value used within the generated processing script, so there is no ambiguity or guesswork. The proc script is cleanly and clearly organized with the same hierarchy as the *afni_proc.py* command itself and contains detailed comments. Thus, the proc script fulfills two roles: it is both an exact, searchable specification of analysis, and it is a learning tool. If there are any questions about any aspect of the procedure, one can verify each step directly. Furthermore, when using the "-execute" option, a log file of all terminal text is also saved, for later reference (users are also encouraged to create their own log files, if not using this option).

Additionally, AFNI itself has inherent provenance tracking within its programs on a file-by-file basis. When a dataset is output, the command used to create it is added to its header. Therefore, the data contains an accumulating history of the commands used to produce it, including the AFNI code version and date+time of creation. This can be viewed with the full header information via AFNI's 3dinfo program, and specifically queried by adding the "-history" option. This useful information is still included even in NIFTI files output by AFNI, via the AFNI extension, as well as NIML datasets such as surface files.

The relatively compact *afni_proc.py* command (typically 20-50 lines, as vertically spaced and aligned in Fig. 1A) can readily be published in a paper's Appendix or in an online repository (GitHub, OSF, etc.). The processing can be reproduced by using the same command with the original code version or in a container. To simplify comparisons of different *afni_proc.py* commands, several options within the program itself exist:
- "-compare_opts .." compares a user's command against a pre-defined *afni_proc.py* example from the help examples;
- "-compare_example_pair .." compares two sets of predefined commands;
- "-compare_opts_vs_opts .." compares two full commands;
- and "-show_example .." displays the full afni_proc.py command from the help file, which includes many examples from publications (we include the relevant command at the start of each example description, below).

For example, users could add "-compare_opts 'example 6b'" to their afni_proc.py command to display a comparison with that help file-enumerated example, or "-compare_opts 'AP publish 3b'" to compare against Ex. 2's processing command from this paper. To display the basic surface-based processing command used in AFNI Bootcamp courses, one could use "*afni_proc.py -show_example 'AP class 3'*".

We note that the program has also developed with minimal external dependencies, to facilitate stability, compatibility and reproducibility over time to the greatest extent possible. See Appendix B for more discussion of development.

Quality Control and Understanding Data Through Processing

A primary goal of the *afni_proc.py* program (and of the entire AFNI platform) since its inception has been to help users "stay close to their data," meaning that they understand the dataset from its raw state through all stages of processing. As part of this, *afni_proc.py* creates a "*.results" directory for each single subject analysis, which contains copies of the original data, many of the intermediate datasets and the final outputs, so that all stages of the processing can be verified after running the script.

The *afni_proc.py* pipelines also generate several automatic and helpful QC features to review many aspects of the single subject (ss) processing; the recent FMRI Open QC Project illustrated the numerous benefits of integrating both qualitative and quantitative QC items with full preprocessing, as demonstrated in both AFNI and several other neuroimaging software packages (see Taylor et al., 2023a and *op cit.*). During *afni_proc.py*'s FMRI processing, relevant "basic" quantities are calculated and reported to the researcher at the end, such as subject motion summaries, degree of freedom (DF) counts, TSNR, and more; these are saved in a text file and can be displayed via the *@ss_review_basic* script in each results directory. Several potential pitfalls of analysis are also checked automatically, with results stored in "warning" files, such as not removing pre-steady state volumes or having collinear regressors of interest (see Taylor et al., 2024 for a more complete list).

Since many processing steps either require or are greatly helped by visual verification, there is also a script created called *@ss_review_driver,* which will open viewing panels and the AFNI GUI to guide the user through visually verifying steps such as alignment, motion censoring and model fitting. Each step of this review is commented with a pop-up GUI guide, as well. A script called *@epi_review.${subj}* (where ${subj} is replaced by the actual subject ID in the filename) is created, which opens up the AFNI GUI graph viewer and image panels, with time-scrolling on, to allow the user to assess the quality of the original EPI time series.

Finally, *afni_proc.py* generates a QC document that can be opened in a web browser to review important features from the pipeline in a single, navigable report: the APQC HTML (Taylor et al., 2024). It incorporates systematic images, automated warning checks and several additional features, and in conjunction with a review of the "basic" quantities may be considered an efficient, minimal QC source for each single subject analysis. The HTML can also now be run using a local Web server, so users can save QC ratings as they scroll through, saving comments and opening up the datasets themselves interactively. These features have all been shown to be useful in understanding the FMRI data and quality issues (Reynolds et al., 2023).

It is worth emphasizing the importance of *afni_proc.py's* quality control features and having them integrated with processing. FMRI data can have a wide array of issues, and many of these can be subtle. While the primary focus of processing is typically to make it more appropriate for later analysis (removing artifacts, aligning datasets, etc.), it allows for underlying properties to be probed in different ways. The QC procedures in AFNI generally and in *afni_proc.py* specifically have been developed with the goal of taking advantage of these intermediate stages to understand the data itself more completely (Reynolds et al., 2023; Taylor et al., 2023a). This is also one reason why intermediate datasets are created and retained as part of *afni_proc.py* outputs, to allow for more detailed QC as needed. As such, quality control should not just be viewed as filtering out "bad" data, but as determining if one can have confidence in it, through its full range of properties.

Pipeline Management

The outputs of other programs can be directly input into *afni_proc.py* scripts. One typical case is using the anatomical T1w-based results of FreeSurfer's *recon-all* (Fischl et al., 2002). Common usages include incorporating the volumetric segmentations for defining tissue-based regressors in the "regress" block or the SUMA-standardized surface meshes (Argall et al., 2006) for projecting EPI data in surface-based analysis via the "surf" block. Nonlinear warps between the subject anatomical and a reference template can also be estimated beforehand, with the results passed along to *afni_proc.py* to concatenate with other transforms. Those warps can be calculated, for example, using AFNI's *3dQwarp* (Cox and Glen, 2013), *sswarper2* (a new, preferred method that incorporates *3dQwarp* for performing both anatomical skullstripping and nonlinear warping simultaneously; see, e.g., Taylor et al. (2018)), or *@animal_warper* (similar set of functionality as *sswarper2* but specifically designed for processing non-human datasets; see Jung et al., 2020). Physiological regressors can be included, after RETROICOR (Glover et al., 2000) time series estimation with programs such as *RetroTS.py* or the newer *physio_calc.py* (Lauren et al., 2023). A benefit of this modularity and separation is that, if the FMRI processing with *afni_proc.py* needs to be rerun for any reason, these precursor steps do not need to be redone; this can also save a lot of time and resources for computing intensive programs such as *recon-all* and *sswarper2*.

Some external programs with specific processing have been integrated within afni_proc.py. This has been the case for some workflows with the increasingly popular multi-echo (ME) FMRI data, in which several (typically 3-5) T2* weighted volumes are acquired per time point. This information can be combined in various ways to boost signal-to-noise (SNR) in the analyzed BOLD signal. Such data can be processed in *afni_proc.py*, using the "combine" block to specify processing choices. The command contains its own optimal combination (OC) formulation from Posse et al. (1999). Additionally, options can be provided so it calls one of the multi-echo independent component analysis (MEICA) routines from either the original Kundu et al. (2015) or the newer DuPre et al. (2021) *tedana* (TE-dependent analysis) codebases. The QC page of the latter is even integrated into *afni_proc.py*'s own APQC HTML. Note that tedana must be installed separately.

In the end, an example *afni_proc.py* workflow for single subject analysis can look like that in Fig. 1C. To incorporate this in a full group analysis, one can simply loop over a list of subjects, changing the input file names but keeping the remaining *afni_proc.py* blocks and options the same. This keeps the scripting simple if, for instance, subjects have different randomly-generated stimulus timing files for a task-based FMRI analysis. Input volumes can be either NIFTI (Cox et al., 2004) or BRIK/HEAD format, and *afni_proc.py* also works directly with BIDS-formatted collections (Gorgolewski et al., 2016). Fig. 3 shows a pseudocode example of running a group level processing in this way: for each subject ID and session ID, do the afni_proc.py processing contained in a script. This framework runs efficiently on a BIDS data structure, as well as on any reasonably organized one. To use JSON sidecar information, AFNI's abids_*.py programs can also be incorporated in the central script. In some cases, processing may need to be rerun on a subgroup of subjects—for example, to fix imprecise EPI-anatomical alignment—and *afni_proc.py* can easily be rerun for either a subgroup or entire group within a data collection.

The results directory created by *afni_proc.py* contains relevant outputs in a standardized structure and naming convention, including copies of the input data, several stages of intermediate files, a QC directory, and the final datasets. There is also a reference dictionary of key datasets and quantities ("out.ss_review_uvars.json"), which is both parsed by the QC generation programs and can serve as a useful reference of important outputs for the user ("uvars" stands for "user variables"; see Taylor et al., 2024, for details). Some additional quantities relevant for group analysis can also be calculated by *afni_proc.py* output scripts, and then used later. This includes estimating smoothness of noise autocorrelation functions (ACF) for group-level clustering (Cox et al., 2017); see Example 2, below. All of these estimated QC quantities might be compared or evaluated for data-dropping criteria with AFNI's *gen_ss_review_table.py* (Reynolds et al., 2023).

The derived outputs can be used for further analysis within AFNI or any other software. The key quantities and datasets from processing are known from the keys defined for the "uvar" and "basic review" dictionaries. This output structure is inherently mappable to ones from analogous software tools in other packages or to BIDS-Derivatives. We have recently added functionality to create an additional output directory that follows the BIDS-Derivatives (v1.9.0) file structure and naming convention for FMRI processing, for the subset of *afni_proc.py* outputs that currently have definitions there. The user implements this by specifying "-bids_deriv yes". An example of this is provided in the supplementary Ex. 9 (see Appendix C).

**Schematic code for processing a group of subjects**

```
# run a script to do afni_proc.py processing
# for a group of subjects in a data tree
# such as BIDS

topdir   = group_level_directory
all_subj = [list of subject IDs]

for subj in all_subj :
    cd ${topdir}/${subj}
    all_ses = [list of session IDs]

    for ses in all_ses :
        cd ${topdir}/${subj}/${ses}
        tcsh do_ap_cmd.tcsh ${subj} ${ses}
    end
end
```

Figure 3. Pseudocode for running single subject processing at a group level, looping over a list of subject IDs and one or more sessions for each. At the heart of the second loop is the action to do the subject processing: here, to run a theoretical shell script ("do_ap_cmd.tcsh") that contains an afni_proc.py command and just needs the subject and session ID values provided as arguments. This runs easily on a BIDS-formatted data collection (though some BIDS trees do not contain a session-level ID or directory structure, and so the second loop would be omitted).

Description of example datasets used here
In the next section we present four examples of *afni_proc.py* commands for various FMRI processing. The full processing scripts for the workflows described here (including *physio_calc.py*, *sswarper2*, FreeSurfer's *recon-all*, etc.) are publicly available[1], including all of the supplementary examples. We first briefly describe the MRI datasets, which are publicly available[2] and representative of standard acquisitions. Each was collected at 3T field strength.

1 https://github.com/afni/apaper_afni_proc
2 The input datasets are available here:
https://afni.nimh.nih.gov/pub/dist/tgz/demo_apaper_afni_proc_rest.tgz,
https://afni.nimh.nih.gov/pub/dist/tgz/demo_apaper_afni_proc_task.tgz

Examples 1, 3 and 4 use the same set of resting state FMRI data ("sub-005"), which has been acquired with ME-FMRI. Exs. 1 and 3 just use a single echo, while example Ex. 4 utilizes all the echos during processing. The acquisition and original application of this dataset is detailed in Gilmore et al. (2019) and in Gotts et al. (2020). Briefly, there is one ME-FMRI resting state run with: TR=2200 ms; TE=12.5, 27.6, 42.7 ms; voxels=3.2x3.2x3.5 mm$^3$; matrix=64x64x33; N=220 volumes; slice acceleration factor = 2 (ASSET). A short, single-echo (SE) reverse encoded EPI set (N=10, TE=27.6ms) was acquired for B0-distortion correction. There is also a standard T1w anatomical volume (1 mm isotropic voxels), and both cardiac and respiratory data were collected. This subject's unprocessed dataset is organized in a BIDS-ish manner: the NIFTI file naming and directory structure match BIDS v1.3.0, but there are minimal differences such as not having associated JSON sidecar files.

Example 2 presents processing for one subject of task-based FMRI ("sub-10506"), using the publicly available Paired Associates Memory Task-Encoding (PAMENC) data collection (Poldrack et al., 2016; https://openneuro.org/datasets/ds000030/versions/1.0.0). The stimulus paradigm during scanning was a two-part memory task involving pairs of words and their pictures. For the "task" trials, participants were shown pairs of words for 1s and then additional corresponding images were also shown for 3s more; one image was drawn in a single color and the other was in black-and-white, and they were instructed to press Button 1 or 2 when the color image was on the left or right, respectively. For the "control" trials, pairs of scrambled stimuli were shown, again one in a single color and one in black-and-white, and users were to similarly indicate the side of the color image with a button push. At the same time, they were instructed to try to remember word pairs, as they would later be asked about how sure they were that two particular words were presented together. Briefly, there is one single echo EPI run with: TR=2000 ms; TE=30 ms; voxels=3.0x3.0x4.0 mm$^3$; matrix=64x64x34; N=242 volumes. There were two FMRI stimulus classes, 40 trials of "task" and 24 trials of "control". A standard T1w anatomical volume (1 mm isotropic voxels) was also acquired. This subject's unprocessed dataset is organized in a complete BIDS v1.3.0 structure.

Overview of processing examples

Ex. 1 presents a special use case of *afni_proc.py*, that of only performing the spatial transformation steps of processing. This subset of steps might be used to test options in setting up alignment, or to investigate only specifically spatial properties of the data. The remaining Examples 2-4 all demonstrate full single subject processing, through regression. Each is aimed at standard voxelwise analysis, as each includes blurring (spatial smoothing) during processing. Further examples or variations are provided in Appendix C.

Most voxelwise studies include blurring with the goal of increasing TSNR locally and to likely increase group overlap in the face of imperfect alignment and significant structural variability (particularly in the human cortex). Blurring can be applied in different ways, and we highlight some of these below. To use these examples as starting points for ROI-based analysis, one could simply remove the "blur" processing block and associated "-blur_*" options. Blurring should generally *not* be included when analyses will include averaging within ROIs, because it will spread signals across the region boundaries, artificially boosting local correlations and likely weakening distant ones. Moreover, the time series averaging within ROIs later on will still play the role of boosting local TSNR.

The packaged outputs of each example run of afni_proc.py (including proc script, terminal log and results directory) are available here: https://osf.io/gn7b5/.

**Ex. 1: QC images of skullstripping and nonlinear warping via sswarper2**

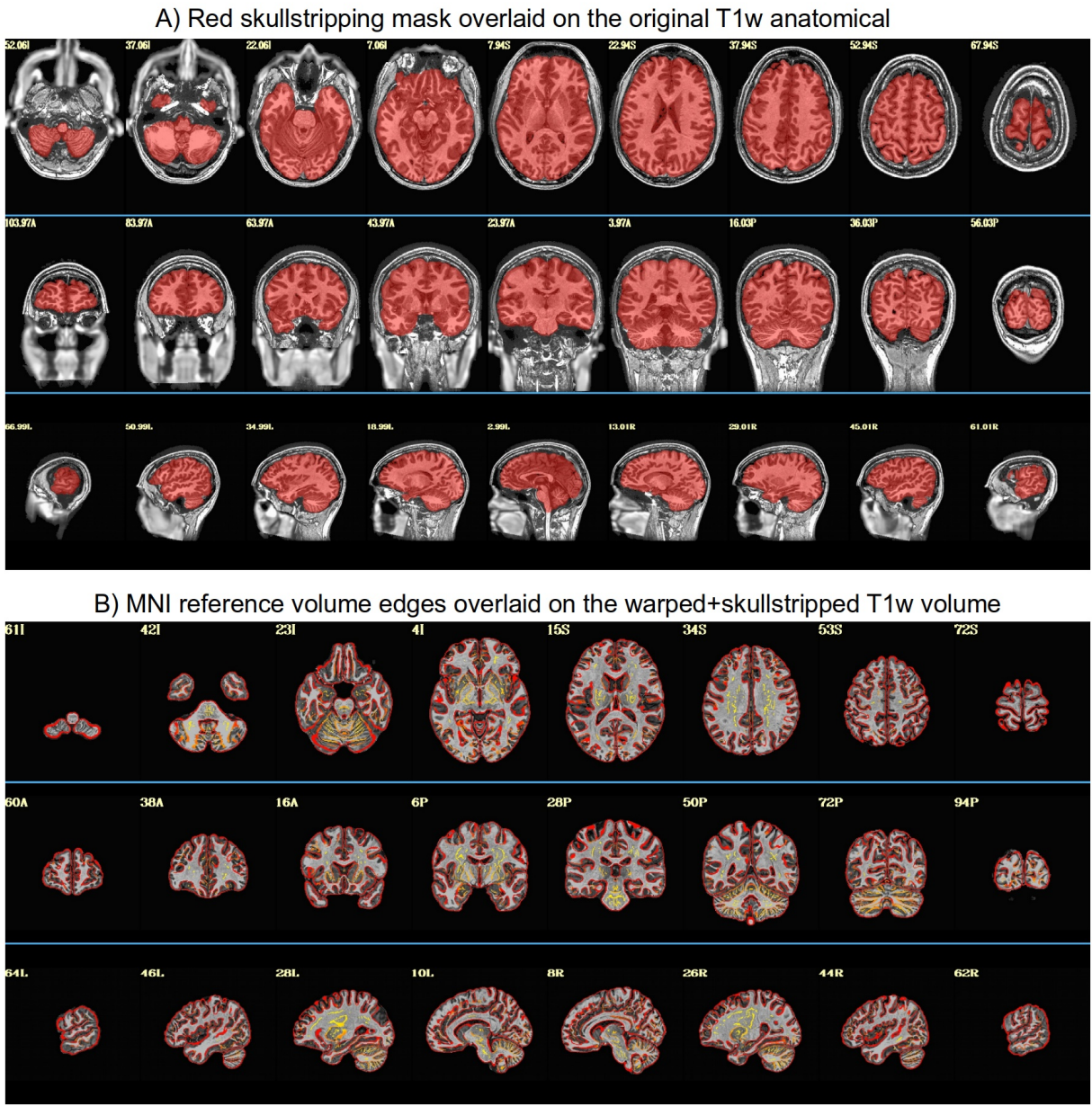

*Figure 4.* QC images generated by AFNI's sswarper2, as it both skullstrips an anatomical volume (panel A) and calculates its nonlinear warp to template space (panel B). Both brainmasking and alignment with the template appear to be generally strong throughout the brain. The outputs of this program (or analogous ones, such as AFNI's @animal_warper) can be used directly in afni_proc.py. Here and in axial/coronal images below, image left is subject left.

In these examples, some separate programs were run prior to *afni_proc.py*. This is sometimes done to derive useful information from supplementary datasets, like tissue maps from the anatomical volume, or physiological based regressors when cardiac and respiratory traces have been acquired during scan time, etc. The results of these steps are then included as additional inputs to *afni_proc.py*. We note these briefly here for each example, as well as in each table that shows an example commands:

- **Example 1:** AFNI's *sswarper2* was run on the anatomical T1w volume to simultaneously skullstrip (SS) it and estimate its nonlinear warp to standard space. Examples of its useful brainmask identification and alignment to standard space are shown in Fig. 4.
- **Example 2:** AFNI's *sswarper2* was run on the T1w volume for skullstripping and nonlinear warp estimation; AFNI's *timing_tool.py* was used to create stimulus timing files from the provided events TSV files.
- **Example 3:** FreeSurfer's recon-all was run on the anatomical T1w volume to estimate CSF and ventricle maps, for estimating local tissue-based regressors with ANATICOR (Jo et al., 2010), as well as ROI parcellations; AFNI's *physio_calc.py* (Lauren et al., 2023) was run on cardiac and respiratory traces, for creating physiological-based regressors with RETROICOR (Chang and Glover, 2009); AFNI's *sswarper2* was run on the T1w volume for skullstripping and nonlinear warp estimation.
- **Example 4:** FreeSurfer's *recon-all* was run on the T1w volume to estimate the anatomical surface, as well as ventricle maps; AFNI's *@SUMA_Make_Spec_FS* was run on the FreeSurfer meshes to convert them to GIFTI format and to create standard meshes (Argall et al., 2006).

When processing included aligned datasets to a volumetric standard space template (i.e., when a script contained the tlrc processing block and a "${template}" dataset[3]), the MNI-2009c-Asym space (Fonov et al., 2011) was used. Specifically, AFNI's MNI152_2009_template_SSW.nii.gz version of the template was used, which has multiple subvolumes of information utilized by *sswarper2* (and its predecessor, *@SSwarper)*. Note that processing does not require use of a template space: final volumetric results could be native subject space; surface-based results typically end up on the mesh estimated from the subject's own anatomical (but if using one of SUMA's standard meshes, these could be equivalently displayed on any surface mesh).

**RESULTS**

We present four *afni_proc.py* example commands and their results, describing the processing choices made in these examples, as well as other ones that could be used. It is impossible to provide a comprehensive set of examples[4], and the ones presented here have been chosen to highlight various features. The order of options within the command does not matter (NB: the order of *blocks* specified within the "-blocks .." option does matter), but is typically chosen for grouping of relevant options and for clarity of purpose. Several of the images shown below come directly from the systematic views provided within the APQC HTML created by the given processing. These were run with AFNI version 24.2.02.[5]

Ex. 1:  Partial processing, warping-only case: spatial transformations

---

[3] The script can include the path to the chosen template, or if it is located in a directory where the AFNI program @FindAfniDsetPath can locate it, then just the name is enough to specify it.
[4] The *afni_proc.py* program help file contains over 30 examples at present. AFNI's online documentation also contains the AFNI Codex, a set of code examples related to publications, and many of these contain *afni_proc.py*:
https://afni.nimh.nih.gov/pub/dist/doc/htmldoc/codex/main_toc.html
The Codex pages contain descriptions, as well as links to the papers and commented scripts and/or repositories. Appendix B describes additional demos.
[5] Interested users can use build_afni.py to locally build a specific code version, as well as use the associated, tagged Docker image, available here: https://hub.docker.com/r/afni/afni_dev_base/tags

In addition to building complete processing pipelines, *afni_proc.py* can be used to perform subsets of processing. This can be simpler than writing a separate script to carry out the task, because *afni_proc.py* includes several convenience features. In this example, we focus on the subset of alignment-related features in standard FMRI processing. Note that the "regress" block is merely included so that the APQC HTML is created; no regression options are used here, and those will be discussed in subsequent examples along with other non-alignment considerations.

```
# AP Example 1, which was preceded by running:
# + sswarper2: creates skullstripped anatomical ${anat_cp}, and
#              nonlinear warps ${dsets_NL_warp}

afni_proc.py                                                                    \
    -subj_id                   ${subj}                                          \
    -dsets                     ${dset_epi}                                      \
    -copy_anat                 ${anat_cp}                                       \
    -anat_has_skull            no                                               \
    -anat_follower             anat_w_skull anat ${anat_skull}                  \
    -blocks                    align tlrc volreg regress                        \
    -blip_forward_dset         "${epi_forward}"                                 \
    -blip_reverse_dset         "${epi_reverse}"                                 \
    -tcat_remove_first_trs     4                                                \
    -align_unifize_epi         local                                            \
    -align_opts_aea            -cost lpc+ZZ -giant_move -check_flip             \
    -tlrc_base                 ${template}                                      \
    -tlrc_NL_warp                                                               \
    -tlrc_NL_warped_dsets      ${dsets_NL_warp}                                 \
    -volreg_align_to           MIN_OUTLIER                                      \
    -volreg_align_e2a                                                           \
    -volreg_tlrc_warp                                                           \
    -volreg_warp_dxyz          3
```

Figure 5. *The afni_proc.py command for Ex. 1 (warping-only, single echo FMRI). The options and any arguments are vertically spaced for readability. Here and throughout, items starting with "$" are variable names, which are typically file names or control options. ${sub} = the subject ID; ${anat_cp} = the input anatomical dataset (here, that has been skullstripped by sswarper2); ${anat_skull} = a version of the input anatomical dataset that still has its skull, for reference during processing; ${dset_epi} = the input EPI dataset (which is a single echo, here the second one from the ME-FMRI acquisition); ${epi_forward} = an EPI volume with phase encoding in the same direction as the main input FMRI datasets, to be used in alignment-based B0-inhomogeneity correction; ${epi_reverse} = an EPI volume with phase encoding in the opposite direction as ${epi_forward}, for B0-inhomogeneity correction; ${template} = name of reference volume for final space (here, the MNI template). Running this command produces a commented script of >450 lines, encoding the detailed provenance of all processing.*

In the Ex. 1 command (Fig. 5, or run: afni_proc.py -show_example "AP publish 3a"), we use *afni_proc.py* to perform four alignment processes:
- EPI with forward phase to EPI with reverse phase alignment ("blip" block, which is implicitly included when the "-blip_*_dset .." options are present), which is estimated using a restricted nonlinear alignment using AFNI's "3dQwarp -plusminus ..";
- EPI-to-anatomical alignment ("align" block), which is estimated with a linear affine transform;
- anatomical-to-template alignment ("tlrc" block), which has been estimated nonlinearly here using AFNI's *sswarper2* so that anatomical skullstripping is also included;

- EPI-to-EPI volumetric motion estimation and correction ("volreg" block), which is estimated with rigid-body alignment across the input FMRI dataset.

From the short command provided in Fig. 5, the created proc script will first estimate each of the alignments individually. It will then conveniently concatenate the transformations into one comprehensive warp and apply that as a single transformation to the raw, input EPI dataset (Fig. 6A). This concatenation minimizes the blurring necessarily incurred by regridding and interpolation, while simultaneously motion correcting and warping the FMRI time series to the template space. The *afni_proc.py* command is an extremely compact way to perform this procedure (the produced "proc" script is over 450 lines long, including comments), and the programmatic details of concatenation can all be checked for educational or verification purposes within the output proc script.

## Concatenation of alignment steps: examples within afni_proc.py

### A) Alignment steps in Ex. 1

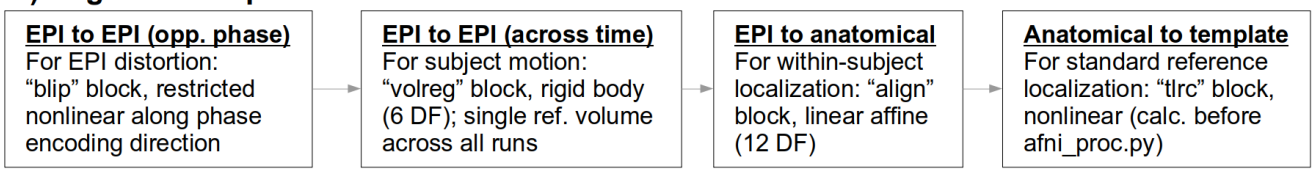

### B) Alignment steps in Ex. 2 and 3

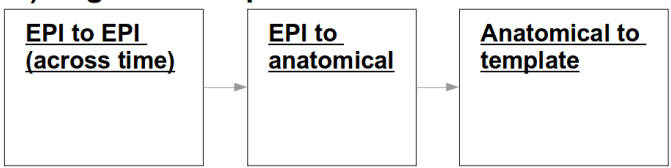

### C) Alignment steps in Ex. 4 (followed by projection to standard mesh surfaces)

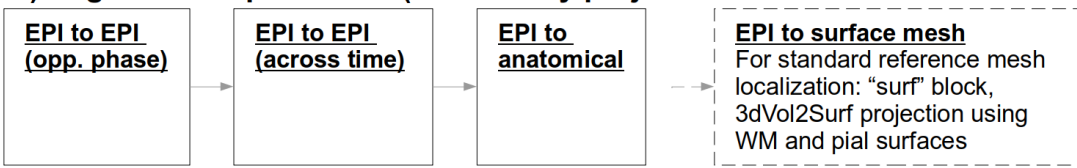

### D) Example variation of "A" (Ex. 1) to use intermediate per-run motion estimation, by adding:

```
-volreg_post_vr_allin      yes
-volreg_pvra_base_index    MIN_OUTLIER
```

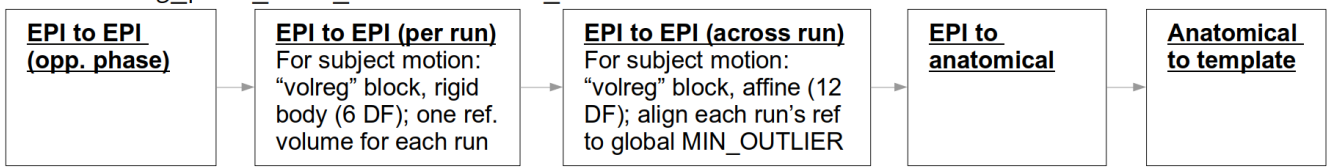

*Figure 6.* Schematics of the various alignment steps within each example's afni_proc.py command. Details are shown for the first time a particular step is presented. Alignment is calculated separately for each step, but then concatenated within the afni_proc.py script before applying to the EPI data. This tends to minimize extra blurring that would be incurred by multiple regridding and interpolation processes, if the stages were applied separately. In C (Ex. 4), after the concatenated warp is applied, the EPI data are projected onto a standardized surface mesh with 3dVol2Surf. Case D displays a variation of how to handle motion estimation when multiple runs are input, particularly if one might expect more differences between runs.

We note some additional points on the input EPI and anatomical data (which are provided with "-dsets .." and "-copy_anat ..", respectively). Firstly, these datasets can be in either NIFTI or BRIK/HEAD format, since AFNI reads and writes both. By default, the input anatomical volume would be skullstripped with a simple method. But when that procedure has already been performed, the user can deactivate that by including "-anat_has_skull no", as has been done here. Multiple single echo EPI datasets can be input after the single "-dsets .." option, such as when there are multiple runs per session to be analyzed simultaneously. The final output from processing will be derived from one concatenated dataset, and one would typically then include an option for "per run regressors" within the regress block, to appropriately handle the breaks in the time series during regression (see Ex. 2-4, below). Finally, we note that it is possible to remove initial time points from the input EPI datasets as they are copied at the start of processing, using the "-tcat_remove_first_trs .." option; this is often done if pre-steady state volumes remain in the data, as shown here to remove the first 4 time points. The *afni_proc.py* script also contains an automatic check for potential pre-steady state volumes, which is included in the "warns" section of the APQC HTML.

In addition to the main EPI time series data, reverse-phase encoded EPI datasets, which are also known as "blip up/blip down" datasets, are input in this example via "-blip_forward_dset .." and "-blip_reverse_dset ..". The "blip" processing block[6] is therefore implicitly included in the processing block list, as noted in Table 1; here, it occurs just prior to the "align" block in the generated processing script. This pair of blip datasets will be mutually aligned and produce a warp that "meets in the middle", which reduces the geometric effects of B0 inhomogeneity (Chang and Fitzpatrick, 1992; Andersson et al., 2003; Holland et al., 2010). This has been shown to improve the matching of structures between subject EPI and anatomical data (e.g., Hutton et al., 2002; Hong et al., 2015; Irfanoglu et al. 2019; Roopchansingh et al., 2020). In Fig. 7, one can see the reduced geometric distortion for the EPI volume particularly by comparing sagittal slice views in "A" (e.g., slice 17.03R) with the post-alignment EPIs underlaying the anatomical edges in "B" (e.g., slice 19R): the stretching of the former along the AP axis is greatly reduced in the latter. Note that some bright CSF can still be observed outside the anatomical edges.

In the anatomical-to-template block ("tlrc"), nonlinear alignment is switched on with "-tlrc_NL_warp", here specifying an MNI template as a reference base ("-tlrc_base .."). By default, this transform would be calculated using AFNI's auto_warp.py, which calls *3dQwarp* (Cox and Glen, 2013). However, in this example we have already run AFNI's *sswarper2* program to both skullstrip the anatomical and generate the nonlinear warp to template space prior to running *afni_proc.py*. This is a useful approach for running and evaluating a computationally expensive procedure before follow-on processing. If we end up running the *afni_proc.py* processing more than once or in a parallel processing stream, the warp estimation for this particular T1w dataset need not be recalculated. Fig. 7C shows the QC images for the anatomical-to-template alignment in the APQC HTML, which in this case match those of *sswarper2* (Fig. 4B). The pre-calculated warps have been passed to *afni_proc.py* with "-tlrc_NL_warped_dsets ..". For nonhuman primate (NHP) and other animal processing, the *@animal_warper* program for skull removal and nonlinear warp estimation (Jung et al., 2021) can be integrated in exactly the same way.

---

6 Instead of using reverse phase-encoded datasets, one could estimate warp from a phase map dataset, such as using AFNI's epi_b0_correct.py. The resulting warp could be included in afni_proc.py via "-blip_warp_dset", and would similarly populate the implicit blip processing block, etc.

## Ex. 1: QC images of raw EPI data and alignment stages via afni_proc.py

A) The original EPI volume (reference for motion correction and EPI-anatomical alignment)

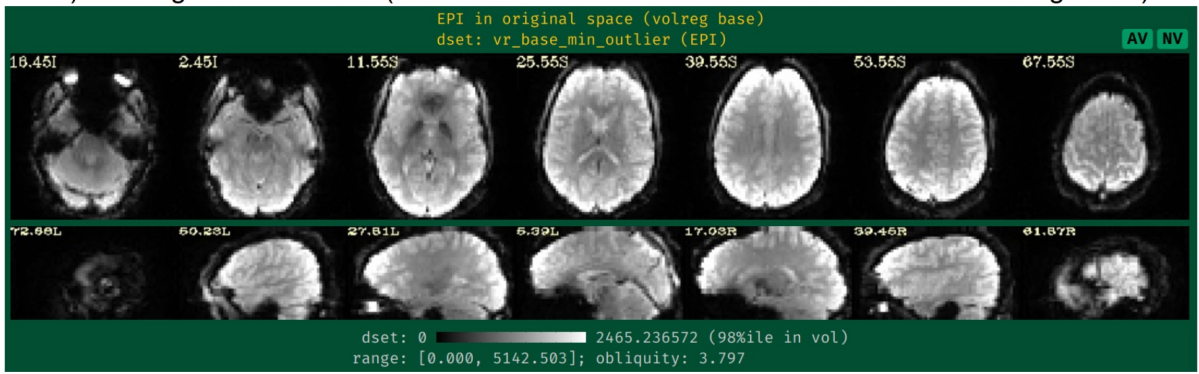

B) EPI-anatomical (affine) alignment: T1w anatomical edges overlaid on the EPI volume

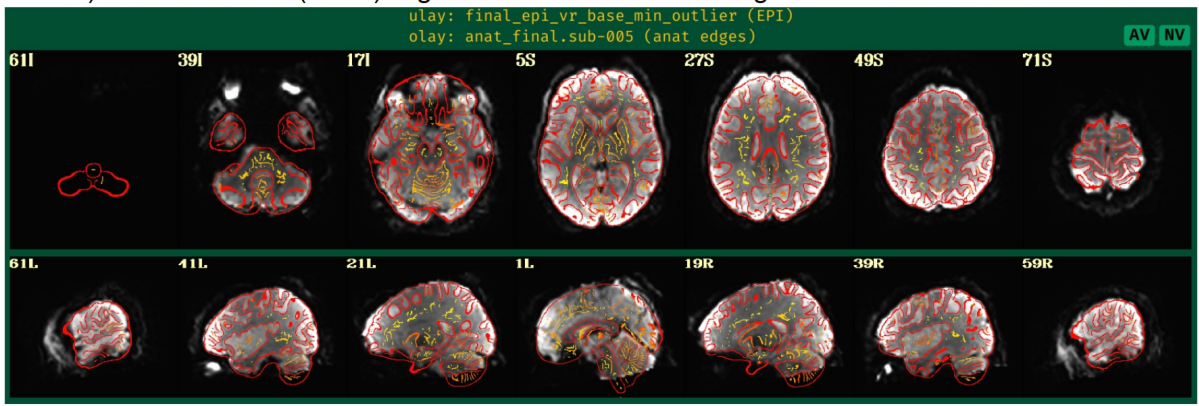

C) MNI reference volume edges overlaid on the nonlinearly warped+skullstripped T1w volume

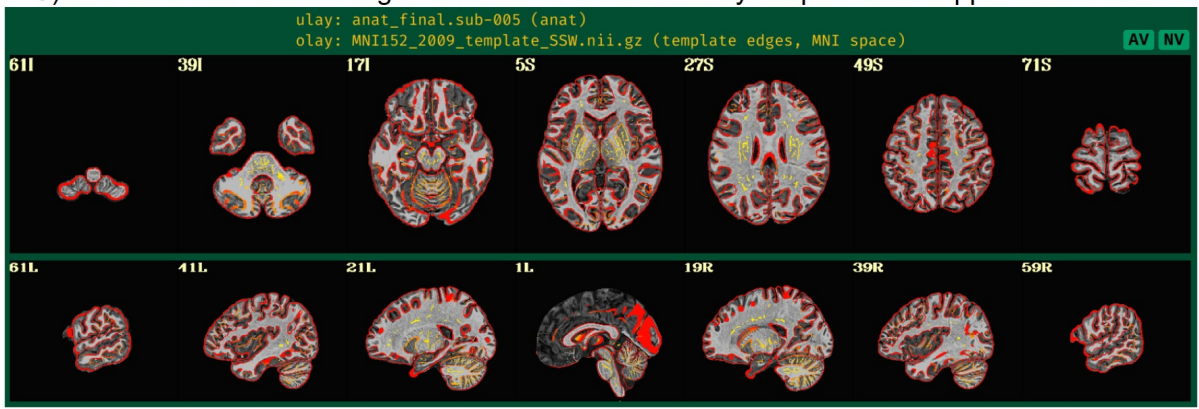

*Figure 7.* A selection of QC images generated by afni_proc.py for Ex. 1, which focuses on alignment-related steps of preprocessing. Panel A shows one EPI volume in original view (specifically, the one used as a reference for motion correction and EPI-anatomical alignment) to check coverage, tissue contrast, etc. Panel B shows the underlaid EPI and overlaid edges of the anatomical volumes after affine alignment. Here, after blip up/down correction, the EPI shows greatly reduced B0 inhomogeneity distortion along the AP axis (cf the sagittal views; some of the bright regions are CSF), and the general matching of the sulcal and gyral features and other tissue boundaries is strong. Panel C shows the anatomical (underlaid) and reference template (overlaid, edges) volumes after nonlinear alignment. There can be local structural differences expected (particularly in situations where there are differing numbers of sulci and gyri), but again the general matching of structural features is quite high.

In the volume registration block ("volreg"), each EPI time point is aligned to a reference using rigid-body alignment with *3dvolreg*. The resulting motion parameter time series will be concatenated with the other transforms later in the processing. In the full processing examples below, we describe how it can also be used for both motion censoring criteria and in the regression modeling. The "-volreg_align_to .." option allows the user to specify which EPI time point should be used as a reference volume. There are many considerations for this, but in general one would like to ensure that the reference volume itself is not corrupted by motion. To accomplish this in a general fashion, we recommend using the "MIN_OUTLIER" keyword with this option, so that the time point selected is that which has the fewest outliers in the input EPI time series; that choice seems to be the most generally reliable, with minimal risk of being corrupted in a given subject. The same volume selected for motion correction reference is also used for EPI-to-anatomical alignment. If the user would like to use a reference volume from a dataset other than the input EPI (e.g., using a separate, pre-steady state volume with higher tissue contrast), then the "-volreg_base_dset .." option could be used instead. Finally, we note that the output spatial resolution for the processed data can be specified here, with the "-volreg_warp_dxyz .." option. If this option is not used, the data will be output at an isotropic spatial resolution slightly higher than the input EPI's minimal voxel dimension.[7]

In the EPI-to-anatomical alignment block ("align"), linear affine registration with 12 degrees of freedom is performed between the anatomical dataset and the same EPI reference volume used in motion correction. These volumes typically have differing/opposite tissue contrast, so an appropriate cost function must be selected to drive the alignment optimization. AFNI's local Pearson correlation (lpc) cost function, specified here with "-align_opts_aea .." has been shown to generally provide excellent alignment for these cases (Saad et al., 2009); the "+ZZ" provides extra stability, since EPIs can be variously noisy, inhomogeneous and distorted. In some protocols, applied contrast agents such as MION can alter the EPI tissue contrast so that it matches that of the T1w volume. In such cases of matching contrast, the related "lpa" or "lpa+ZZ" cost function would be recommended. In some scenarios where the EPI volume has minimal tissue contrast, the "nmi" cost function may be a useful alternative. For other specialized scenarios, there are other cost functions that can be tried.

In addition to distortions, EPI volumes can have other non-ideal properties that affect alignment. Recently, *3dLocalUnifize* was developed and added to AFNI, to help deal with EPI volumes that have notable brightness inhomogeneity patterns that effectively change or greatly reduce tissue contrast. This program creates a brightness-homogenized version of a dataset while still preserving structural patterns, creating an intermediate dataset that is used for the alignment only. It is invoked within *afni_proc.py* by "-align_unifize_epi local", as used here. We find that this option provides general stability and typically improves co-registration for human datasets even if they are non-inhomogeneous; for non-human datasets, it is not currently recommended, as there tend to be more significant non-brain features present in the FOV with which this functionality interacts. The "-giant_move" sub-option opens the parameter search space for alignment parameters, which is useful when the EPI and anatomical do not have a strong initial overlap (even though in this case, the datasets overlay well). The "-check_flip" sub-option is implemented to check for instances when the EPI and anatomical might be relatively left-right flipped to each other; while this sounds like an odd concern, this functionality has found such orientation errors in datasets contained in major public repositories such as FCON-1000, ABIDE and OpenfMRI/OpenNeuro (Glen et al., 2020; Reynolds et al., 2023).

---

[7] In general we would not recommend upsampling to a much smaller grid size. It might make the final results appear smoother or to have finer features, but one cannot create new information for more detailed features by decreasing voxel size for a subject. It also incurs a large cost of computing resources.

Ex. 2:  Full task-based FMRI processing, with amplitude modulation

In this example, we demonstrate the full subject-level processing of a task-based FMRI dataset—that is, through the regression modeling block—with *afni_proc.py* (Fig. 8, or run: afni_proc.py -show_example "AP publish 3b"). The processing here would apply to a standard voxelwise analysis, given the presence of the "blur" block. The main facets of the "volreg", "align" and "tlrc" block were described above in Ex. 1, and apply equivalently here. As in Ex. 1, *sswarper2* was run prior to *afni_proc.py*, and both its skullstripping and warping results are imported. An alignment-based difference from Ex. 1 is that this dataset does not have a pair of opposite phase encoding datasets for "blip" block geometric adjustment (though it could, if such were available for this subject).

Since this example will focus more on time series modeling, the "-volreg_compute_tsnr yes" option has been added to include an image of the TSNR immediately after motion correction has been applied to the EPI time series, prior regression modeling; the TSNR after the GLM is already included by default. An example of the utility of this can be seen in Fig. 9B, where ghosting in the dataset is apparent. This means that one must be particularly cautious interpreting statistical results in the frontal regions where signal overlaps onto the brain. As noted above, just masking data would hide this circumstance and potentially lead to erroneous conclusions from just viewing GLM outputs. Panel A of the same figure shows raw EPI data, which indeed shows an extremely tight FOV for the acquisition; typically more empty space around the brain will help prevent such issues.

```
# AP Example 2, which was preceded by running:
# + timing_tool.py : creates stimulus timing files times.*.txt from TSV
# + sswarper2      : creates skullstripped anatomical ${anat_cp}, and
#                    nonlinear warps ${dsets_NL_warp}

afni_proc.py                                                              \
    -subj_id                   ${subj}                                    \
    -dsets                     ${dset_epi}                                \
    -copy_anat                 ${anat_cp}                                 \
    -anat_has_skull            no                                         \
    -anat_follower             anat_w_skull anat ${anat_skull}            \
    -blocks                    tshift align tlrc volreg mask blur scale   \
                               regress                                    \
    -radial_correlate_blocks   tcat volreg regress                        \
    -tcat_remove_first_trs     0                                          \
    -tshift_opts_ts            -tpattern alt+z2                           \
    -align_unifize_epi         local                                      \
    -align_opts_aea            -giant_move -cost lpc+ZZ -check_flip       \
    -tlrc_base                 ${template}                                \
    -tlrc_NL_warp                                                         \
    -tlrc_NL_warped_dsets      ${dsets_NL_warp}                           \
    -volreg_align_to           MIN_OUTLIER                                \
    -volreg_align_e2a                                                     \
    -volreg_tlrc_warp                                                     \
    -volreg_warp_dxyz          3.0                                        \
    -volreg_compute_tsnr       yes                                        \
    -mask_epi_anat             yes                                        \
    -blur_size                 6                                          \
    -blur_in_mask              yes                                        \
    -regress_stim_times        ${sdir_timing}/times.CONTROL.txt           \
                               ${sdir_timing}/times.TASK.txt              \
    -regress_stim_labels       CONTROL TASK                               \
    -regress_stim_types        AM1                                        \
    -regress_basis_multi       'dmUBLOCK(-1)'                             \
    -regress_motion_per_run                                               \
    -regress_censor_motion     0.3                                        \
    -regress_censor_outliers   0.05                                       \
    -regress_compute_fitts                                                \
    -regress_fout              no                                         \
    -regress_opts_3dD          -jobs 8                                    \
                               -gltsym 'SYM: TASK -CONTROL'               \
                               -glt_label 1 T-C                           \
                               -gltsym 'SYM: 0.5*TASK +0.5*CONTROL'       \
                               -glt_label 2 meanTC                        \
    -regress_3dD_stop                                                     \
    -regress_reml_exec                                                    \
    -regress_make_ideal_sum    sum_ideal.1D                               \
    -regress_est_blur_errts                                               \
    -regress_run_clustsim      no                                         \
    -html_review_style         pythonic
```

*Figure 8. The afni_proc.py command for Ex. 2 (task-based, single echo FMRI, full processing). Options with gray background have already been described earlier in Ex. 1 here, and any variables described in the captions of Fig. 5. ${blur_size} is the FWHM size of applied blur, in mm; ${sdir_timing} is the*

*directory containing stimulus timing files. Running this command produces a commented script of >640 lines, encoding the detailed provenance of all processing.*

**Ex. 2: QC images of raw EPI data and TSNR (prior to GLM)**

A) Original EPI volume, which shows a tight FOV and potential ghosting

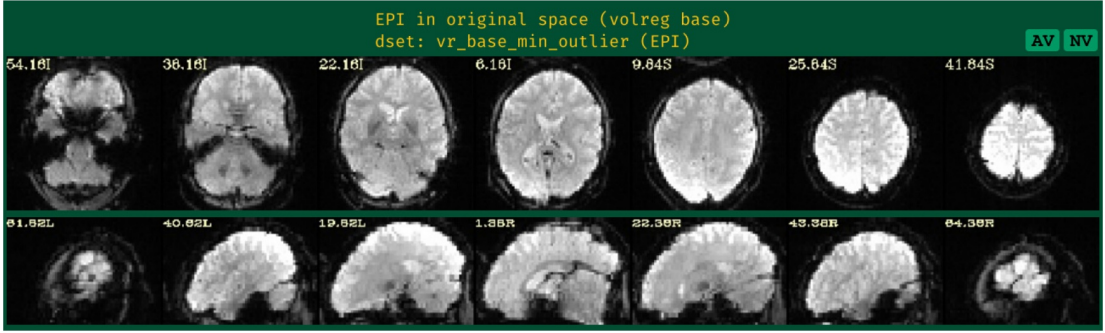

B) TSNR (after volreg processing block; before GLM), which more clearly shows ghosting

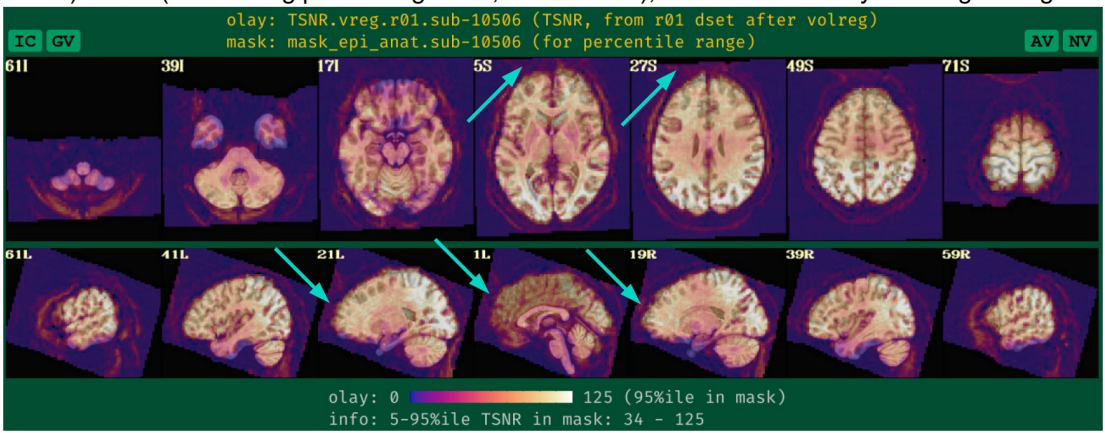

*Figure 9. QC images generated by afni_proc.py for Ex. 2, showing: A) the raw EPI volume in native space; and B) the unmasked TSNR after the volreg processing block, prior to regression modeling. The unmasked TSNR image shows evidence for ghosting artifact overlapping into the brain (cyan arrows); as described in Fig. 2, this shows the benefits of not masking data during processing to understand it better and more reliably evaluate it. The fact that the EPI has been acquired with such a tight FOV (see panel A) likely contributes to the presence of ghosting. The TSNR map also shows the presence of EPI distortion (the anterior TSNR pattern extends beyond the anatomical boundaries, even though structural alignment is good).*

This afni_proc.py command includes some additional QC-specific options, and these will add or modify features in the output APQC HTML. For example, "-radial_correlate_blocks .." specifies a list of blocks for which images will be created of local radial correlation (via AFNI's @radial_correlate). That is, a dataset will be calculated where each voxel's time series will be correlated with the Gaussian-weighted average of surrounding voxels' times series over a large radius (20 mm half width at half max, by default); such a dataset has been shown to be useful in revealing scanner- or motion-related artifacts (Taylor et al., 2024). The "-html_review_style" option allows the user whether to use an older, simple program for line plots within the HTML ("basic" style), or a more modern version that contains extra information but requires having Python's Matplotlib library installed ("pythonic" style). At present, the program will check Python dependencies to be able to run the latter style, to provide more informative plots by default.

When slice timing information is available in the EPI data, it can be applied to shift the input time series datasets appropriately by including the "tshift" block, as in this example. The user can supply various options to the 3dTshift command that will be used in the proc script via "-tshift_opts_ts ..". In this case, the timing pattern is specified. With other "-tshift_* .." options, users can control the interpolation method or the part of the TR to which shifted times should be aligned.

The "mask" block leads to the estimation of a whole brain mask, or as close to one as the data allow. Importantly, the mask is not applied directly to the FMRI data (zeroing out much of the field of view, FOV), but instead is reserved for use with various calculations. While most studies focus their investigations within the brain only, it is quite helpful to see EPI data across the whole FOV in order to be aware of possible distortions, noise values, and other properties within the data (see Taylor et al. (2023b), for example). One might calculate mean quantities within the mask, such as average TSNR to report, as well as combine it with similar masks across all subjects to create a group-level mask. The "-mask_epi_anat yes" option added here tightens the EPI mask by intersecting it with the anatomical mask, which is typically done to improve specificity.

The "blur" block is also included in the processing, which is standard for FMRI data that will be analyzed voxelwise, in order to boost local SNR though at the cost of spatial specificity. While the selected amount of blur can vary based on application (including not blurring, in the case of ROI-based analysis), a general guideline for single echo FMRI inputs might be to use a "-blur_size .." that is about 1.5-2 times the minimum voxel dimension. The present EPI voxels are 3x3x4 mm3, and therefore the selected blur has 6 mm FWHM. For multi-echo FMRI data, one might prefer minimal blurring, because the TSNR after combining echos tends to be much higher. There are many different styles of blurring that can be applied (see Ex. 4 below for surface-based smoothing, and supplementary Ex. 6 in Appendix C for blurring data to an average amount).

Including the "scale" block leads to an important time series feature: the coefficients (or effect estimates) from the regressors of interest will then have meaningful units of BOLD percent signal change based on per-voxel baseline scaling. This method of voxelwise scaling has been shown to be useful in interpreting results and for promoting more detailed comparisons across studies (Chen et al., 2017). Note that other software may provide other formulations for scaling, which will have different interpretations and properties. We typically recommend including the scale block, particularly in task-based FMRI studies. This is in line with both the aforementioned paper and Chen et al. (2022), who further argued that doing so provides results with more valuable information both for quality control (QC) and analysis.

The "regress" block contains a number of important processing options related to the subject-level general linear model (GLM). This block typically contains the largest number of detailed specifications for the processing, as it produces the main outputs at the single-subject level. In the current block, there are two criteria set for censoring time points during regression—that is, removing specific time points from influencing the model (which is referred to as "scrubbing" in some software). First, an outlier-based criterion is used via "-regress_censor_outliers 0.05", so that volumes whose brain masks contain more than 5% temporal outliers from the input time series will be censored. Additionally, a motion censoring criterion is based on the Euclidean norm (Enorm) of the first difference of the EPI motion parameter time series, in this case where the magnitude of change in Enorm>0.3, which has approximate units of mm ("-regress_censor_motion 0.3"). Since this motion criterion is based on the difference of parameters, the volumes at both flagged time points are censored for suprathreshold estimates. The Enorm is the square root of the sum of squares (or L2-norm) of the motion parameter differences, similar to how standard distance metrics are formulated, making it more sensitive to a large change in any single component than an L1-norm, such as the framewise displacement (FD) parameter. Fig. 10A shows the Enorm and outlier plots from the APQC HTML, as well as part of the individual motion parameter plots; censor thresholds (cyan lines) and suprathreshold locations (red fields) are both displayed on those plots and on subsequent line plots within the HTML.

In the present task paradigm, the modeling of the memory and button-response includes duration modulation (DM), to allow for stimulus events of varying duration, whose values are also encoded in the stimulus timing files. There are two stimulus classes here, with the timing files and labels provided respectively by "-regress_stim_times .." and "-regress_stim_labels ..". Each event duration here is convolved as a boxcar with the BOLD impulse response function to create event responses of varying shape and magnitude within the idealized hemodynamic response function (HRF). Amplitude modulation (AM; also referred to as "parametric modulation" in some software) might be an alternate choice. The duration modulation function has two specifications:[8] its shape, which here is "BLOCK"; and its length scale, which for the present task is 1s (the negative sign is a syntax convention, see the online description in the footnote); by convolution, a stimulus of 1 s duration would then have unit magnitude in the units of the scaled regressor, and longer responses would have a larger magnitude, up to the limits of the basis function. The "idealized" response curve for each stimulus from the APQC HTML is shown in Fig. 10C, along with the location of any censoring (see next paragraph). Note that each stimulus class can have its own basis function (which is what the "multi" in the "-regress_basis_multi .." option refers to). There are many choices for how to model events, such as whether to use a fixed HRF or to fit the shape from the data, and these can greatly affect results; this is described more in the Discussion.

Beyond just modeling the events, one can also evaluate hypotheses about their effects as general linear tests, using "-regress_opts_3dD" to specify the desired tests for 3dDeconvolve to run. In this analysis, the two general linear tests evaluated are a basic "Task - Control" contrast and the mean "0.5 (Task + Control)" response, given the labels "T-C" and "meanTC," respectively. We note that the topic of HRF modeling itself is quite large, and it depends heavily on the study and task details. More basis function considerations and options are presented in the Discussion. Fig. 11 shows the regression modeling results as displayed in the APQC HTML. Statistical maps are used as threshold datasets, and where possible, effect estimates are used as the overlay (color) dataset. To retain useful information while thresholding and to better perform QC, transparent thresholding is applied to highlight regions of high significance while still showing results throughout the full field of view, even outside the brain (Allen et al., 2012; Taylor et al., 2023b).

Finally, there are three other points to note about the regress block options. Even though there is only one time series in this dataset, we include the "-regress_motion_per_run" flag, to highlight that in cases of concatenating multiple runs one can better account for motion-correlation variance. This option has no particular effect when there is only one input time series, but we typically leave it in as a practical default to not forget it in other scenarios. Secondly, 3dREMLfit is used to account for serial correlation in the time series residuals, using a generalized least squares estimation of temporal autocorrelation ("-regress_reml_exec").  This generalizes the default estimation functionality of 3dDeconvolve, allowing simultaneous estimation of the beta coefficients in the model with estimating temporal correlation and variance. Finally, "-regress_est_blur_errts" flags the processing script to estimate smoothness of the residual time series (the "errts*" dataset) with *3dFWHMx*, which can be useful for QC considerations as well as for some group analyses that use clustering.

---

8 More details on the DM block choices and the dmUBLOCK() function are provided in AFNI's online documentation here:
https://afni.nimh.nih.gov/pub/dist/doc/htmldoc/statistics/deconvolve_block.html

**Ex. 2: QC images of motion plots and stimulus modeling**

A) Enorm and outlier fraction across time, with censoring (in red)

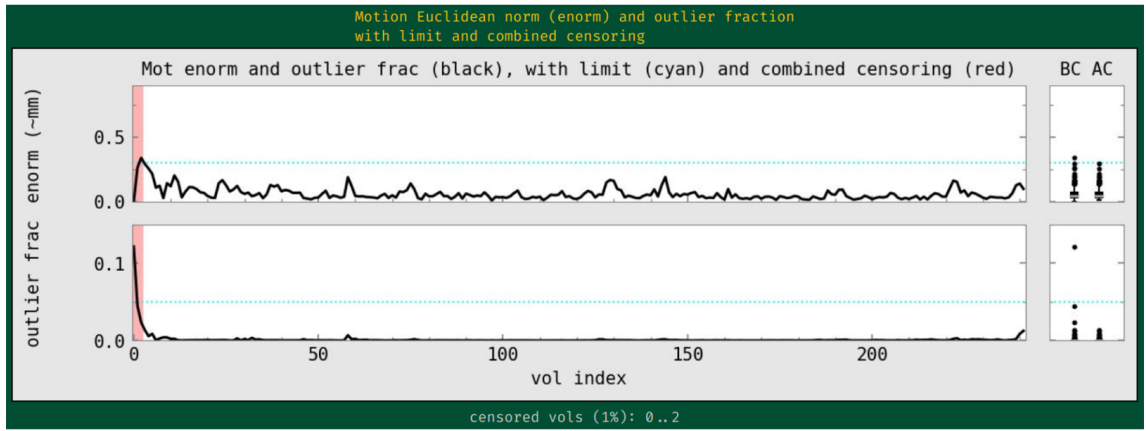

B) Individual motion parameter plots across time (showing only 2 of 6 here)

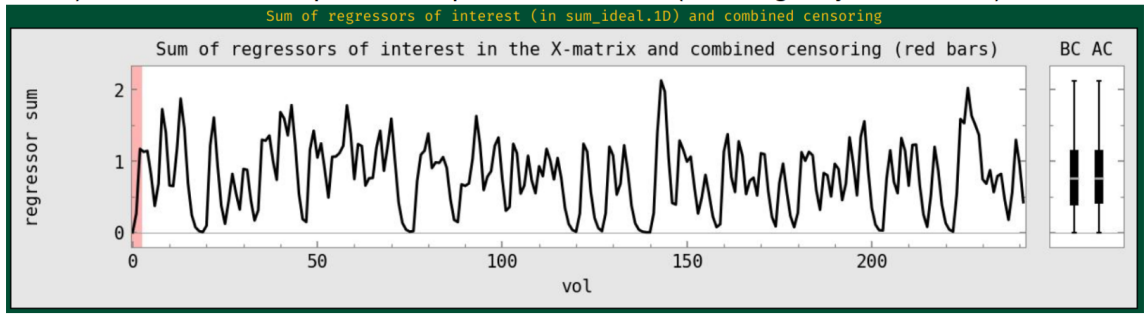

C) Convolved (ideal) task regressors, for each stimulus

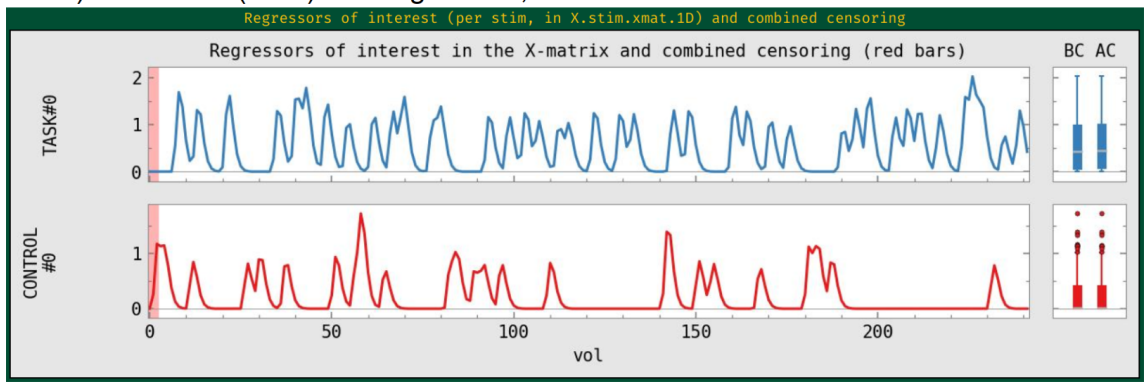

*Figure 10. QC images generated by afni_proc.py for Ex. 2, focused on the motion and regression model setup when processing task-based FMRI. Panel A shows the Enorm and outlier fraction plots across time, which are used for time point censoring. The dashed lines show the thresholds for each quantity, and the red bands highlight the location of any volumes to be censored (here, only 3 volumes are censored). The "BC" and "AC" boxplots show distributions of each plotted parameter before and*

*after censoring, respectively. The lower two panels show the "ideal" stimulus response based on the timing and chosen hemodynamic response function (HRF): B shows the sum of responses, and C shows each individual stimulus class. The red band of censoring is also displayed here, to reveal any cases of stimulus-correlated motion (which is also checked automatically in the "warns" section of the APQC HTML).*

### Ex. 2: QC images of regression modeling: effect estimates and statistics

A) Full F-stat of the regression modeling (both overlay and translucent threshold)

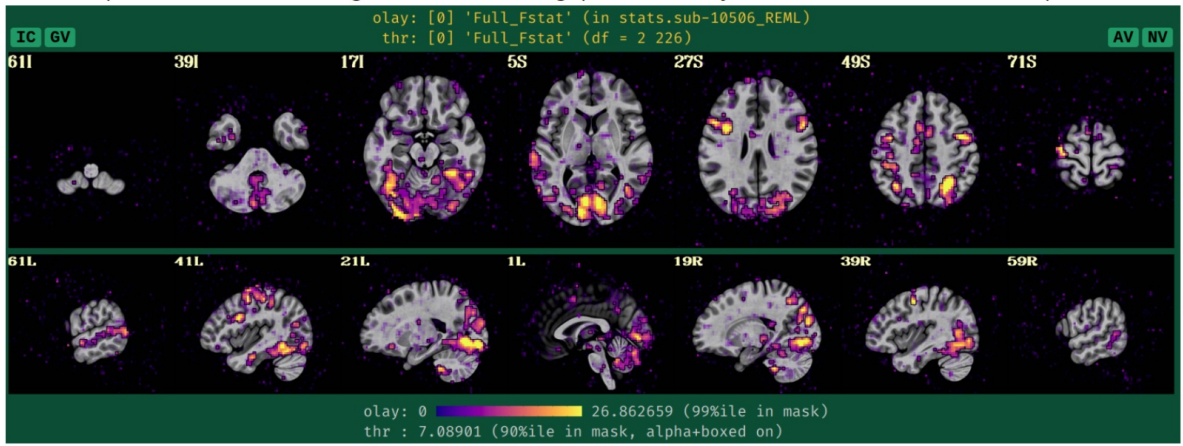

B) GLT contrast "Task – Control" effect estimate (overlay) and T-stat (transparent threshold)

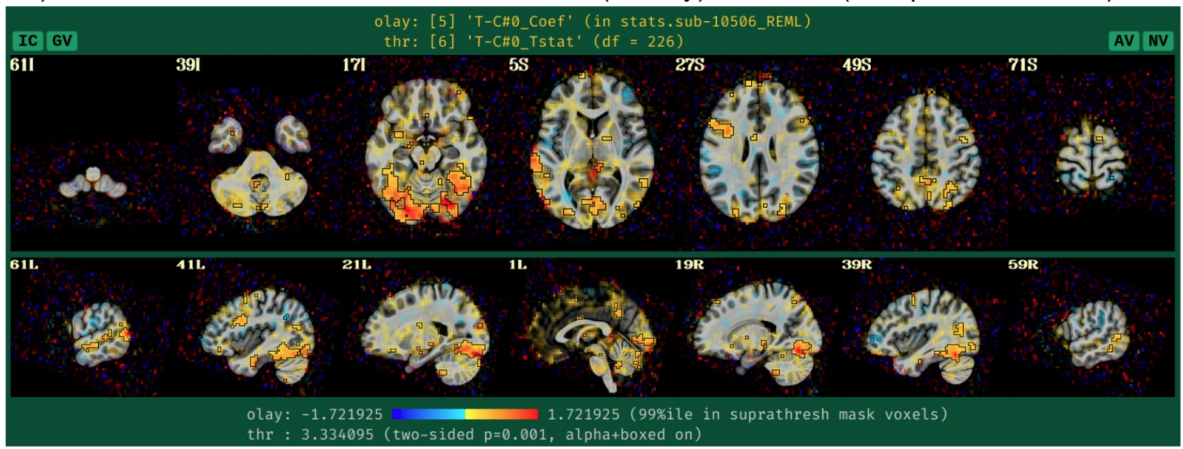

C) GLT contrast "0.5(Task+Control)" effect estimate (overlay) and T-stat (transparent threshold)

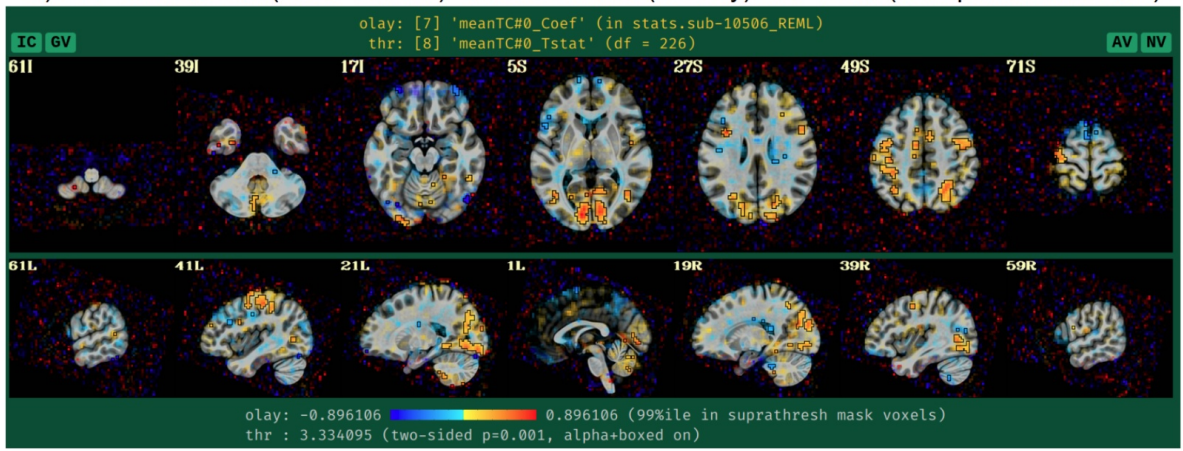

*Figure 11. QC images generated by afni_proc.py for Ex. 2, focused on evaluating the task-based regression modeling results. In each panel, the statistic value is used for thresholding in a translucent fashion: suprathreshold locations are opaque and outlined, and subthreshold locations are increasingly translucent. The overlay color is the accompanying effect estimate coefficient where available (panels B and C). Panel A exhibits the full F-stat, which shows the relative quality of model fit. Panels B and C show the two contrasts specified in the afni_proc.py command. In all cases, modeling results outside the brain are shown, for more complete evaluation and understanding of the processing results (Taylor et al., 2023b).*

```
# AP Example 3, which was preceded by running:
# + FreeSurfer's recon-all: estimates subject-specific parcellation
# + @SUMA_Make_Spec_FS    : creates NIFTI format parcellation, and tissue-
#                           based subsets like GM regions ${roi_gmr_2009}
# + sswarper2             : creates skullstripped anatomical ${anat_cp},
#                           and nonlinear warps ${dsets_NL_warp}
# + physio_calc.py        : estimates physiological (respiratory and
#                           cardiac) regressors ${physio_regs}

afni_proc.py                                                              \
    -subj_id                        ${subj}                               \
    -dsets                          ${dset_epi}                           \
    -copy_anat                      ${anat_cp}                            \
    -anat_has_skull                 no                                    \
    -anat_follower                  anat_w_skull anat ${anat_skull}       \
    -anat_follower_ROI              aagm09 anat ${roi_gmr_2009}           \
    -anat_follower_ROI              aegm09 epi  ${roi_gmr_2009}           \
    -ROI_import                     BrodPijn ${atl_brod}                  \
    -ROI_import                     SchYeo7N ${atl_sy7n}                  \
    -blocks                         ricor tshift align tlrc volreg mask blur \
                                    scale regress                         \
    -radial_correlate_blocks        tcat volreg regress                   \
    -tcat_remove_first_trs          4                                     \
    -ricor_regs                     ${physio_regs}                        \
    -ricor_regs_nfirst              4                                     \
    -ricor_regress_method           per-run                               \
    -align_unifize_epi              local                                 \
    -align_opts_aea                 -cost lpc+ZZ -giant_move -check_flip  \
    -tlrc_base                      ${template}                           \
    -tlrc_NL_warp                                                         \
    -tlrc_NL_warped_dsets           ${dsets_NL_warp}                      \
    -volreg_align_to                MIN_OUTLIER                           \
    -volreg_align_e2a                                                     \
    -volreg_tlrc_warp                                                     \
    -volreg_warp_dxyz               3                                     \
    -volreg_compute_tsnr            yes                                   \
    -mask_epi_anat                  yes                                   \
    -blur_size                      5                                     \
    -regress_motion_per_run                                               \
    -regress_make_corr_vols         aegm09                                \
    -regress_censor_motion          0.2                                   \
    -regress_censor_outliers        0.05                                  \
    -regress_apply_mot_types        demean deriv                          \
    -regress_est_blur_epits                                               \
    -regress_est_blur_errts                                               \
    -regress_compute_tsnr_stats     BrodPijn 7 10 12 39 107 110 112 139   \
    -regress_compute_tsnr_stats     SchYeo7N 161 149 7 364 367 207        \
    -html_review_style              pythonic
```

*Figure 12. The afni_proc.py command for Ex. 3 (resting state, single echo FMRI, full processing). Options with gray background have already been described in earlier examples here, and any variables described in the captions of Figs. 5 and 8. ${sdir_timing} is the directory containing stimulus timing files. Running this command produces a commented script of >740 lines, encoding the detailed provenance of all processing. Two additional atlases are imported here, for extracting ROIs for checking TSNR and shape properties: "BrodPijn" is the Brodmann atlas (1909) digitized by Pijnenburg et al. (2021); and*

*"SchYeo7N" is the refined version of the 7-network, 400 parcellation Schaefer-Yeo atlas (Schaefer et al., 2018; Glen et al., 2021).*

Ex. 3:  Full resting state FMRI: volumetric, voxelwise analysis

We now show an example of full resting state processing with single-echo FMRI data, for voxelwise analysis (Fig. 12, or run: *afni_proc.py -show_example "AP publish 3c"*). Again, both the skullstripping and nonlinear warping from sswarper2 have been integrated. Unlike previous examples, this command includes local regressors that were estimated from separately measured physiological time series, as described below. It also imports results from FreeSurfer's recon-all on the subject's anatomical T1w dataset. The reverse phase encoding datasets from Ex. 1 could easily be applied here with the same "-blip_*" options, as well. We leave the blip correction out in this case and users can view the difference between applying the B0 distortion correction and not doing so; note that the amount of difference including vs ignoring this step will have for a dataset will depend strongly on the scanner and acquisition details being used.

There are several situations where it can be useful to add pre-calculated masks or ROI atlas maps into the processing stream, each of which ends up in the final space (here, MNI) on either the EPI or anatomical grid, as specified. Here, the "-anat_follower_ROI .." option is used to bring anatomical parcellation datasets FreeSurfer's recon-all into the processing stream. For each imported dataset, the user assigns a brief label for working with the dataset within the code and also designates the final grid. Here, we are bringing in the gray matter (GM) map from FreeSurfer's "2009" parcellation (Destrieux et al., 2009), which is used twice: one copy will have "epi" grid spacing (label = "aegm09") and one will have "anat" grid spacing (label = "aagm09"). These could be imported as part of processing for an ROI-based analysis, for example (in which case we would remove the blur block in the afni_proc.py command), but in the present case they will be used only for QC-related purposes.

The "ricor" block is used to include regressors that have been estimated from physiological time series, which were measured during the FMRI acquisition. These typically include slicewise RETROICOR regressors (Glover et al., 2000; Chang and Glover, 2009), as well as volumetric respiration volume per time (RVT; Birn et al., 2006) regressors. In the present case, both cardiac and respiratory traces were acquired, so that 8 slicewise regressors were calculated along with 5 volumetric ones (shifted copies of RVT). These sets of regressors were calculated with AFNI's physio_calc.py prior to running afni_proc.py. Fig. 13A displays the QC image of peak- and trough- estimation of a respiratory time series during that processing. Panel B in the same figure shows the 8 cardiac and respiratory regressors based on RETROICOR that get applied to the FMRI data (5 RVT regressors, not shown, are also included). Because RETROICOR contains slicewise information, its regressors are applied to the un-warped data, so "ricor" is typically one the earliest processing blocks. Additionally, users can specify whether the regressors should be applied "per run" or "across runs (the former is selected here), which applies when multiple EPI datasets are present. The relative variance of the combined physiological regressors is shown in Fig. 13C, where the largest amounts are in the subcortical/inferior regions, as expected.

**Ex. 3: Incorporating physiological regressors from physio_calc.py ("ricor" block)**

A) Peak and trough estimates for respiratory time series (blue/red highlight shorter/longer intervals)

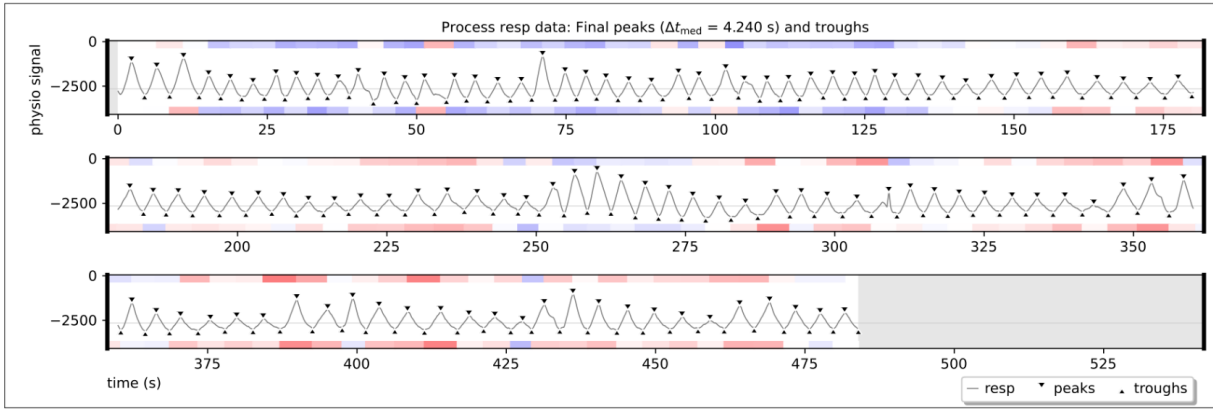

B) RETROICOR regressors for respiratory and cardiac time series

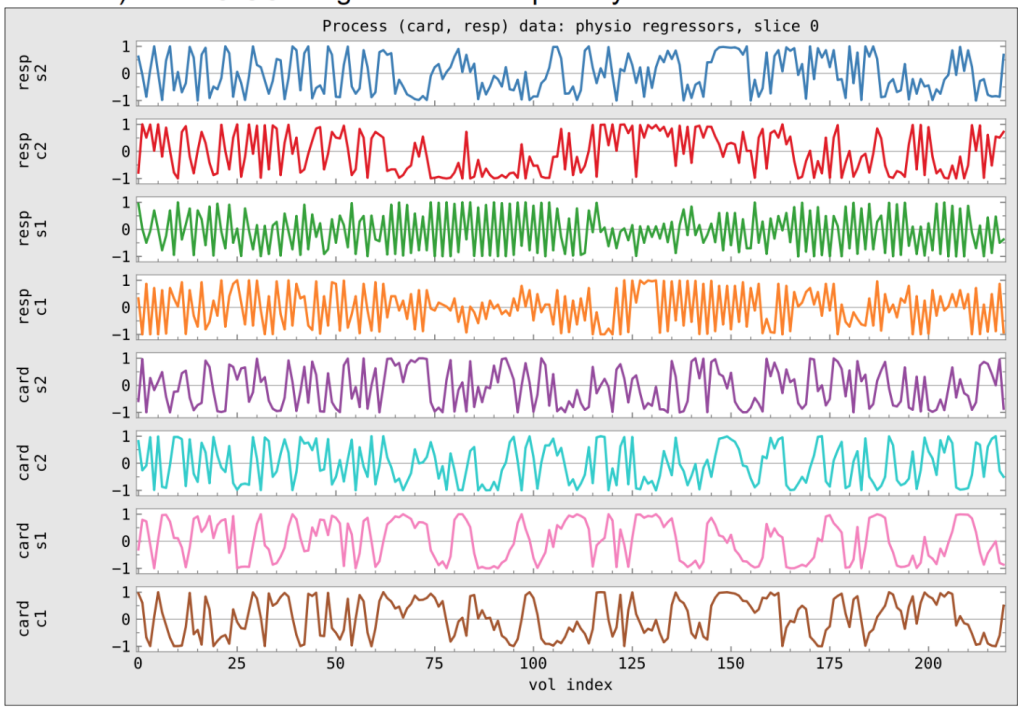

C) Relative variance explained by respiratory and cardiac physiological regressors ("ricor" block)

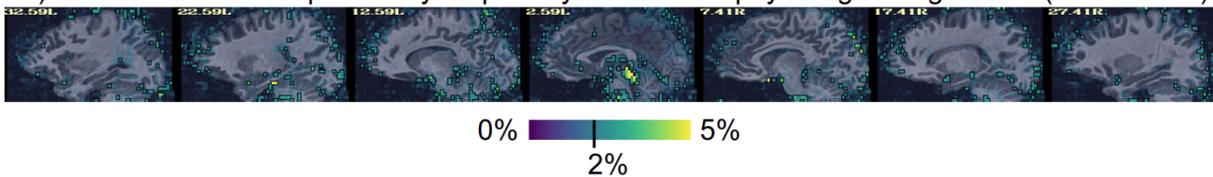

*Figure 13. Aspects of processing related to having respiratory and cardiac time series data included in FMRI processing in Ex. 3. The AFNI program physio_calc.py was run on these physiological time series, for which peak and/or trough detection is a first key step, shown in panel A for the respiratory data. The QC image shows the estimated peak and trough locations with triangles (which can be edited in the program's interactive mode, if necessary); the blue and red color bands reflect the relative intervals between pairs of each, which can help highlight potential algorithm problems. Panel B shows the final RETROICOR regressors estimated by physio_calc.py. These are included in a slice-wise manner within early afni_proc.py processing, along with 5 RVT regressors from the same program (not*

*shown). Finally, panel C shows a map of the fractional variance explained using the 13 physiological regressors, with highest values around the subcortical and inferior regions.*

It is worth noting that the way the physiological regressors are implemented relies on the ability of AFNI's 3dREMLfit to work with voxelwise regressors. Most FMRI regression modeling uses a single set of regressors across the whole volume, but inherently the RETROICOR regressors are calculated per slice. This, and other features such as the voxelwise regressors of ANATICOR (Jo et al., 2010), which are described in Appendix C, rely on this extended functionality.

This example includes both outlier- and motion-based censoring, similar to Ex. 2. In this case, a slightly stricter motion criterion (Enorm>0.2) is utilized, since resting state FMRI tends to be more susceptible to motion-based artifacts than task-based data (as long as the motion is not strongly stimulus-correlated).  Fig. 14A displays the Enorm and outlier fraction estimates for this subject, for which there were only 2 time points that reached threshold values for censoring. As is common for resting state FMRI, we also include both the estimated motion profiles and their derivatives in the regression model ("-regress_apply_mot_types demean deriv"), spending slightly more degrees of freedom (DFs) to try to reduce motion effects. The degree of freedom bookkeeping for this regression model is displayed in Fig. 14B), organized by category. It is worth noting that these EPI data display very little motion, and therefore censoring uses up very few degrees of freedom (<1%). However, in many cases censoring can use up a sizable fraction of degrees of freedom, and one must take care in overall model design to not use up too many degrees of freedom. The APQC HTML includes automatic checks for this.

**Ex. 3: QC images of motion and regression modeling tables**

A) Subject motion (Enorm) and outlier fraction estimates, with thresholds and censoring

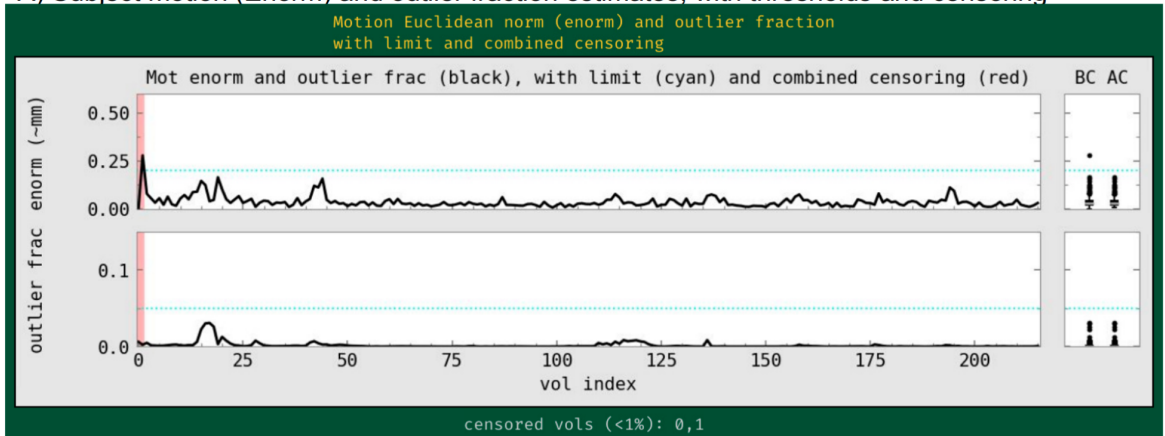

B) Statistical DF summary

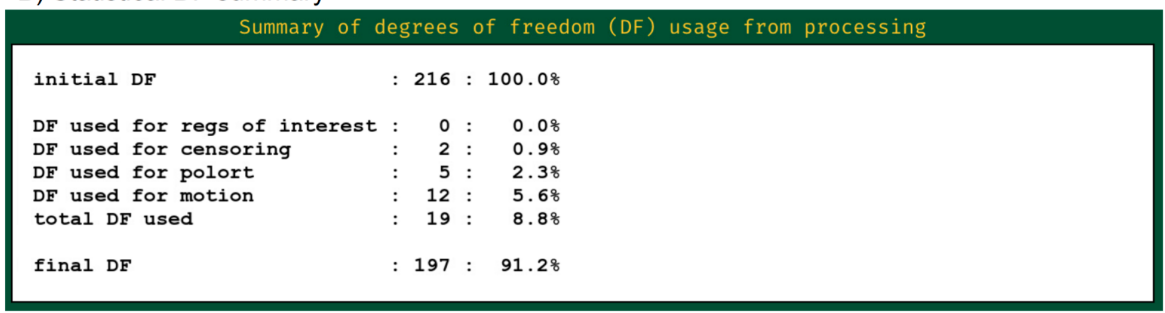

C) Statistical DF summary, with bandpassing included

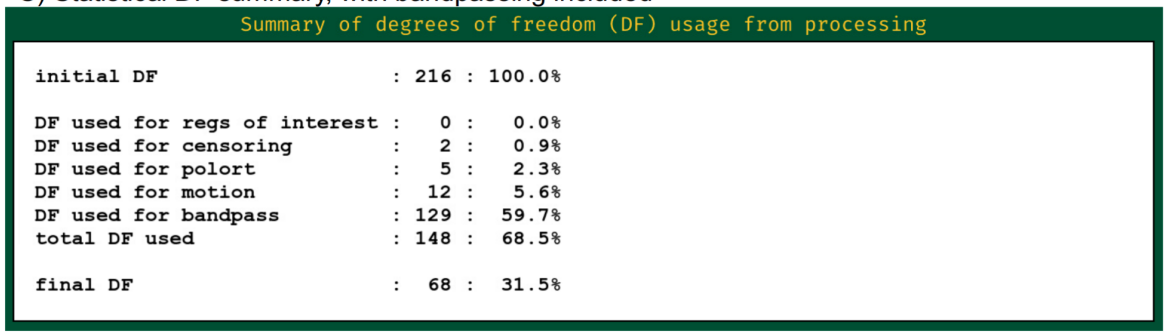

*Figure 14. QC images generated by afni_proc.py related to motion effects and regression modeling in Ex. 3 processing. Panel A shows the primary quantities that are used to assess subject motion and its effects: Enorm (Euclidean norm), which is approximately the amount of subject motion between time points, in mm; and outlier fraction. Users typically set thresholds for these quantities (horizontal blue lines) to determine which time points should be censored (highlighted in red). Panel B shows the degree of freedom bookkeeping for the regression model, organized by category of regressor. During modeling, data analysts must balance the removal of motion and other non-neuronal effects with the reduction of the statistical DF count. This example did not include bandpassing in processing, but Panel C shows the DF count if it were (see supplementary Ex. 5, in Appendix C). Note that bandpassing itself reduces the DF count by 60% of the original amount. Bandpassing can be problematic, particularly in cases of more subject motion.*

It is worth highlighting that this processing example does not include bandpassing, even though doing so to restrict the analyzed time series to "low frequency fluctuations" (LFFs) has historically been a widely implemented choice in resting state FMRI. The LFF band is typically 0.001-0.1 Hz or 0.01-0.1 Hz. It is certainly possible for afni_proc.py to include this within the regression model, using a single option to specify the interval of the frequency band to keep (e.g., "-regress_bandpass 0.01 0.1", in units of Hz). In fact, one can specify multiple bands to keep, simply by specifying multiple pairs of boundaries. But there are several caveats that should be noted about the typical bandpassing to the LFF range. Many of them deal with counting degrees of freedom (DFs) while processing, which is something that is unfortunately often overlooked in the field, along with how software must take care with how bandpassing is performed within the processing; see Appendix A and Caballero-Gaudes and Reynolds (2017).

The fractional loss of degrees of freedom for FMRI bandpassing can be approximated by either of the following simple formulas:

$$DF_{loss} = 1 - 2*TR*(ftop - fbot), \quad (1a)$$
$$DF_{loss} \approx 1 - 2*TR*ftop, \quad (1b)$$

where TR is the acquired EPI data's repetition time (in s), and ftop and fbot are respectively the chosen upper and lower bounds of the band (in Hz); when fbot is much smaller than ftop, as would be common for standard LFF bands, then the simpler form in Eq. 1b applies. In most resting state papers that bandpass to LFFs, ftop = 0.1 Hz, so that for TR = 2s one loses 60% of the degrees of freedom of the input data solely from the bandpass regressors. For FMRI with faster temporal sampling, the amount of loss grows: for TR = 1s, one loses 80% of the DFs from the standard bandpassing alone. These are huge fractions of the time series DFs to remove, even before considering the further removal of DFs via motion, baseline and censoring regressors. We note that afni_proc.py's processing applies bandpassing in a mathematically consistent way within the regression model. It also tabulates these features and reports them so that users are aware, with various warning levels. If such accounting is not done, users might not be aware of the severe loss of DFs, even using up more than 100% of them, which is often a risk due to subject motion.

For resting state and naturalistic FMRI, the main output of interest is the residual or "error" time series (errts) from the regression model. This is a notable difference from task-based data, where the effect estimates and statistics from the regressors of interest are the main output. In that case, the residuals should mostly contain the "noise" and other non-modeled parts of the data, and they are typically just used to help judge the quality of data fit. In resting state processing, evaluating the modeling job is more difficult because the conceptual separation of "signal" and "noise" does not materialize in the outputs. To help evaluate the final time series, the afni_proc.py QC HTML displays seed-based correlation maps of several major resting state networks[9] (Taylor et al., 2024): default mode network (DMN), visual and auditory networks (Fig. 15A), when the template space is known. These can reveal the presence of artifacts, signal dropout, problems with motion or regressors, and more. As above, transparent thresholding is applied. Additionally, effects of B0 inhomogeneity distortion are visible in the anterior regions (axial slices at Z=5S and 27S), which may be reduced by including distortion correct via fieldmaps or reverse-phase encoding-derived warps (see Ex. 1, above).

---

9 Seed locations for DMN, visual and auditory networks for several recognized template spaces, including MNI, Talairch-Tournoux, India Brain Templates (Holla et al., 2020), Haskins Pediatric (Molfese et al., 2020), and a number of non-human ones. This list is growing, and users can help to integrate others, as well.

### Ex. 3: QC images of regression modeling

A) Seed-based correlation map of the DMN (axial views only)

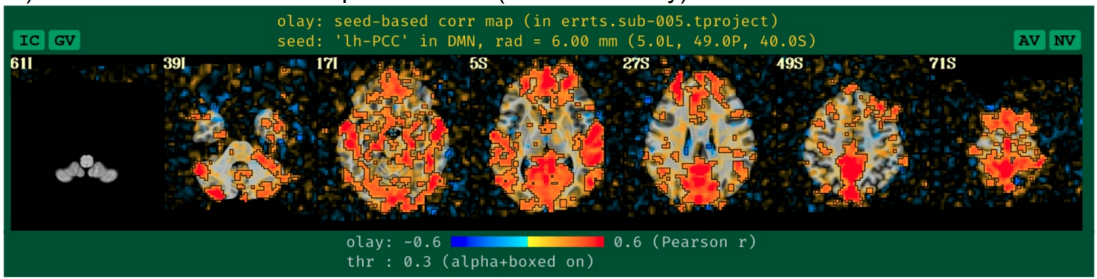

B) Seed-based correlation map of the visual network (axial views only)

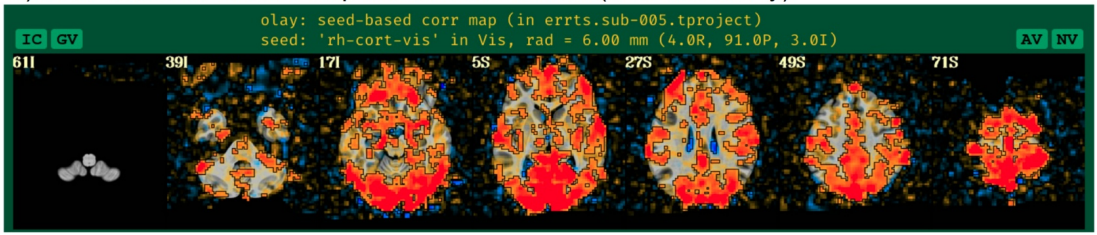

C) Seed-based correlation map of the auditory network (axial views only)

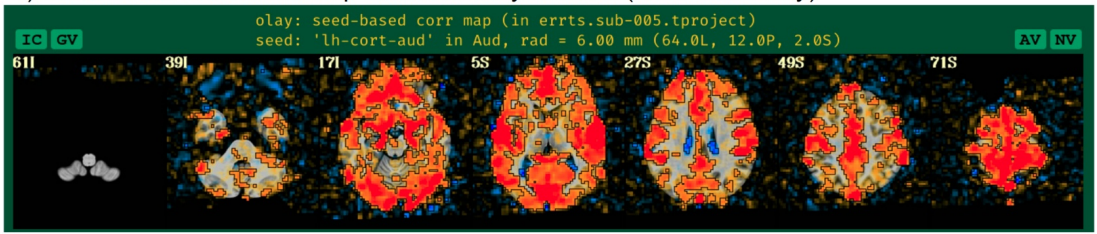

D) TSNR map (after regress block GLM)

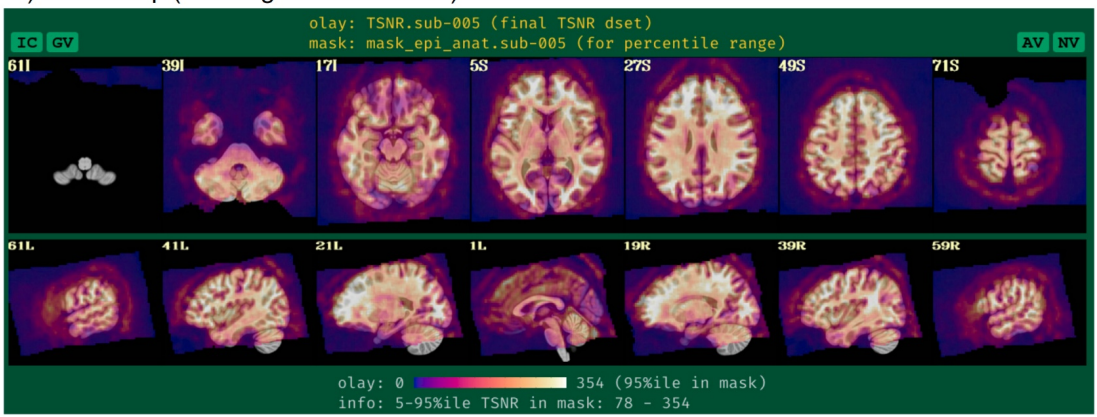

*Figure 15. QC images of statistical output for resting state time series, for which the residuals are the time series of interest (Ex. 3). Panels A-C show axial maps the three seed-based correlation maps shown in the APQC HTML when the final space is a known template: for the default mode network (DMN), the visual network and the auditory network. These allow for checks for artifacts and other potential problems from processing. Panel D displays the TSNR for this data, which can help distinguish regions of strong signal coverage from those with dropout or artifact.*

# Ex. 3: ROI TSNR and shape properties in APQC HTML warns block

A) TSNR/shape table for user-selected ROIs in Brodmann_pijn_afni.nii.gz atlas

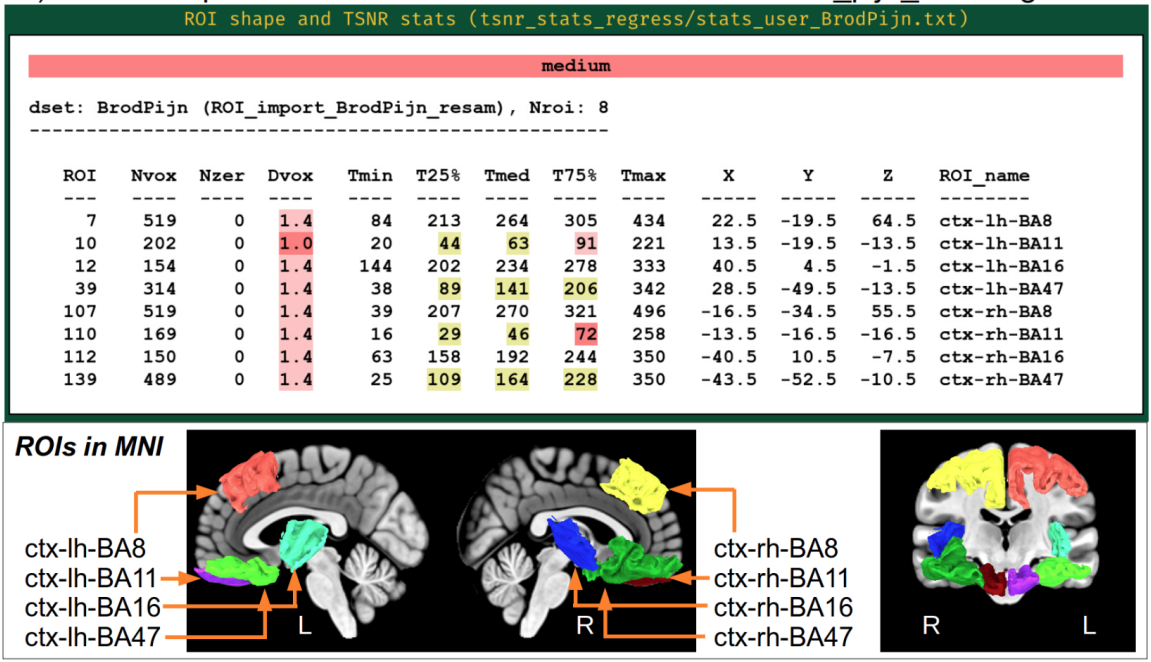

B) TSNR/shape table for user-selected ROIs in Schaefer_7N_400.nii.gz atlas

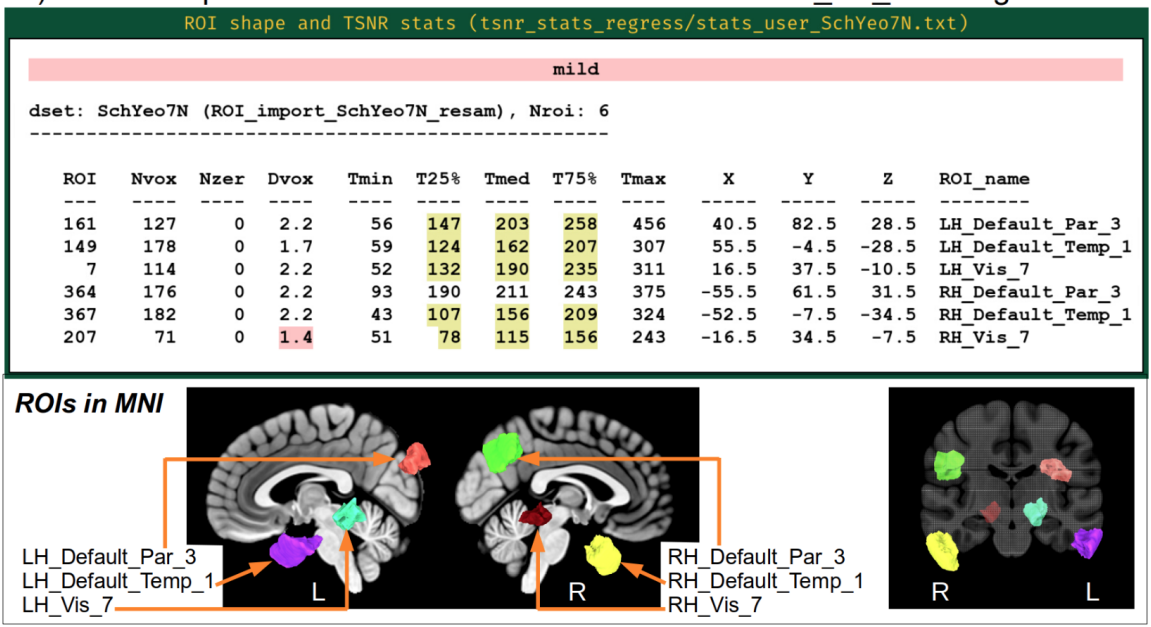

Figure 16. QC tables of ROI shape and TSNR properties of user-defined regions of interest (Ex. 3). The regions in panel A are defined in the Brodmann Atlas digitized by Pijnenburg et al. (2021), and those in panel B are from the refined version of the 7-network, 400 parcellation Schaefer-Yeo atlas (Schaefer et al., 2018; Glen et al., 2021). See Taylor et al. (2024) for details on the columns and warning levels, such as for narrow ROIs and low/unstable TSNR. Briefly: ROI = integer value of the region in the dataset; Nvox = total number of voxels in the ROI; Nzer = number of zero-valued voxels in the region (e.g., due to masking or limited FOV); Dvox = maximum depth, counted in voxels; Tmin, T25%, Tmed, T75%, Tmax = the minimum, lower quartile, median, upper quartile and maximum TSNR

*values in the ROI; X,Y,Z = RAI coordinates of maximum-depth location; ROI_name = string label of ROI.*

Additionally, viewing the temporal SNR (TSNR) of the time series can be useful for judging the regions of the brain with reasonable coverage for analysis (see Fig. 15D). Even within the acquired field of view, distortions and dropout can occur that greatly reduce the measured signal, and this can affect the ability of the researcher to investigate particular regions. For example, medial frontal and subcortical regions can be particularly affected by dropout, and differentially across subjects depending on brain-to-slice angle, reducing the ability to accurately study specific locations within the brain. We note that Fig. 7 of Reynolds et al. (2023) contains an illustrative set of TSNR maps of varied degrees of quality, which are discussed in detail in the text.

Studies have various paradigms, designs and areas of interest, while also having unique scanner properties, sequence protocols, subject populations and more. To help users judge the appropriateness of the data at hand for their study, *afni_proc.py* lets users provide one or more sets of ROIs to evaluate for signal strength and spatial sampling. A table of TSNR and ROI shape properties for those regions is then added to the "warns" block of the APQC HTML, following the format of the report's automatically-generated table of regions in known template spaces (Taylor et al., 2024). In this example, two "-ROI_import .." options are used to load a pair of atlases: "BrodPijn" is the Brodmann atlas (1909) digitized by Pijnenburg et al. (2021); and "SchYeo7N" is the refined version of the 7-network, 400 parcellation Schaefer-Yeo atlas (Schaefer et al., 2018; Glen et al., 2021). Then, a corresponding pair of "-regress_compute_tsnr_stats .." options are used to provide a specific subset of regions of interest for each. Note that one can all use a keyword "ALL_LT" to select all regions in a labeled atlas dataset. Fig. 16A shows the table and locations of the Brodmann atlas and B shows the same for the refined Schaefer-Yeo atlas. A detailed table description is provided in Sec. 2.3.7 and Fig. 15 of Taylor et al. (2024). Briefly, warning levels reflect ROIs with empty voxels or very narrow shapes (that are susceptible to partial voluming or misalignment), as well as TSNR values that are either generally low or unstably changing across the ROI. In the Brodmann set, two ROIs have particularly low TSNR (in frontal/inferior regions, which often have signal strength challenges), as well as one whose maximum depth is only 1 voxel.

Finally, we note that the regress option "-regress_make_corr_vols  aegm09" makes use of the anatomical follower ROI dataset information that was added earlier in the command. Specifically, afni_proc.py will generate a whole brain correlation map using the average time series of all nonzero regions of the referenced dataset. In this case, "aegm09" is the anatomical gray matter parcellation on the final EPI grid. We note that this option can take several arguments, so multiple correlation maps from anatomical followers can be made in parallel.

Ex. 4:  Full multi-echo FMRI resting state: surface-based processing and analysis

This example presents another case of resting state processing for the same sub-005 participant as in Ex. 1 and 3. However, here we use all echos in the ME-FMRI acquisition and also include surface-based processing (Fig. 17, or run: afni_proc.py -show_example "AP publish 3d"). As in Ex. 1, a pair of reverse-blip EPI datasets are also utilized for B0-inhomogeneity distortion correction. The "ricor" block could be added in the same form demonstrated in Ex. 3, if desired. Several of the other processing options have been discussed in previous examples, as well (highlighted in Fig. 17).

```
# AP Example 4, which was preceded by running:
# + FreeSurfer's recon-all : estimates anatomical surface mesh datasets
# + @SUMA_Make_Spec_FS     : creates standard-mesh versions of surfaces and
#                            specification files ${suma_specs} and NIFTI-
#                            format anatomical surface volume ${suma_sv}
# + sswarper2              : creates skullstripped anatomical ${anat_cp},
#                            and nonlinear warps ${dsets_NL_warp}

afni_proc.py                                                              \
    -subj_id                   ${subj}                                    \
    -dsets_me_run              ${dsets_epi_me}                            \
    -echo_times                12.5 27.6 42.7                             \
    -copy_anat                 ${anat_cp}                                 \
    -anat_has_skull            no                                         \
    -anat_follower             anat_w_skull anat ${anat_skull}            \
    -blocks                    tshift align volreg mask combine surf      \
                               blur scale regress                        \
    -radial_correlate_blocks   tcat volreg                                \
    -tcat_remove_first_trs     4                                          \
    -blip_forward_dset         "${epi_forward}"                           \
    -blip_reverse_dset         "${epi_reverse}"                           \
    -tshift_interp             -wsinc9                                    \
    -align_unifize_epi         local                                      \
    -align_opts_aea            -cost lpc+ZZ -giant_move -check_flip       \
    -volreg_align_to           MIN_OUTLIER                                \
    -volreg_align_e2a                                                     \
    -volreg_warp_final_interp  wsinc5                                     \
    -volreg_compute_tsnr       yes                                        \
    -mask_epi_anat             yes                                        \
    -combine_method            m_tedana                                   \
    -surf_anat                 ${suma_sv}                                 \
    -surf_spec                 ${suma_specs}                              \
    -blur_size                 4                                          \
    -regress_motion_per_run                                               \
    -regress_censor_motion     0.2                                        \
    -regress_censor_outliers   0.05                                       \
    -regress_apply_mot_types   demean deriv                               \
    -html_review_style         pythonic
```

*Figure 17. The afni_proc.py command for Ex. 4 (resting state, multi-echo FMRI with surface analysis, full processing). Options with gray background have already been described in earlier examples here, and any variables described in the captions of Figs. 5, 8 and 12. ${sv_suma} is the surface volume dataset; ${suma_specs} are the surface specification files in the SUMA directory; ${dsets_epi_me} is a set of a single run of EPI datasets with different echo times. This example's "-radial_correlate_blocks ..." option does not include "regress", because that stage of processing occurs on the surface and radial correlation QC has not yet been implemented there (but it will be added in the future). Running this afni_proc.py command produces a commented script of >650 lines, encoding the detailed provenance of all processing.*

# Ex. 4: Results from combined volume-surface processing

A) Seed-based correlation map of DMN

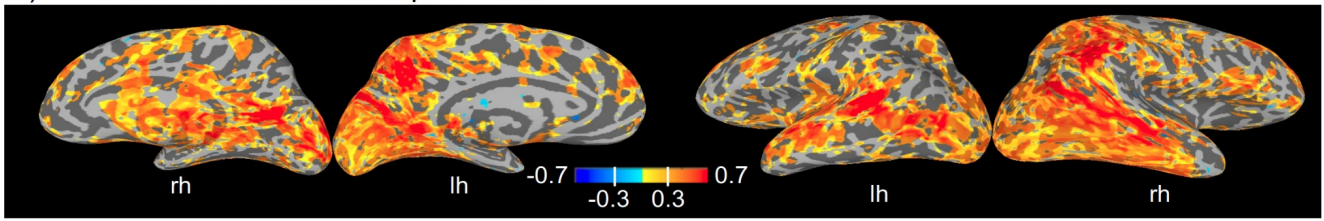

B) Seed-based correlation map of visual network

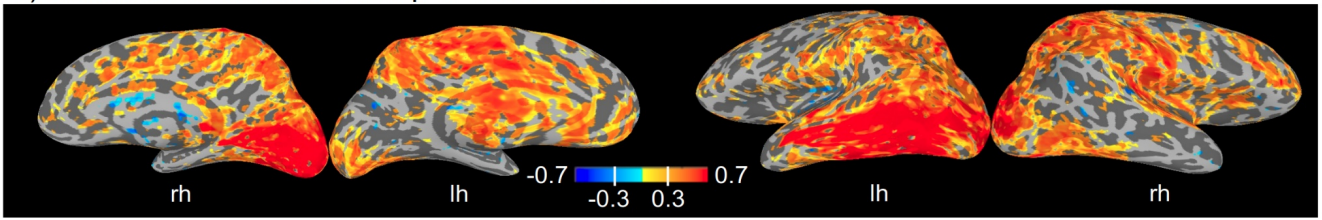

C) Seed-based correlation map of auditory network

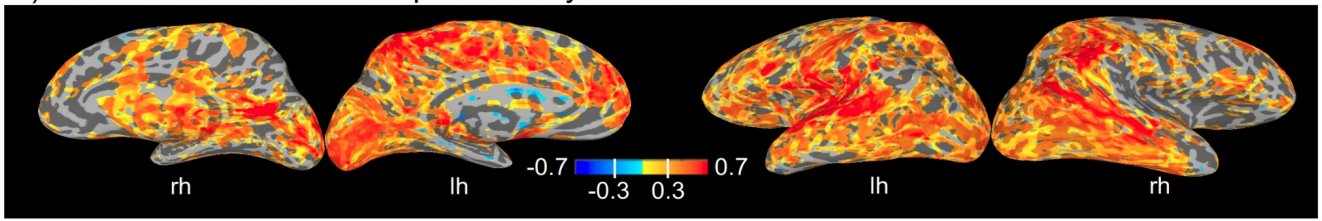

D) TSNR map (after regress block GLM)

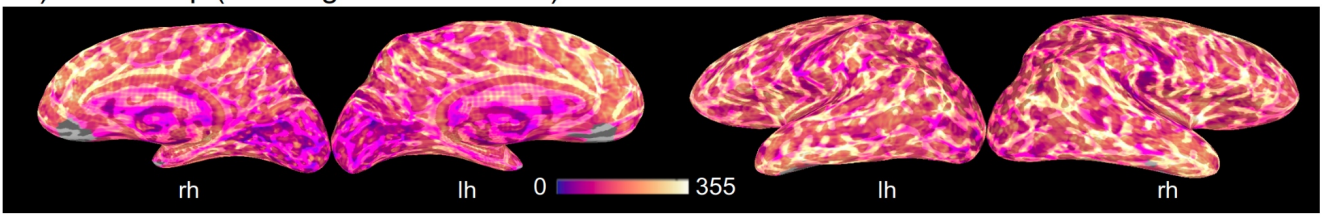

*Figure 18. Some results generated by afni_proc.py for Ex. 4, which uses surface-based processing for resting state ME-FMRI data. Images are displayed using SUMA (Saad et al., 2004). Panels A-C show seed-based correlation maps for the same seed locations used in the standard APQC HTML reports when purely volumetric processing is used (cf. Fig. 15A-C, showing QC images for Ex. 3). Panel D shows the TSNR across the cortical surface, for ME-FMRI data which has been processed using MEICA-estimated regressors. Some empty patches in the TSNR maps reflect the fact that the utilized MEICA requires brainmasking and occurs before surface projection.*

The primary reason for acquiring multiple echos is to combine their information in some manner to boost EPI SNR, which is a major benefit. ME-FMRI can typically be acquired without increasing standard TR times, though some combination of multiband and slice acceleration is typically required. As long as using the acquisition setup does not lead to artifacts (which should be checked as part of the piloting and quality control), this style of acquisition may be widely considered as beneficial. In the present case, a single run of three echos was acquired with TR = 2.2s. The set of echos for a given EPI run are input with "-dsets_me_run .." and a corresponding set of echo times with "-echo_times ..". It is possible to input and process multiple ME, simply adding a "-dsets_me_run .." option for each additional run; all runs must have the same set of echo times, so the option providing those values need only be provided once.

Because we are processing ME-FMRI data here, we can add two options to try to minimize temporal and spatial blurring during processing to a maximal degree that were not used in earlier examples. In both cases, the choice is to specify a sinc-function kernel for interpolation, since mathematically this kind of interpolation kernel preserves edges (high frequency features) to the greatest degree possible, from signal processing theory. First, "-tshift_interp -wsinc9" specifies using a high-order sinc function (instead of the default quintic spline) during the early step of time shifting slices to the same temporal origin. Similarly, "-volreg_warp_final_interp wsinc5" specifies using a sinc (instead of the default cubic spline) for spatial interpolation when applying the concatenated spatial transform to the input EPI datasets. The cost to be paid for these relatively minor reductions in blurring are slightly increased computing time and the introduction of ringing to the interpolated domain (an unavoidable trade-off with sinc functions). In theory, the spatial sinc kernel could introduce some GM-like correlations artificially into non-GM in some cases, which should be checked. Since single-echo EPI is typically blurred more notably as part of processing (as noted before, ME-FMRI will tend to have inherently higher TSNR from combining echos), we generally don't include these options due to their diminished impact. But for ME-FMRI, their benefits may be more noticeable in processing.

The "combine" block controls the details for the ME-FMRI processing. There are several potential formulations of echo combination, and a large number can be specified via afni_proc.py with "-combine_method ..". The most commonly applied ones include:
- "OC", to use the straightforward optimal combination method of Posse et al. (1999), which is implemented within AFNI.
- "m_tedana", to use the open source tedana package version of MEICA (DuPre et al., 2021), which is selected in this example.
- "tedana", to use the earlier version of MEICA from Kundu et al. (2015), which is distributed in AFNI (but is no longer updated).

There are further variations and combinations of these methods, which can be chosen with separate keywords and controlled with other "-combine_*" options, which the *afni_proc.py* help file describes in detail. Each method has its own requirements, assumptions and potential benefits. For example, OC is the simplest: for each voxel, a weighted average is calculated to optimize BOLD contrast, and this provides a straightforward boost to local SNR. MEICA requires more processing and algorithmic choice, but aims to remove noise components from the dataset, potentially providing "clean up" as well as SNR boost. We note that the "mask" block is specifically placed before "combine" in this example, because the selected tedana MEICA currently requires a brainmask for processing (with OC, it would not be included, so more of the FOV results could be seen for artifact checks; see supplementary Ex. 8).

Inclusion of the "surf" block leads to the EPI data being projected onto the subject's anatomical surface mesh dataset, which was initially calculated by FreeSurfer's recon-all here. After running AFNI's @SUMA_Make_Spec_FS on the FreeSurfer results for format conversion and some additional processing, the final meshes are standardized in the sense that nodal correspondence exists among subjects (Argall et al., 2006), to the degree the mesh creation was accurate. @SUMA_Make_Spec_FS creates two different sized meshes: the "std.141.*" datasets have 198812 nodes, and the "std.60.* ones have 36002 nodes. The higher resolution mesh corresponds to approximately 1mm spacing between nodes (analogous to a standard T1w spatial resolution at 3T), while the lower resolution mesh corresponds to approximately 3mm spacing (analogous to a typical EPI resolution at 3T). Users may choose whichever mesh is most appropriate for their analyses. These meshes are input to afni_proc.py by providing the name of the corresponding "specification" file(s), such as std.141.${subj}_lh.spec and std.141.${subj}_rh.spec. Additionally, one inputs the reference anatomical for volume-to-surface coordinate mapping via "-surf_anat ..". Using these datasets, the EPI values are mapped to the surface using the average value along the line segments connecting corresponding nodes of the smooth white matter and the pial surfaces.

Importantly, "surf" precedes the "blur" block, because one primary benefit of surface-based processing is that blurring can be constrained to locally relevant GM (as opposed to the more indiscriminate volumetric blurring, which can blur in WM or GM from other gyri). The same "-blur_size .." option specifies the size of blurring to apply, but it will be applied along the surface mesh here. Since this is a multi-echo FMRI data, which should have inherently better TSNR than a corresponding single-echo acquisition, blurring should likely be kept smaller than what would be applied for single-echo EPI data. It remains useful to still have some small blurring applied, with the primary purpose being to accommodate anatomical variability remaining after surface-based template registration; humans in particular have relatively high variability of sulcal/gyral patterns.

Essentially the same "regress" block modeling is performed here as in Ex. 3. Again, LFF bandpassing is not included, but the mean and derivative of motion parameters are. Fig. 18A-C shows the seed-based correlation maps for the same networks shown in Fig. 15A for Ex. 3's volumetric processing. In each case, the seed location was projected onto the nearest mesh node using AFNI's Surf2VolCoord, and then 3dTcorr1D applied to estimate the correlation map. Fig. 18D shows the TSNR after final regression (cf volumetric case in Fig. 15B).

Computation notes
For the present work, afni_proc.py was run on a computing cluster (the NIH's Biowulf), which currently uses a Linux Rocky 8 operating system. While not necessary for this small set of examples, we offer this demonstration of "swarming" the processing (here via a slurm workload manager) for those analyzing a larger collection of data. The GitHub repository for this work contains both the HPC scripts (in the "scripts_*_biowulf/" directories) and the desktop versions (in "scripts_*_desktop/").

For each Biowulf example, we requested 8 CPUs and 10 GB RAM. More resources could be used as desired: several AFNI programs—particularly the alignment ones—have inherent parallelization with OpenMP to make use of multiple CPUs. We chose to run with resource levels that might mirror a reasonable desktop or laptop for processing. The runtimes for each command, with final results directory disk space usage, were:
- Ex. 1: 0.8 GB,  23 min
- Ex. 2: 1.4 GB,  36 min
- Ex. 3: 1.1 GB,  19 min
- Ex. 4: 2.8 GB,  20 min

Ex. 4's disk space is the largest because of the multiple echo datasets that exist through several of the initial processing stages. This resource usage seems widely applicable. Note that processing time and disk space will vary depending on the dataset properties, such as voxel size, number of time points and/or runs, etc.

Some of the preliminary programs took more time: FreeSurfer's *recon-all* took 3h 38min (using 1 CPU, since using "-parallel" did not always succeed), and *sswarper2* took 1h 25min (using 16 CPUs).

**DISCUSSION**

FMRI studies are complex, with many fundamental decisions to make, each of which can strongly affect the outcome. What field strength of scanner should be used? How many echoes should the EPIs have, and with what flip angle and voxel size? How many subjects should be scanned, and how many sample trials should be acquired during each run, or how many time points? The processing is also complex, with many specifications tied in with study design: what blur size to use, whether to process on the surface or volume, and how strict to be with motion censoring. In the end, no FMRI study can be made without careful design and lots of choices made at all stages.

The purpose of pipeline-generation tools is well known: they aim to allow the user to perform analysis by specifying just input data and a set of options, rather than doing their own scripting and code-level interaction. This should generally reduce difficulty for the user, as well as the chance of errors leaking into the analysis stream, as individual pipelines can be widely tested. The specification of the pipeline command itself serves as a condensed and readable summary of important processing choices, which can be shared and interpreted more easily than a full processing script. Each of these aspects should improve stability, reproducibility, applicability, shareability and extensibility (i.e., expanding the processing to include more runs) of the analysis.

AFNI's *afni_proc.py* was designed to have the above features, as well as a particular emphasis to allow users to have detailed control over their processing and a close understanding of each step. To facilitate this, the pipeline script is readable and commented, and outputs facilitate visualization and checks. The researchers should be able to design the analysis that they feel is appropriate for their study, within bounds of mathematical/algorithmic correctness; to help verify the latter, the created pipelines include several checks and warnings during processing.

It is true that with greater flexibility, complexity also increases. To balance this, the *afni_proc.py* specification is organized modularly and hierarchically, so the user can first organize their main processing steps (the blocks), and then provide details for each with option names that typically begin with that block name, for readability. To assist with constructing a pipeline, one can use one of the many starter or published examples, and one can also programmatically compare options between two provided commands and/or pre-defined examples. Because *afni_proc.py* creates a commented script of all processing steps, one can be sure of exactly how each choice is implemented, which is data provenance at the level of code specificity. Importantly, this helps the researcher deeply understand their processing choices and see them in action. Being aware of the details may require more work, but it can also greatly improve the consistency and interpretation of the results.

Learning more about FMRI processing and using afni_proc.py

As noted above, FMRI data and its processing are complicated. Users must be sure of their study goals, and then have the ability to carry them out as successfully as possible during their analysis; successful processing helps to validate (or to uncover issues) in both. While there are many educational resources available for FMRI, we present some here that the present authors have contributed to (and the authors note that they continue to learn and improve their own understanding over time, as well). These cover both general FMRI processing principles and more specific afni_proc.py usage.

The afni_proc.py program help contains a large set of commented examples, as well as descriptions of each option and several comments on topics (e.g., bandpassing considerations in resting state). This can be viewed from the terminal ("-help"), in a text editor ("-hview") or in a chapter-formatted online version (https://afni.nimh.nih.gov/pub/dist/doc/htmldoc/programs/alpha/afni_proc.py_sphx.html#ahelp-afni-proc-py).

Several other processing examples with afni_proc.py and other AFNI programs are provided in the Code Examples (Codex) portion of the online documentation (https://afni.nimh.nih.gov/pub/dist/doc/htmldoc/codex/main_toc.html). These provide examples and descriptions of processing, links to the related paper, and either the scripts or pointers to the associated repositories. Codes here tend to be commented, in order to provide descriptions of processing choices. We strongly encourage all users to provide the afni_proc.py and other processing commands in supplementary files and/or associated code repositories, to benefit both their colleagues, trainees and even likely themselves in future studies.

AFNI contains several demos that comprise both executable code and data. For example, we provide a full demo implementing several variations of processing and echo combination for resting state FMRI via afni_proc.py. This can be installed with @Install_APMULTI_Demo1_rest (Taylor et al., 2022).

The present paper provides a combination of afni_proc.py code examples and option description, as well as general FMRI processing comments (above, and see below in the Discussion). It also contains an associated GitHub repository of scripts and downloadable data. Taylor et al. (2018) provided an earlier description of afni_proc.py processing options and considerations. For examples of code development and validation of several steps used within afni_proc.py, see the next section as well as the larger online list of AFNI methods and data resources (https://afni.nimh.nih.gov/pub/dist/doc/htmldoc/published/citations.html).

Non-human imaging has been increasing over time, along with the availability of public datasets such as through PRIMatE Resource Exchange (PRIME-RE; Messinger et al., 2021) and combined data and resource exchange (PRIME-DRE; Milham et al., 2022). As part of these public resources, we created processing demos, with data and full processing pipelines combining AFNI's @animal_warper and afni_proc.py programs, for both task and resting state processing using the NIMH Macque Template and CHARM atlases (Jung et al., 2020). These can be respectively downloaded and installed with: @Install_MACAQUE_DEMO and @Install_MACAQUE_DEMO_REST.

For over 20 years, the AFNI Bootcamp is a long-running educational course that focuses on both theory and hands-on practice, from initial data checks through group analysis and results reporting. This course is taught without charge both at NIH and host institutions around the globe. Recorded lectures from some of these workshops are available online (https://afni.nimh.nih.gov/pub/dist/doc/htmldoc/educational/main_toc.html), along with the associated demo data. More recently, an expanded (and still growing) set of lectures have been created on the AFNI Academy YouTube Channel (https://www.youtube.com/c/afnibootcamp), which is also freely available.

Finally, the developers maintain active discussions about a wide range of data processing topics on the AFNI Message Board (https://discuss.afni.nimh.nih.gov/). These often involve discussing code, such as actively commenting on and developing afni_proc.py commands. This can also involve suggestions or queries that become new afni_proc.py functionality.

Method development and validation

There are several layers to afni_proc.py, which are discussed in more detail in Appendix B. Briefly, we note that the Python program "afni_proc.py" was started in 2006 by R. Reynolds, who has been its primary designer and developer since then. The program's scripts make use of the "afnipy" Python library as well as the larger AFNI code base itself, and its functionality has grown as new tools have been developed. It also integrates with several external software projects, such as FreeSurfer, tedana's MEICA and NiiVue, and has benefited from input, bug fixes and suggestions by the wider neuroimaging community. Throughout its development, its testing infrastructure has also grown, so that it is currently tested using more than 200 executions across 135 scripts.

Because *afni_proc.py* is part of an actively developed software package, many new features have been added over time (see Appendix B for a summary timeline). These additions variously arise from user requests, to adapt to new acquisition techniques, or simply as developer-based ideas to improve processing. Several of these features have been tested and validated within their own sub-studies. For example, the "lpc" cost function used to align EPI and T1w anatomicals was developed specifically to improve alignment between images with differing contrast while also focusing on local features (Saad et al., 2009), and the local EPI unifizing was recently added as an additional improvement when EPIs are strongly inhomogeneous. The quality of nonlinear warping results using *3dQwarp* (which also underlies *@SSwarper*, *sswarper2*, *auto_warp.py* and *@animal_warper*) was demonstrated using both structural ROIs and FMRI data (Cox and Glen, 2013). The benefits of reducing EPI distortion using either AFNI's reverse phase-encoding with 3dQwarp or phase map unwarping with epi_b0_correct.py were demonstrated in (Roopchansingh et al., 2020). To address Eklund et al. (2016)'s observation that the field-wide assumption of Gaussian spatial smoothness was not very accurate, Cox et al. (2017) developed and validated a new, non-Gaussian "mixed" ACF that 3dFWHMx uses. As some concerns had been raised about voxel resampling instabilities in other software packages, Cox and Taylor (2017) validated and demonstrated the stability of AFNI's existing resampling methods for preserving smoothness. The left-right flip check was added to the quality control to guard against DICOM conversion and other data errors, and has proved useful in warning of problems in a surprising number of public datasets (Glen et al., 2020).

The benefits of afni_proc.py's built-in quality control features, particularly the APQC HTML, have been demonstrated in detail on a large number of public datasets by Reynolds et al. (2023), Birn (2023), Lepping et al. (2023) and Teves et al. (2023), as part of the FMRI Open QC Project (Taylor et al., 2023a). These showed a variety of useful features for combining quantitative and qualitative QC checks, which applied to a large number of observed features in real data. These included: FOV issues, upside down brains, mismatched datasets, motion-related issues, variance-line artifacts, non-physiological spatiotemporal patterns observed with InstaCorr, and more. Several new features were subsequently integrated into the APQC HTML, for user interaction and faster evaluation of in-depth exploration, as described in Taylor et al. (2024).

Independent groups have performed validations of various features across software packages. Temporal autocorrelation modeling methods were compared across major software packages, and AFNI's *3dREMLfit*, which is used by *afni_proc.py*, was found to perform the best (Olszowy et al., 2019). The recapitulation of the importance of including bandpassing as a single step within the linear regression model (as done in *afni_proc.py*) as opposed to being done separately[10] was given by Hallquist et al. (2013). Among motion estimation tools of major packages, AFNI's *3dvolreg* was found to be tied for most accurate motion parameters, while having the additional benefit of introducing the least smoothing and being the fastest of those tested (Oakes et al., 2005).

Flexibility of HRF modeling and timing file formats

Research designs and questions can vary greatly in both their setup and assumptions. How best to model the coupling between observed BOLD and underlying neuronal firing remains an open but important question. Utilizing an appropriate HRF is an important part of FMRI processing, and there are many options for modeling the hemodynamic response function in *afni_proc.py*, primarily based on the functionality offered by AFNI's *3dDeconvolve*.

Researchers can choose among fixed-shape basis functions (such as GAM, BLOCK, SPMG1, TWOGAM, WAV and MION), fully variably-shaped basis functions (such as TENT and CSPLIN) intermediate functions (such as SPMG2 and SPMG3), and many more. Most of the fixed-shape functions even offer detailed control over the parameters used, such as GAM, TWOGAM and WAV. EXPR allows one to specify an arbitrary linear expression in entirety. One can also provide externally generated regressors, akin to those based on motion parameters.

Additionally, one can control aspects such as event duration (for those basis functions that allow duration convolution) and amplitude modulation, where the expected magnitude of individual events varies based on external parameters. Event durations can be fixed for a stimulus class or be allowed to vary per event, which we refer to as duration modulation and is most commonly applied via dmUBLOCK(-value). Amplitude modulation can be applied with one or more modulators for each stimulus class, and amplitude and duration modulation can be applied together. There is also an individual modulation option (IM), which allows each event to generate a single regressor (or a single set of them). IM allows the regression to generate a time series of beta weights, measured across events. *3dDeconvolve's* program help contains further details.

These aspects are specified in *afni_proc.py* by providing the input timing file (usually via "-regress_stim_times"), a basis function for each stimulus class ("-regress_basis" or "-regress_basis_multi") and the type of stimulus timing to apply ("-regress_stim_types"). The latter can be "time" for simple times, "AM1" or "AM2" for fixed or estimated amplitude modulation, as well as duration modulation, "IM" for individual modulation, or "file" for a simple regressor file. Single or multi-row contrasts can be specified in the format directly readable by 3dDeconvolve.

As noted above, regression modeling is generally set up by *3dDeconvolve*, formulating the global linear regression matrix. After that, the actual regression is done by either *3dDeconvolve*, the more commonly used *3dREMLfit*, or even *3dTproject*, for a simple projection of nuisance regressors. In cases of slicewise or voxelwise regressors (e.g., for RETROICOR or ANATICOR), *3dREMLfit* must be used. To explicitly use *3dREMLfit*, *afni_proc.py* provides the "-regress_reml_exec" option.

---

10 Technically, it is possible to perform the GLM after bandpassing, as long as all of the regressors have been bandpassed prior to modeling. However, one must still include the lost number of DFs in the regression model count, so that one is not actually performing an invalid model, which can easily happen.

Finally, we note that there are a large number of timing file formats used by various recording tools and software packages across the field. While *afni_proc.py* is set up to handle AFNI-style stimulus timing files, *timing_tool.py* can be used to convert FSL-formatted files (tables of onset, duration, modulator) or BIDS-style TSV files (with column headers like "onset", "duration", "stim_class", or others) into AFNI timing format.

"Simple" afni_proc.py commands and quick quality control

Full single-subject FMRI processing for a study requires the choice of many parameters, such as blur size, motion censoring thresholds, a list of regressors to include and more. Many of these are tailored carefully to fit appropriately with the study design, aims and data collection cohort. For example, motion thresholds might differ between studies of healthy adults and adolescent participants with ADHD; blur values might differ based on voxel size; and stimulus response models will likely vary widely by task design. When preparing for a careful analysis, these choices can have strong effects on outcomes.

However, there are other times where a quick analysis without detailed attention to many of those choices is appropriate. For example, many meaningful QC features of the data and processing can be checked without these considerations, such as EPI FOV, TSNR, variance line artifacts, signal saturation, subject motion and alignment success. For this reason, two "simple" wrappers for afni_proc.py that require essentially no options except for the names of data files have been developed: *ap_run_simple_rest.tcsh* and *ap_run_simple_rest_me.tcsh*, for single- and multi-echo FMRI analysis, respectively.

These commands are designed to be general enough to be integrated into general data acquisition protocols, so that APQC HTMLs and further quantitative information can be automatically available soon after scans (e.g., as part of an XNAT or similar platform). This facilitates making regular QC checks as soon as possible after acquiring data, greatly reducing the chance for data waste due to the undetected presence of scanner artifacts or accidental protocol changes for a large number of subjects. These could even be run while participants are still present at the scanner, so that they could be rescanned if QC evaluations deem their data unreliable, rather than simply losing their data later.

Fig. 19 shows examples of running these commands for the single- and multi-echo FMRI data included here. When a template name is provided ("-template ..."), quick affine alignment with the subject anatomical is performed, so that approximate but still systematic seed-based network maps are output in the APQC HTML. If no template is input, two seed locations for correlation maps are still chosen, in central but left-right offset locations. The option to remove initial time points ("-nt_rm ...") exists to remove pre-steady state time points from the EPI, if necessary (default: remove the first two time points). The single echo version can now also be run as a BIDS App[11].

These quick QC evaluations can also be integrated with complementary programs of more purely quantitative data checks. AFNI's *gen_ss_review_table.py* can quickly sort through the basic review

11 Created in collaboration with Y. Halchenko: https://github.com/afni/afni_proc_simple_bids_app/. D. Nielson also led an earlier project that created a more general BIDS App for afni_proc.py: https://github.com/bids-apps/afni_proc

quantities stored by *afni_proc.py* during processing. Moreover, *gtkyd_check* ("getting to know your data" check) both creates its own quantitative summaries of raw, unprocessed datasets and wraps around *gen_ss_review_table.py* to compare datasets for consistency. These are described in more detail in Taylor et al. (2024) and Reynolds et al. (2023), with examples. They provide systematic and scriptable tools for evaluating data properties and appropriateness.

```
# These AP examples were not preceded by any processing.

# "simple" AP command example for single echo-FMRI data
ap_run_simple_rest.tcsh                                                 \
    -subjid                 ${subj}                                     \
    -epi                    ${dsets_epi_me}                             \
    -anat                   ${dset_anat_00}                             \
    -template               ${template}                                 \
    -nt_rm                  4                                           \
    -run_ap

# "simple" AP command example for ME-FMRI data
ap_run_simple_rest_me.tcsh                                              \
    -subjid                 ${subj}                                     \
    -epi_me_run             ${dsets_epi_me}                             \
    -echo_times             12.5 27.6 42.7                              \
    -anat                   ${dset_anat_00}                             \
    -template               ${template}                                 \
    -run_ap
```

Figure 19. *Examples of "simple" afni_proc.py commands, using wrapper programs in AFNI for both single- and multi-echo EPI input. Each performs a quick, volumetric analysis of the provided input data, treating the input like resting state FMRI with essentially no detailed options required. This convenient processing still produces useful outputs for informative QC evaluations of data. These commands are general enough to be applied as part of a standard data acquisition, so APQC HTMLs could be created and checked automatically and even while a subject is still present. Some simple processing options that might be useful are: "-nt_rm ...", to provide the number of initial time points to remove; or "-template ...", to specify a reference template for quick, approximate (affine) alignment.*

Variety of pipeline tools

At present, there are many available pipeline tools for FMRI processing across the field, in addition to *afni_proc.py*, which was created in 2006. These include the Conn Toolbox (Whitfield-Gabrieli and Nieto-Castanon, 2012), C-PAC (Craddock et al., 2013), DPARSF/DPABI (Yan CG and Zang, 2010; Yan et al., 2016), ENIGMA's HALFpipe (Waller et al., 2022), fMRIPrep (Esteban et al., 2019), FSL FEAT, FsFast from FreeSurfer (Fischl and Dale, 2000), SPM12 (Ashburner et al., 2012) and others. These tools have the same general goals, but differing underlying methods, packages, languages, philosophies and even endpoints. For example, some do not include regression modeling at the subject level, while others do; some offer flexibility of options, while others prefer less variety of choice; some perform analogous steps in notably different ways (such as local vs grand mean or other scaling); some are built for minimal (pre)processing (no smoothing, regression, …), while others aim to be more comprehensive. Some pipelines overlap in the specific programs used (particularly for tools that are wrappers of existing software packages; many of the above use AFNI programs), and some can optionally substitute in programs across various packages.

It is not practicable to perform a comprehensive comparison across all software and all varieties of analyses. Some forms of comparisons do exist in the literature. For example, Bowring et al. (2022) compared similarly-structured AFNI, FSL and SPM pipelines with fMRIPrep in various combinations for three different data collections; most results showed notable overlaps, and they found that fMRIPrep's pipeline results were most similar to AFNI's (set up with *afni_proc.py*). For one collection, the largest difference was observed due to how different packages formulated their signal model (choices related to the task regressor hemodynamic response and/or parametric modulations), and in another collection the largest difference was due to differences in the noise modeling of temporal autocorrelation structure.

One must also be careful how comparisons are made, in order to not bias results. In the NARPS study (Botvinik-Nezer et al., 2020), different teams analyzed a single task FMRI data collection with any software (including AFNI, fMRIPrep, FSL, SPM and/or other packages) and processing settings of choice to answer 9 region-specific hypotheses. While the abstract mentioned "sizeable variation" in results, the degree of difference/similarity depended strongly on the manner of comparison performed, so that it might be viewable as a stronger comment on meta analysis variability than on processing variability per se (Taylor et al., 2023b). Applying thresholding to statistics volumes *before* comparisons led to relatively notable apparent variability. However, when thresholding was *not* injected before comparisons there was a much greater similarity—see the predominantly high similarity matrix values of Fig. 2 of Botvinik-Nezer et al. (2020), as well as those of their Extended Data Fig. 2. Taylor et al. (2023b) explored the comparisons further, and showed how the kind of variability observed in the results was predominantly that of varied *strength of agreement*, rather than disagreement, across the set of tools and analysis setups. Only looking at thresholded data creates a bias toward disagreement and perceived variability, as dichotomization bifurcated results. Using more complete data in comparisons, as well as in visualizations, provides a better basis for more informative meta analysis.

Furthermore, more than the software choice itself, variations tend to arise more from purposeful decisions made by the researchers. For example, choosing whether or not to orthogonalize regressors, or the method of doing so, or implementing amplitude modulation, can understandably lead to differing effect estimates and statistics, esp. in relative magnitude. While software packages overlap in many processing steps, some may be unique to one or another, hence leading users to choose a toolbox that is most appropriate for their experimental design of choice. Another practical factor of difference is a combination of familiarity with software and flexibility of options:  when some step along the pipeline fails or has improvable results, the ability of the researcher to enact improvement depends on both of these attributes.

Similarly, the choice of smoothing (or blurring) radius during processing will certainly affect results, though choices like this tend to be software independent. While some obviously inappropriate values for parameters can be recognized (e.g., a blur radius of 100mm for human FMRI data), many have what might be termed a "semi-arbitrary" interval: there exists a range of reasonable values without any obviously optimal one. For example, blurring an EPI dataset that has 3mm isotropic voxels by anywhere within a range of 4-6mm seems reasonable; and in some cases *not* blurring at all is reasonable.  Importantly, the degree of difference this kind of processing choice makes can only be accurately assessed without inserting an artificial dichotomization step by thresholding results before comparison, since doing so heavily biases results toward disagreement (Chen et al., 2022; Taylor et al., 2023b).

Design choices in afni_proc.py

Since its inception, the primary goal of *afni_proc.py* is to allow the user to perform the FMRI analysis that is most appropriate for their study design and research goals. There are a large number of study design configurations and paradigms, and therefore there are a large number of options available and very few default settings. This flexibility has allowed *afni_proc.py* to be readily adapted to non-human primate and other animal imaging studies, such as rodents. It also means that single- and multi-echo analyses, or volume- or surface-based, or voxelwise or ROI-based can all be accomplished with a basic framework and by the adjustment of a small number of options. By specifying more options and relying less on default options (that might change over time with design or dependencies), both reproducibility and clarity are also enhanced.

While this can produce a relatively large learning curve to start analysis, we have tried to reduce this by having several "vanilla" processing examples described within the *afni_proc.py* help. There are also various downloadable demos and online repositories with published papers (e.g., the AFNI Codex, mentioned above), which provide useful references as possible starting points for many analyses. The newer *run_ap_simple\*.tcsh* programs also require almost no options to run, but they still generate a full set of outputs to examine (including the APQC HTML), as well as full *afni_proc.py* commands to learn from or to further adapt. Appendix B provides a more detailed list of demos and available example commands.

Another benefit of the flexibility is to allow users to update and adapt their analyses easily. FMRI data are known to be noisy and susceptible to various distortions, and differences in scanner, sequence, pre-/post-scanner software update, voxel size, acceleration factor, population age, study design and more can affect data properties significantly. One may want to adapt a pipeline to process new datasets that use contrast agents such as MION (e.g., in animal imaging). Within a data collection, some subjects may require tweaks to the cost function or alignment parameters to overcome poor initial overlap or brightness inhomogeneities. *afni_proc.py* can easily be run over subsets of datasets that need to be reprocessed. Public data from different sites can vary greatly in properties, requiring variations of processing. Having a single framework for all this makes comparing different methodologies or approaches easier (e.g., the effect of smoothing on data), as well as adapting to new data (such as going from single to multi-echo).

Moreover, *afni_proc.py* has a strong focus on the user understanding all stages of the analysis as well as being able to verify the success (or otherwise) of them. To that end, the results directory contains many intermediate datasets. This does create a relatively large footprint of disk space,[12] but these have continually proved useful to answer questions that arise when checking back about the data (and these can be removed when the user is assured of their processing). The user has access to the provenance of all processing steps in multiple levels of detail: first, through the *afni_proc.py* options themselves; then through the complete, commented processing script that is created; and finally through the accumulated command history contained within the header of each processed dataset.

---

12 There is the "-remove_preproc_files" option flag to remove a large fraction of intermediate datasets when the processing completes, to save disk space. However, using this can make full evaluation of processing more difficult. Users could instead remove the intermediate datasets after checking results fully.

While great flexibility is allowed in processing, *afni_proc.py* also contains automated checks for mathematical and other practical problems. Examples of these include checks for: collinearity in model regressors (e.g., due to accidentally providing same file for two different stimulus classes, or to poor study design); left-right flipping between EPI and anatomical (discussed above, possibly due to DICOM conversion error); pre-steady state volumes left in the EPI (due to inexact data knowledge); high censor fraction overall or within a stimulus class (subject motion issues, possibly related to task design); high usage of degrees of freedom (through either processing choices like bandpassing, or subject motion, or both); through-plane lines of high variance (scanner artifacts); and more. Additionally, users can query a dictionary of other quantitative outputs automatically after *afni_proc.py* using *gen_ss_review_table.py*, further facilitating automated evaluations of the datasets.

Even with those quantitative checks in place, data visualization remains a major part of processing, assessment and understanding. Indeed, *afni_proc.py* includes the systematic APQC HTML with interactive features directly with the view that quality control and evaluation of intermediate procedures are not separate from data processing but indeed part of it (see Reynolds et al., 2023; Taylor et al., 2023a). Some derived datasets in the results directory and QC HTML have been developed over time to troubleshoot scanner coil issues and other potential artifacts, such as the radial correlation and the corr_brain (correlation map of global brain average signal). All time series in the FOV are analyzed, not just within the brain mask, to help illuminate possible artifacts. Transparent highlighting is used in the APQC HTML to be able to highlight the regions of greatest magnitude, while still allowing features that may be sub-threshold to be seen, since those often contain useful information for understanding the data more completely (Allen et al., 2012; Taylor et al., 2023b).

How do these features facilitate understanding and evaluating their FMRI data? Much of this comes from the intermediate datasets that are calculated and useful checks throughout (unsurprisingly, many of these are also integrated into the APQC HTML and summary dictionary). The intermediate EPI time series help provide an understanding of each of those main steps, particularly when viewed in linked *afni* GUI controllers as "before and after pictures," such as done in the AFNI Bootcamp course (see the AFNI Academy videos). TSNR maps can show scanner artifacts, signal dropout or reasonable coverage; they might help explain low test-retest variability. For task FMRI, seeing images of both the full F-statistic and individual stimulus modeling results helps a researcher judge whether the data are appropriate for a study, including aspects of reasonable subject performance. In the Bootcamp course, there is an example of using motion plots with censoring to reveal likely task-correlated motion that, in turn, helps explain odd statistical results. For resting state FMRI, GCOR is a useful parameter for gauging data quality, but it is the combination of seed-based correlation, corr_brain and radcor images that provides helpful interpretations of *why* the value is what it is, because it shows the underlying spatiotemporal relationships in the data. These can reveal aspects of motion or scanner artifacts lurking in data, as can the interactive InstaCorr feature of the APQC HTML. Showing original data can reveal anatomical variability that may affect alignment, suggest geometric distortions due to B0 inhomogeneity or other sources, or disambiguate alignment questions that bright CSF can cause. Many of these have been discussed above and in associated quality control papers (Reynolds et al., 2023; Taylor et al., 2024), as well as in the educational sources listed previously.

Considerations when setting up FMRI pipelines

There are a large number of factors to consider when setting up an FMRI processing pipeline. One has to make sure that processing details are consistent with both the theoretical and practical aims of the study design. That is, many analysis choices are closely linked with the acquired data's properties, and therefore some of the most important "processing" decisions that researchers make are actually ones about the acquisition itself. These choices should be integrated into pilot data acquisitions and test evaluations. We list some (but by no means all) important questions that researchers should ask themselves when designing an acquisition and processing pipeline.

The afni_proc.py help file contains a concise list of important choices to consider when starting an analysis.[13] These form a kind of checklist of items to consider and choices to make. For example, what is the type of analysis: task or resting/naturalistic? Also, what is the domain of analysis: volume or surface, and will it be ROI-based? Please see the help file for the full set, which will likely grow over time, along with brief notes about each point. Below, we note additional factors, many of which also connect study design planning and the analysis implementation.

**What are the main regions of interest?** This will help determine appropriate acquisition settings and voxel sizes. Whether performing voxelwise or ROI-based analysis, one will have a primary set of locations of interest to study. It is key to make sure that the FMRI signal there is acquired stably and reliably. This can be done by including follower ROIs (via "-anat_follower_ROI", "-ROI_import" and "-mask_segment_anat" options) and checking their shape and TSNR properties, making sure that both the EPI spatial resolution is fine enough to capture the region well and that the TSNR distribution is stable. The APQC HTML table (calculated via AFNI's *compute_ROI_stats.tcsh*) greatly assists this. Inferior frontal and subcortical regions, as well as the temporal lobe, may require special acquisition parameters to avoid sinus-driven signal dropout and distortion. Studying ROIs that are small and/or contain narrow features might require high-resolution EPI. If acquiring one's own data, one can verify that the current sequence is adequate for the study. If using already acquired data, voxel size and TSNR coverage might constrain the areas of the brain that can be reliably studied and the parcellation regions that can be used. If the data do contain notable distortions, one can check the degree to which they affect the regions of interest and whether the dataset is still suitable.

**Should I use single- or multi-echo FMRI?** ME-FMRI can provide useful TSNR increases, just from averaging the multiple echoes via optimal combination (OC). MEICA methods may further remove non-physiological features, also boosting TSNR. These processing techniques can be used directly within *afni_proc.py* via the "combine" block (using either AFNI or the tedana version of MEICA, respectively). ME-FMRI acquisitions typically require using some multiband or slice-acceleration to preserve TR of around 2s, but often one can keep these acquisition factors low to help reduce artifacts. As long as ME-FMRI sequences do not introduce artifactual features such as cross-slice correlations (which the APQC HTML's InstaCorr buttons can help check; Song et al. 2017; Taylor et al., 2023a), then this may be a useful way to increase signal strength. For a practical comparison of single- and multi-echo FMRI results in a naturalistic study, see Gilmore et al., 2022.

**How can I reduce EPI distortion?** Among other acquisition adjustments, using phase images (field maps) or acquiring an opposite phase-encoded EPI are two common ways, and these can be integrated directly into processing with afni_proc.py. Neither can make the data appear exactly as it would if it were acquired without any distortion, but each do help and add negligible time to a typical scan session. In practice, using opposite phase-encoded EPI may have slight advantages in most software (see e.g., Roopchansingh et al., 2020). Acquiring 5-10 reverse encoded volumes is recommended, to reduce the odds of subject motion ruining the complete set, and this still only takes 10-20 s total in most cases. One can then add the forward- and reverse-encoded pair to the *afni_proc.py* command easily with "-blip_forward_dset" and "-blip_reverse_dset".

---

13 See https://afni.nimh.nih.gov/pub/dist/doc/htmldoc/programs/alpha/afni_proc.py_sphx.html#setting-up-an-analysis

**Should I blur/smooth the FMRI data?** When processing data for voxelwise studies, it is common to do some blurring (in *afni_proc.py*, via the "blur" block). For single echo FMRI, one might blur 1.5-2 times the minimum voxel size. For ME-FMRI, which has higher TSNR, one might blur just slightly above voxel dimension. Other options for special cases of blurring have been discussed above and in the appendices (e.g., "-blur_to_fwhm", etc.). When performing ROI-based studies, blurring should *not* be applied (and one should not include the "blur" block), so that ROI averages used for correlations are not corrupted from outside the ROIs.

**What should the final space be?** If there is a particular atlas of interest, that will often imply a final template space. Otherwise, there are many reasonable options to use within a pipeline, depending on study assumptions and design. When choosing between a volume or a surface, one must make sure that the latter includes all regions of interest (e.g., the subcortex or brainstem are not part of cortical meshes). Additionally, smoothing and any clustering are primary considerations: using a surface restricts both to local cortical gray matter as much as possible. (Visualization need not be a determinant, because volumetric data can be projected onto a surface.) For group-level statistics, volumes or surfaces each offer "standard elements", whether at the voxel/node or ROI level, but users may have a preferred approach. If using a volumetric template, there is typically a choice to make; it is best to use a closely representative template for the study cohort (e.g., age appropriate); and the choice of using a particular atlas can be a decisive factor, are the regions of interest well defined for the template space? One might choose each subject's own anatomical dataset as a final space, which is common when using FreeSurfer parcellations that are defined from those volumes or in clinically-focused scans. Finally, it is possible to use the subject's own EPI, such as to minimize blurring from regridding, particularly in cases where subject movement is expected to be minimal and/or at high spatial resolution.

**Should I use tissue-based regressors in the processing?** Non-GM, tissue-based regressors are often applied in FMRI processing to try to maximize the removal of non-neuronal BOLD features from EPI time series, particularly when processing resting state or naturalistic data. These approaches are applied on the assumption that non-GM signals contain *only* non-neuronal effects, and therefore they are useful proxies of motion or other non-signal effects. These approaches include making regressors of no interest from time series averages, principal components or local components from WM and ventricles (e.g., via "-regress_fast_anaticor", "-regress_ROI_PC", "-regress_ROI_PC_per run"). When considering these methods, such as with the goal of removing motion artifacts, one must be sure that the non-GM tissue maps do *not* intersect with actual GM; this is one reason applied tissue maps for these options in *afni_proc.py* are eroded during or before processing, due to the potential of partial voluming. This can be particularly tricky in the presence of EPI distortion or other artifacts that spread signals around. Furthermore, recent work looking in detail at BOLD signals in non-GM tissue have suggested that assumptions of non-GM-BOLD-like signal might not be obvious: Gore et al. (2019) provide a review of early work for WM, and see also Wang et al. (2022); Gonzalez-Castillo et al. (2022) have shown that signals in ventricles can correlate strongly with physiological measures and even GM. Chen et al. (2023) used local HRF modeling to show that WM signals are typically not null and can carry useful information. Therefore, while using local tissue regressors can help reduce some artifacts (e.g., Jo et al. 2020), care should be taken with assumptions of non-GM signals, and likely more work in the field will be required for this topic.

**What steps should be done before running the FMRI pipeline?** This list can vary widely depending on study design and paradigm, but a few items:
- It is important to check the validity and consistency of initial properties of the data, what we have termed "getting to know your data" (GTKYD; see Reynolds et al., 2023). One can use AFNI's *gtkyd_check* to create systematic tables of dataset properties and compare them (Taylor et al., 2024). Verifying the properties of non-FMRI data are also useful, such as stimulus timing files or physiological response regressors.
- Obliquity in datasets (that is, a stored rotation of the FOV relative to scanner coordinates) can be handled in various ways by software. Both in processing and particularly in visualization, one must decide to either apply or ignore it (with respective tradeoffs in apparent smoothness or location, respectively). We typically recommend removing obliquity from the anatomical datasets before processing, especially if using multiple software packages (such as running FreeSurfer before *afni_proc.py*). This can be done in a way to both preserve coordinate origin and avoid interpolative smoothing, using AFNI's *adjunct_deob_around_origin*. However, obliquity can (and likely should) be left within the EPI data, to be navigated during processing. Using *gtkyd_check*, above, will inform about the presence of obliquity, as would simply running *3dinfo -obliquity*.
- Nonhuman datasets are often acquired in "sphinx" position. Therefore, they should be reoriented so that standard viewing planes (axial, sagittal and coronal) are correct. This can be done with *3drefit* or the new wrapper program *desphinxify*, but special care must be taken to ensure that left and right are correct at the end of this step (other directionality is easily visually verifiable). Using test data with clear left-right delineation, such as a vitamin E tablet, greatly helps this process.
- Having reasonable coordinates so that the participant's brain datasets have reasonable or relatively close overlap with the template space (and with each other) helps processing. In many nonhuman datasets, achieving this may require resetting the coordinate origin (x, y, z) = (0, 0, 0) to be relocated to be in the brain (e.g., center of mass or near the anterior commissure). Occasionally, this is required for human datasets. AFNI's *3dCM* and *@Align_Centers* programs facilitate shifting individual and/or groups of datasets to more useful coordinates.
- If performing final analyses in a standard space, we typically recommend estimating the nonlinear alignment of the anatomical dataset to the template before running *afni_proc.py*. That prevents needing to run the computationally expensive process more than once if analyzing multiple tasks or if (or when) reprocessing the FMRI data. Moreover, using either *sswarper2* (for human data) or *@animal_warper* (for nonhuman data) accomplishes the further task of skullstripping the anatomical. The outputs of this processing are then simply provided to *afni_proc.py* with options (e.g., "-tlrc_NL_warped_dsets", "-copy_anat" and "-anat_has_skull no").
- Other programs that might likely be run before afni_proc.py, with the results passed along, include FreeSurfer's *recon-all*. One might provide the anatomical parcellations via "anat_follower_ROI", or the surface datasets via the "surf" block with "-surf_spec" and "-surf_anat". If using RETROICOR, the physiological time series can be turned into FMRI regressors using AFNI's *physio_calc.py*.

Future work

Over the past 18 years, this program has expanded to include new options and functionality. Improvements have been made to algorithms, as well as fixes added whenever necessary. These trends will surely continue. We plan to add further automated checks and warnings to the quality control. Furthermore, we plan to add a layer around *afni_proc.py* to facilitate running analyses across a group of subjects, as well as more easily facilitating QC checks across them. This process is made easier by the presence of BIDS-formatted inputs or any systematic structures; moreover, users will still be able to specify their analyses in detail through shared scripts. While a large amount of flexibility of HRF modeling exists within afni_proc.py, recent work has shown how much variability there is across the brain (Chen et al. 2023) and potentially across subjects and tasks; we will continue to explore new ways to integrate developments in this area of active research. Finally, in earlier times, there was a GUI interface to help manage and set up *afni_proc.py* commands. The underlying dependency that made that possible (the PyQt4 module) was dropped from some Python distributions, so it could no longer be supported; in the future, we plan to work on a replacement.

**CONCLUSIONS**

We have described several aspects of AFNI's main program for creating a full FMRI processing pipeline, *afni_proc.py*. The program is organized around specifying major processing blocks for a given subject's data, and then adding desired processing details to those. We demonstrated some of the considerable flexibility of analyses that can be run using it here. The user's *afni_proc.py* command is easily and openly sharable, and the pipelines have high reproducibility. More deeply, *afni_proc.py* allows users to control a large number of details about the processing, as well as to examine all the details of the processing provenance within the commented script that it generates. The goal of this integrated design is to facilitate understanding and remove surprises from the analysis stage since FMRI data and their processing are notably complicated. Details matter. For similar reasons, the program also facilitates efficiently verifying both intermediate processing steps and final outcomes through the dictionary of diagnostic quantities and the APQC HTML it creates. This program has expanded and adapted to a variety of new needs over the many years it has existed, in large part due to input from users and collaborators across the neuroimaging community, supported by the modularity of its architecture. We expect it to keep doing so, to meet the continually growing needs and requirements of neuroscience researchers across the field.

**ACKNOWLEDGMENTS**

We would like to thank the myriad users across NIMH, NIH and the globe who have made useful suggestions and provided helpful feedback, all of which has greatly contributed to afni_proc.py. The research and writing of the paper were supported by the NIMH Intramural Research Programs (ZICMH002888) of the NIH (HHS, USA). ZS Saad's contributions were made while employed at the NIMH/NIH through July, 2015. This work utilized the computational resources of the NIH HPC Biowulf cluster (http://hpc.nih.gov).

Gonzalez-Castillo J, Fernandez IS, Handwerker DA, Bandettini PA (2022). Ultra-slow fMRI fluctuations in the fourth ventricle as a marker of drowsiness. Neuroimage 259:119424.

Gore JC, Li M, Gao Y, Wu TL, Schilling KG, Huang Y, Mishra A, Newton AT, Rogers BP, Chen LM, Anderson AW, Ding Z (2019). Functional MRI and resting state connectivity in white matter - a mini-review. Magn Reson Imaging 63:1-11.

Gorgolewski KJ, Auer T, Calhoun VD, Craddock RC, Das S, Duff EP, Flandin G, Ghosh SS, Glatard T, Halchenko YO, Handwerker DA, Hanke M, Keator D, Li X, Michael Z, Maumet C, Nichols BN, Nichols TE, Pellman J, Poline JB, Rokem A, Schaefer G, Sochat V, Triplett W, Turner JA, Varoquaux G, Poldrack RA (2016). The brain imaging data structure, a format for organizing and describing outputs of neuroimaging experiments. Sci Data 3:160044.

Gotts SJ, Gilmore AW, Martin A (2020). Brain networks, dimensionality, and global signal averaging in resting-state fMRI: Hierarchical network structure results in low-dimensional spatiotemporal dynamics. Neuroimage 2020 205:116289. doi: 10.1016/j.neuroimage.2019.116289.

Hallquist MN, Hwang K, Luna B (2013). The nuisance of nuisance regression: spectral misspecification in a common approach to resting-state fMRI preprocessing reintroduces noise and obscures functional connectivity. Neuroimage. 82:208–225. doi:10.1016/j.neuroimage.2013.05.116

Hasson U, Malach R, Heeder DJ (2010). Reliability of cortical activity during natural stimulation. Trends Cogn Sci 14(1):40-8.

Holla B, Taylor PA, Glen DR, Lee JA, Vaidya N, Mehta UM, Venkatasubramanian G, Pal P, Saini J, Rao NP, Ahuja C, Kuriyan R, Krishna M, Basu D, Kalyanram K, Chakrabarti A, Orfanos DP, Barker GJ, Cox RW, Schumann G, Bharath RD, Benegal V (2020). A series of five population-specific Indian brain templates and atlases spanning ages 6 to 60 years. Hum Brain Mapp 41(18):5164-5175.

Holland D, Kuperman JM, Dale AM (2010). Efficient correction of inhomogeneous static magnetic field-induced distortion in Echo Planar Imaging. Neuroimage 50(1):175-83.

Hong X, To XV, Teh I, Soh JR, Chuang KH (2015). Evaluation of EPI distortion correction methods for quantitative MRI of the brain at high magnetic field. Magn Reson Imaging 33(9):1098-1105.

Hutton C, Bork A, Josephs O, Deichmann R, Ashburner J, Turner R (2002). Image distortion correction in fMRI: A quantitative evaluation. Neuroimage 16(1):217-40.

Irfanoglu MO, Sarlls J, Nayak A, Pierpaoli C (2019). Evaluating corrections for Eddy-currents and other EPI distortions in diffusion MRI: methodology and a dataset for benchmarking. Magn Reson Med 81(4):2774-2787.

Jo HJ, Saad ZS, Simmons WK, Milbury LA, Cox RW (2010). Mapping Sources of Correlation in Resting State FMRI, with Artifact Detection and Removal. Neuroimage 52(2): 571–582.

Jo HJ, Gotts SJ, Reynolds RC, Bandettini PA, Martin A, Cox RW, Saad ZS (2013). Effective Preprocessing Procedures Virtually Eliminate Distance-Dependent Motion Artifacts in Resting State57

**Appendix A: Considerations for bandpassing (or not) in resting state FMRI**

Since the earliest days of resting state FMRI (Biswal et al., 1995), it has been quite common in the field to apply bandpassing in resting state FMRI preprocessing, where "low frequency fluctuations" (LFFs) within an interval of approximately 0.01-0.1 Hz are kept in the EPI data and all frequencies outside that range are filtered out. Beyond historical precedent, additional reasons for such bandpassing typically include reduction of high-frequency noise or an attempt to reduce physiological components, though at least some of the latter get aliased down into the traditional LFF range. However, there are notable reasons to not necessarily include bandpassing.

Firstly, there is still useful signal, not just noise, in BOLD data above 0.1 Hz (Gohel and Biswal, 2015; Shirer et al., 2015). Secondly, there is a tremendous statistical cost that is paid with bandpassing, removing a large number of degrees of freedom from the data. For each frequency removed from the original EPI time series spectrum, *two* degrees of freedom are used up. Simple relations for approximating loss of degrees of freedom are provided in Eq. 1a-b. For typical data with TR=2.0s (such as in the resting state data used above), standard LFF bandpassing to 0.01-0.1 Hz would use up over 60% of the degrees of freedom of the data, *just* from bandpassing, which would reduce the final DF count of approx. 90% in Fig. 13B to just about 30% (Fig. 13C). For data with TR=1.0s, the same bandpassing would use up over 80% of the DFs. Bandpassing has a very large statistical cost to pay.

Additionally, any processing pipeline must take care in how bandpassing is performed. There are multiple ways to perform bandpassing, of which one is the Fourier Transform, but it is mathematically incorrect to include it in preprocessing separate from the regression model (see Hallquist et al., 2013), unless the regression model was similarly bandpassed. One negative consequence of separating it can be to re-introduce frequencies that were supposed to be removed. But far more deleteriously, it can lead to using up all the degrees of freedom present in the original data, *in a way that the analyst does not realize*. As a result, the output of the regression model is purely noise and random fluctuations. In the present example, the censor fraction is quite low, but in practice for many time series (particularly for children and other motion-prone populations), it would be easy for motion and bandpassing to use up all available DFs. So, while one can implement bandpassing in the processing (and, again, it is implemented within the regression model to avoid mathematical inconsistency), one should consider whether it is worth the costs.

**Appendix B: Timeline of selected afni_proc.py features and demo examples**

Over the past 16 years since *afni_proc.py* was first created, many methodological and acquisition developments have occurred. The program has continued to grow, enabling a wider range of FMRI processing functionality within its efficient pipeline-generating framework. Some of these updates are noted for historical record and/or curiosity in Table B1.

```
-----------------------------------------------------------
2006: start of afni_proc.py: FMRI preprocessing through regression modeling
2008: include smoothness estimates (clustering)
2009: enable use of 3dREMLfit (estimating temporal autocorrelation)
2009: turn off masking of EPI results (better QC), though masks estimated
2009: add ricor block (physio regressors)
2009: add anat/EPI alignment via align_epi_anat.py with concatenated
      transformations
```

```
2009: add censoring based on motion parameters
2010: enable amplitude modulation in the linear regression model
2011: auto-create review scripts for QC (basic quantities and driving GUI)
2011: enable surface analysis (typically with SUMA-standardized meshes)
2012: bandpassing (via mathematically correct, single-regression model)
2012: enable tissue-based regression
2013: ANATICOR for rest FMRI (local WM regressors that vary per voxel)
2013: nonlinear align to template (from 3dQwarp; improve spatial specificity)
2013: "MIN_OUTLIER" functionality for principled selection of volreg base
2015: fast ANATICOR
2015: ROI/PC regression
2016: check left/right flip of EPI vs anat
2016: enable reverse blip correction (reduce B0 distortion along phase axis)
2017: Python 3 compatible (maintaining compatibility with Python 2)
2018: enable multi-echo FMRI data compatibility
2018: auto-create QC HTML (built off basic/driver reviews, plus new features)
2019: add more ME-FMRI combinations with tedana from MEICA group
2020: option to compare afni_proc.py commands (help both users and developers)
2021: ap_run_simple.tcsh wrapper: low-option afni_proc.py cmd for quick/scanner QC
2022: local unifize option to help inhomogenous EPI align to anatomical volumes
2022: find_variance_lines.tcsh: detect high-variance I/S lines in EPI
2023: run APQC HTML from local server, for interactive buttons and features
2024: compute TSNR stats across automatic or provided ROIs
2024: ap_run_simple_me.tcsh wrapper: multiecho FMRI version of ap_run_simple.tcsh
2024: enable output of BIDS derivative tree
2024: create BIDS App for ap_run_simple.tcsh processing
2024: enable blip correction via phase map (e.g., processed by epi_b0_correct.py)
----------------------------------------------------------
```

*Table B1:* *A brief history of afni_proc.py, though a selected list of major feature additions.*

There are multiple layers to what we simply refer to as "afni_proc.py." Its history has been a combination of individual leadership, team development and wider community contribution. The top layer is the actual Python language program itself, which parses the command line options and develops the analysis pipeline. R Reynolds started the afni_proc.py file and its primary library db_mod.py in 2006 and has been the primary developer of its growth since then. The next layer is the Python library "afnipy", which is distributed within AFNI. It contains a range of functionality for handling files, parsing text, calling the command line and performing a host of intermediate tasks in library files associated with the large number of Python programs in AFNI. This has been largely developed over time by R Reynolds, the other co-authors here and the members of the principal AFNI development group. The module has been designed to minimize external dependencies, for improved stability over time. The next layer is the larger AFNI code base itself. This was started by Robert Cox in 1994 and has been developed by the AFNI group and an extended set of contributors since then. This contains a wide range of programs for handling data from fundamental processes through specialized tasks in various geometries and formats. The processing script that *afni_proc.py* builds calls this layer of programs. Additionally, it also calls another layer of programs, which are those imported from external software that can be utilized, such as the tedana MEICA tool, surface/parcellation results from FreeSurfer, and QC HTML visualization with NiiVue. Finally, we note that the contents and development of afni_proc.py have also benefited from its neuroimaging user base. Useful ideas have been suggested by collaborators, either via Message Board posts, GitHub Issues, emails or conversations.

The maintenance and testing of afni_proc.py also has multiple layers. There is a set of more than 100 tests run internally to evaluate and maintain its performance over time, running afni_proc.py more than 200 times and currently generating 135 processing scripts, all of which is compared against prior execution. The full AFNI release build will not proceed without the main class example of single subject analysis succeeding. The AFNI GitHub page contains a set of continuation integration tests over programs run whenever the code base is updated. Users (particularly other developers) are able to make pull requests and raise issues via AFNI's GitHub homepage. Any crashes, bug reports, questions or suggestions can also be raised via the Message Board page. The regular use of afni_proc.py in teaching examples also helps maintain its code.

As the list of *afni_proc.py*'s features has grown, so has its range of usage and applications. There are many examples of partial and full commands within the program help, and this list will likely continue to increase. The AFNI Bootcamp data collection contains multiple examples of processing functionality along with input data, for ease of starting to practice with the program; there is also a large amount of educational material in the accompanying afni_handouts directory and AFNI Academy YouTube channel (https://www.youtube.com/c/afnibootcamp), and other educational materials described in the main text.

In all cases, we note that FMRI techniques will adapt over time, as will the usage of various options in processing. These examples and others that will be made in the future will provide useful guides and references for processing choices, but users should always make their final processing choices based on their own study paradigm, goals and data at hand. Progress in acquisition techniques may alter data properties; new options may be created to improve analyses; new understanding may shift thinking in processing strategies. Examples may be updated over time, or changed entirely. As noted in the main text, users can directly and usefully compare their own *afni_proc.py* pipeline commands against existing ones using "-compare_opts .." and "-compare_example_pair ..".

**Appendix C: Supplementary examples of running afni_proc.py**

As noted in the main text, there are a vast array of study designs, technical assumptions and processing options that researchers might adopt. In the main text we provided four fundamental examples with comments on major processing steps (as well as two cases of running "simple" afni_proc.py commands for quick processing). Here we provide a small set of additional afni_proc.py examples, generally offering small variations or alternatives to the primary examples. Importantly, some of these might *not* be recommended choices, but we mention for contrast or for comparison; we highlight such cases clearly.

As with the examples in the main text, each of these supplementary *afni_proc.py* commands is freely available within this paper's associated GitHub repository (https://github.com/afni/apaper_afni_proc).

Since each command here is directly based on one from the main text, we simply highlight the minor point-by-point changes in each case. A copy of the full extra example commands are below in the text. In this case, the highlighting within the command reflects the option differences from the referenced examples in the main text.

The first few supplementary examples are for resting state FMRI processing, based on main Ex. 3. Ex. 5 differs by including bandpassing to the common low frequency fluctuation (LFF) range of 0.01-0.1 Hz, by adding the regress block option "-regress_bandpass 0.01 0.1". While such bandpassing is widely used within resting state processing, aimed at reducing the influence of breathing and heart rate in results, it also carries significant costs. The main text discusses the loss of degrees of freedom in modeling (see main Fig. 14), which might be prohibitive to processing in cases of medium-to-severe motion , when accounted for. Appendix A contains further discussion in the literature about the implications and tradeoffs of including such bandpassing.

Ex. 6 starts from Ex. 3 and adds additional regressors based on non-GM tissue, using WM and CSF/ventricle maps. The implementation in this example relies on adding pre-calculated masks into the processing stream, each of which ends in the final space (here, MNI) on either the EPI or anatomical grid. Here, "-anat_follower_ROI ..." is used to bring some results from running FreeSurfer's recon-all on the subject's T1w anatomical dataset. In each case, the user assigns a brief label for working with the dataset within the code and designates the final grid:
- The ventricle map ${roi_FSvent} will have "epi" grid spacing (label = "FSvent").
- The WM map ${roi_FSWe} will have "epi" grid spacing (label = "FSWe").

The partnered option "-anat_follower_erode FSvent FSWe" means that the ventricle and WM maps will each be eroded by one voxel on their input grid. This erosion is done to reduce the probability that they contain any GM information through partial voluming or imprecise estimation, since they are used to generate tissue-based regressors in the "regress" block, below. Larger erosion levels reduce the odds of including GM signal in the regressors, but quickly shrink the regions.

The first three principal components (PCs) from the (censored time series of the) eroded ventricles are utilized ("-regress_ROI_PC FSvent 3"), and these are included as per-run regressors ("-regress_ROI_PC_per_run FSvent") in case multiple EPI datasets have been input to *afni_proc.py*. This is quite similar to the CompCor approach (Behzadi et al., 2007). Secondly, fast ANATICOR (Jo et al., 2013; Jo et al., 2020) is utilized ("-regress_anticor_fast"), by generating a local regressor for each voxel based on a distance-weighted average over the eroded WM mask ("-regress_anaticor_label FSWe"). We note that the inclusion of both ANATICOR and the earlier full RETROICOR depend on AFNI's ability to include voxelwise regressors, rather than being restricted to those that are constant across the volume.

Additionally, Ex. 6 includes a different style of smoothing the data. Typical smoothing applies a constant kernel size of blur through the entire FOV. EPI datasets have inhomogeneous spatial smoothness, and this variability remains after this process (though everywhere is smoother by a constant amount). In this example, we have added "-blur_to_fwhm", so that the blurring is applied in a *non-constant* manner to create a final output that has *approximately homogeneous* spatial smoothness. This option is particularly useful when combining data from different collections within a study, which is fairly common in resting state studies, since each scanner or site will have different noise and spatial properties. Blurring the data to end up with the same target value is one way to reduce the inhomogeneity of the data. Adding this option changes the interpretation of the specified "-blur_size ..", so that the given value no longer describes the amount of blurring added to the EPI time series, but instead the target amount of blurring that the data should have at the end of the block. Note that at present this option should only be used either after the "mask" block or with "-blur_in_automask", as it is not appropriate to include non-brain regions in the blur estimates.

Example 7 also starts from main Ex. 3, but it does not apply any blur (spatial smoothing). This is appropriate when preparing data for ROI-based analyses, to prevent signal mixing across ROI boundaries before analysis (which would artificially increase correlations). To make this processing change, "blur" is removed from the list of blocks, as is any option starting with "-blur_*". Including the atlases or ROI maps of interest with "-anat_follower_ROI .." is particularly useful in this case, so that desired parcellations for later analysis are automatically mapped to the final EPI space.

It is worth noting that when blurring is not applied, the voxelwise TSNR will be notably decreased. In the final ROI-based analyses, this should not be a problem because the time series are averaged within the individual regions, boosting TSNR in that manner. However, when performing quality control (QC) checks, one has to take into consideration that seed-based correlation maps may look much sparser. Additionally, TSNR distributions within regions of interest will also be lower. One must account for this when evaluating or comparing results. It may be useful to check the data separately with blurring applied (e.g., using *ap_run_simple\*.tcsh*), for a more standard view.

Example 8 is based on the surface-based, ME-FMRI processing in main Ex. 4. The only change in this supplementary example is the choice of technique for combining the EPI's multiple echos. Here, the straightforward optimal combination (OC) method of Posse et al. (1999) is used, by changing "-combine_method m_tedana" to "-combine_method OC". This is a simple but useful weighted average across echos, which should purely boost local TSNR.

Example 9 is an extension of Ex. 3, using multi-echo data and reverse phase-encoded distortion correction. Echo combination is performed using AFNI's optimal combination (OC) method for weighting the echoes. The blur size is reduced from 5mm to 4mm due to the TSNR increase from combining multiple echoes. For simplicity, physiological noise correction and the "-anat_follower_ROI" datasets were omitted, though they could easily be added back as a variation of this example, as the researcher prefers and the data allow.

Example 10 refers to the task-based FMRI processing in main Ex. 2. In this case, no processing option changes, but an additional output directory is created with a subset of standard *afni_proc.py* outputs in BIDS-Derivatives format, by adding the option "-bids_deriv yes". In this case, the new directory will be placed in the afni_proc.py results directory and called "bids_deriv"; if desired, the user can change the "yes" argument to be a path to a new output location and name.

Additionally, the user can provide further information that might be used within the naming and structure of the BIDS-Derivative output via "-uvar .." options. This option is used to provide a key+value pair that is passed along to the dictionary of user variables ("uvars"), described in the main text. For BIDS-Derivatives, one might add in a session ID (if the input data were in a BIDS structure containing that optional level) and/or a taskname, such as with "-uvar ses ses-01" and "-uvar taskname the.task.name", respectively.

Below are the text of the supplementary examples. The are also all included in the main GitHub repository (https://github.com/afni/apaper_afni_proc/tree/main), for easier copy+pasting.

```
-------------------------------------------------------------------------------
# AP Example 5 (based on Ex. 3, but adding LFF bandpassing, whose caveats are
# discussed in the text), which was preceded by running:
# + FreeSurfer's recon-all: estimates subject-specific parcellation
# + @SUMA_Make_Spec_FS   : creates NIFTI format parcellation, and tissue-
#                          based subsets like GM regions ${roi_gmr_2009}
# + sswarper2            : creates skullstripped anatomical ${anat_cp},
#                          and nonlinear warps ${dsets_NL_warp}
# + physio_calc.py       : estimates physiological (respiratory and
#                          cardiac) regressors ${physio_regs}

afni_proc.py                                                              \
    -subj_id                    ${subj}                                   \
    -dsets                      ${dset_epi}                               \
    -copy_anat                  ${anat_cp}                                \
    -anat_has_skull             no                                        \
    -anat_follower              anat_w_skull anat ${anat_skull}           \
    -anat_follower_ROI          aagm09 anat ${roi_gmr_2009}               \
    -anat_follower_ROI          aegm09 epi  ${roi_gmr_2009}               \
    -ROI_import                 BrodPijn ${atl_brod}                      \
    -ROI_import                 SchYeo7N ${atl_sy7n}                      \
    -blocks                     ricor tshift align tlrc volreg mask blur  \
                                scale regress                             \
    -radial_correlate_blocks    tcat volreg regress                       \
    -tcat_remove_first_trs      4                                         \
    -ricor_regs                 ${physio_regs}                            \
    -ricor_regs_nfirst          4                                         \
    -ricor_regress_method       per-run                                   \
    -align_unifize_epi          local                                     \
    -align_opts_aea             -cost lpc+ZZ -giant_move -check_flip      \
    -tlrc_base                  ${template}                               \
    -tlrc_NL_warp                                                         \
    -tlrc_NL_warped_dsets       ${dsets_NL_warp}                          \
    -volreg_align_to            MIN_OUTLIER                               \
    -volreg_align_e2a                                                     \
    -volreg_tlrc_warp                                                     \
    -volreg_warp_dxyz           3                                         \
    -volreg_compute_tsnr        yes                                       \
    -mask_epi_anat              yes                                       \
    -blur_size                  5                                         \
    -regress_motion_per_run                                               \
    -regress_make_corr_vols     aegm09                                    \
    -regress_censor_motion      0.2                                       \
    -regress_censor_outliers    0.05                                      \
    -regress_apply_mot_types    demean deriv                              \
    -regress_bandpass           0.01 0.1                                  \
    -regress_est_blur_epits                                               \
    -regress_est_blur_errts                                               \
    -regress_compute_tsnr_stats BrodPijn 7 10 12 39 107 110 112 139       \
    -regress_compute_tsnr_stats SchYeo7N 161 149 7 364 367 207            \
```

```
       -html_review_style            pythonic
    --------------------------------------------------------------------------------

    --------------------------------------------------------------------------------
# AP Example 6 (based on Ex. 3, but adding ANATICOR and CSF component regressors),
# which was preceded by running:
# + FreeSurfer's recon-all: estimates subject-specific parcellation
# + @SUMA_Make_Spec_FS    : creates NIFTI format parcellation, and tissue-
#                          based subsets like GM regions ${roi_gmr_2009}; here
#                          we also use the ventricles ${roi_FSvent} and white
#                          matter ${roi_FSWe} for ANATICOR and PC regressors
# + sswarper2             : creates skullstripped anatomical ${anat_cp},
#                          and nonlinear warps ${dsets_NL_warp}
# + physio_calc.py        : estimates physiological (respiratory and
#                          cardiac) regressors ${physio_regs}

afni_proc.py                                                              \
    -subj_id                   ${subj}                                    \
    -dsets                     ${dset_epi}                                \
    -copy_anat                 ${anat_cp}                                 \
    -anat_has_skull            no                                         \
    -anat_follower             anat_w_skull anat ${anat_skull}            \
    -anat_follower_ROI         aaseg  anat ${roi_all_2009}                \
    -anat_follower_ROI         aeseg  epi  ${roi_all_2009}                \
    -anat_follower_ROI         aagm09 anat ${roi_gmr_2009}                \
    -anat_follower_ROI         aegm09 epi  ${roi_gmr_2009}                \
    -anat_follower_ROI         aagm00 anat ${roi_gmr_2000}                \
    -anat_follower_ROI         aegm00 epi  ${roi_gmr_2000}                \
    -anat_follower_ROI         FSvent epi  ${roi_FSvent}                  \
    -anat_follower_ROI         FSWe   epi  ${roi_FSWe}                    \
    -anat_follower_erode       FSvent FSWe                                \
    -ROI_import                BrodPijn ${atl_brod}                       \
    -ROI_import                SchYeo7N ${atl_sy7n}                       \
    -blocks                    ricor tshift align tlrc volreg mask blur   \
                               scale regress                              \
    -radial_correlate_blocks   tcat volreg regress                        \
    -tcat_remove_first_trs     4                                          \
    -ricor_regs                ${physio_regs}                             \
    -ricor_regs_nfirst         4                                          \
    -ricor_regress_method      per-run                                    \
    -align_unifize_epi         local                                      \
    -align_opts_aea            -cost lpc+ZZ -giant_move -check_flip       \
    -tlrc_base                 ${template}                                \
    -tlrc_NL_warp                                                         \
    -tlrc_NL_warped_dsets      ${dsets_NL_warp}                           \
    -volreg_align_to           MIN_OUTLIER                                \
    -volreg_align_e2a                                                     \
    -volreg_tlrc_warp                                                     \
    -volreg_warp_dxyz          3                                          \
    -volreg_compute_tsnr       yes                                        \
    -mask_epi_anat             yes                                        \
    -blur_size                 5                                          \
    -regress_motion_per_run                                               \
```

```
    -regress_ROI_PC            FSvent 3                                     \
    -regress_ROI_PC_per_run    FSvent                                       \
    -regress_make_corr_vols    aeseg FSvent                                 \
    -regress_anaticor_fast                                                  \
    -regress_anaticor_label    FSWe                                         \
    -regress_censor_motion     0.2                                          \
    -regress_censor_outliers   0.05                                         \
    -regress_apply_mot_types   demean deriv                                 \
    -regress_est_blur_epits                                                 \
    -regress_est_blur_errts                                                 \
    -regress_compute_tsnr_stats BrodPijn 7 10 12 39 107 110 112 139         \
    -regress_compute_tsnr_stats SchYeo7N 161 149 7 364 367 207              \
    -html_review_style         pythonic
--------------------------------------------------------------------------------

--------------------------------------------------------------------------------
# AP Example 7 (based on Ex. 3, but removing blurring so it is appropriate for
# ROI-based analysis), which was preceded by running:
# + FreeSurfer's recon-all: estimates subject-specific parcellation
# + @SUMA_Make_Spec_FS    : creates NIFTI format parcellation, and tissue-
#                           based subsets like GM regions ${roi_gmr_2009}
# + sswarper2             : creates skullstripped anatomical ${anat_cp},
#                           and nonlinear warps ${dsets_NL_warp}
# + physio_calc.py        : estimates physiological (respiratory and
#                           cardiac) regressors ${physio_regs}

afni_proc.py                                                                \
    -subj_id                   ${subj}                                      \
    -dsets                     ${dset_epi}                                  \
    -copy_anat                 ${anat_cp}                                   \
    -anat_has_skull            no                                           \
    -anat_follower             anat_w_skull anat ${anat_skull}              \
    -anat_follower_ROI         aagm09 anat ${roi_gmr_2009}                  \
    -anat_follower_ROI         aegm09 epi  ${roi_gmr_2009}                  \
    -ROI_import                BrodPijn ${atl_brod}                         \
    -ROI_import                SchYeo7N ${atl_sy7n}                         \
    -blocks                    ricor tshift align tlrc volreg mask          \
                               scale regress                                \
    -radial_correlate_blocks   tcat volreg regress                          \
    -tcat_remove_first_trs     4                                            \
    -ricor_regs                ${physio_regs}                               \
    -ricor_regs_nfirst         4                                            \
    -ricor_regress_method      per-run                                      \
    -align_unifize_epi         local                                        \
    -align_opts_aea            -cost lpc+ZZ -giant_move -check_flip         \
    -tlrc_base                 ${template}                                  \
    -tlrc_NL_warp                                                           \
    -tlrc_NL_warped_dsets      ${dsets_NL_warp}                             \
    -volreg_align_to           MIN_OUTLIER                                  \
    -volreg_align_e2a                                                       \
    -volreg_tlrc_warp                                                       \
    -volreg_warp_dxyz          3                                            \
    -volreg_compute_tsnr       yes                                          \
```

```
    -mask_epi_anat             yes                                       \
    -regress_motion_per_run                                              \
    -regress_make_corr_vols    aegm09                                    \
    -regress_censor_motion     0.2                                       \
    -regress_censor_outliers   0.05                                      \
    -regress_apply_mot_types   demean deriv                              \
    -regress_est_blur_epits                                              \
    -regress_est_blur_errts                                              \
    -regress_compute_tsnr_stats BrodPijn 7 10 12 39 107 110 112 139      \
    -regress_compute_tsnr_stats SchYeo7N 161 149 7 364 367 207           \
    -html_review_style         pythonic
--------------------------------------------------------------------------------

--------------------------------------------------------------------------------
# AP Example 8 (based on Ex. 4, but using AFNI's OC rather than tedana's MEICA for
# combining echos), which was preceded by running:
# + FreeSurfer's recon-all : estimates anatomical surface mesh datasets
# + @SUMA_Make_Spec_FS     : creates standard-mesh versions of surfaces and
#                            specification files ${suma_specs} and NIFTI-
#                            format anatomical surface volume ${suma_sv}
# + sswarper2              : creates skullstripped anatomical ${anat_cp},
#                            and nonlinear warps ${dsets_NL_warp}

afni_proc.py                                                             \
    -subj_id                   ${subj}                                   \
    -dsets_me_run              ${dsets_epi_me}                           \
    -echo_times                12.5 27.6 42.7                            \
    -copy_anat                 ${anat_cp}                                \
    -anat_has_skull            no                                        \
    -anat_follower             anat_w_skull anat ${anat_skull}           \
    -blocks                    tshift align volreg mask combine surf     \
                               blur scale regress                       \
    -radial_correlate_blocks   tcat volreg                               \
    -tcat_remove_first_trs     4                                         \
    -blip_forward_dset         "${epi_forward}"                          \
    -blip_reverse_dset         "${epi_reverse}"                          \
    -tshift_interp             -wsinc9                                   \
    -align_unifize_epi         local                                     \
    -align_opts_aea            -cost lpc+ZZ -giant_move -check_flip      \
    -volreg_align_to           MIN_OUTLIER                               \
    -volreg_align_e2a                                                    \
    -volreg_warp_final_interp  wsinc5                                    \
    -volreg_compute_tsnr       yes                                       \
    -mask_epi_anat             yes                                       \
    -combine_method            m_tedana                                  \
    -surf_anat                 ${suma_sv}                                \
    -surf_spec                 ${suma_specs}                             \
    -blur_size                 4                                         \
    -regress_motion_per_run                                              \
    -regress_censor_motion     0.2                                       \
    -regress_censor_outliers   0.05                                      \
    -regress_apply_mot_types   demean deriv                              \
    -html_review_style         pythonic
```

```
--------------------------------------------------------------------------------

--------------------------------------------------------------------------------
# AP Example 9 (based on Ex. 3 and 4, being a fairly basic multi-echo FMRI
# example run volumetrically), which was preceded by running:
# + sswarper2      : creates skullstripped anatomical ${anat_cp}, and
#                    nonlinear warps ${dsets_NL_warp}

afni_proc.py                                                                    \
    -subj_id                   ${subj}                                          \
    -dsets_me_run              ${dsets_epi_me}                                  \
    -echo_times                12.5 27.6 42.7                                   \
    -copy_anat                 ${anat_cp}                                       \
    -anat_has_skull            no                                               \
    -anat_follower             anat_w_skull anat ${anat_skull}                  \
    -ROI_import                BrodPijn ${atl_brod}                             \
    -ROI_import                SchYeo7N ${atl_sy7n}                             \
    -blocks                    tshift align tlrc volreg mask combine            \
                               blur scale regress                               \
    -radial_correlate_blocks   tcat volreg regress                              \
    -tcat_remove_first_trs     4                                                \
    -blip_forward_dset         "${epi_forward}"                                 \
    -blip_reverse_dset         "${epi_reverse}"                                 \
    -align_unifize_epi         local                                            \
    -align_opts_aea            -cost lpc+ZZ -giant_move -check_flip             \
    -tlrc_base                 ${template}                                      \
    -tlrc_NL_warp                                                               \
    -tlrc_NL_warped_dsets      ${dsets_NL_warp}                                 \
    -volreg_align_to           MIN_OUTLIER                                      \
    -volreg_align_e2a                                                           \
    -volreg_tlrc_warp                                                           \
    -volreg_warp_dxyz          3                                                \
    -volreg_compute_tsnr       yes                                              \
    -mask_epi_anat             yes                                              \
    -combine_method            OC                                               \
    -blur_size                 4                                                \
    -regress_motion_per_run                                                     \
    -regress_censor_motion     0.2                                              \
    -regress_censor_outliers   0.05                                             \
    -regress_apply_mot_types   demean deriv                                     \
    -regress_est_blur_epits                                                     \
    -regress_est_blur_errts                                                     \
    -regress_compute_tsnr_stats  BrodPijn 7 10 12 39 107 110 112 139            \
    -regress_compute_tsnr_stats  SchYeo7N 161 149 7 364 367 207                 \
    -html_review_style         pythonic
--------------------------------------------------------------------------------

--------------------------------------------------------------------------------
# AP Example 10 (based on the task-based Ex. 2, but including an output directory
# of BIDS-Derivative format outputs), which was preceded by running:
# + timing_tool.py : creates stimulus timing files times.*.txt from TSV
# + sswarper2      : creates skullstripped anatomical ${anat_cp}, and
#                    nonlinear warps ${dsets_NL_warp}
```

```
afni_proc.py                                                              \
    -subj_id                 ${subj}                                      \
    -uvar                    taskname task-pamenc                         \
    -dsets                   ${dset_epi}                                  \
    -copy_anat               ${anat_cp}                                   \
    -anat_has_skull          no                                           \
    -anat_follower           anat_w_skull anat ${anat_skull}              \
    -blocks                  tshift align tlrc volreg mask blur scale     \
                             regress                                      \
    -radial_correlate_blocks tcat volreg regress                          \
    -tcat_remove_first_trs   0                                            \
    -tshift_opts_ts          -tpattern alt+z2                             \
    -align_unifize_epi       local                                        \
    -align_opts_aea          -giant_move -cost lpc+ZZ -check_flip         \
    -tlrc_base               ${template}                                  \
    -tlrc_NL_warp                                                         \
    -tlrc_NL_warped_dsets    ${dsets_NL_warp}                             \
    -volreg_align_to         MIN_OUTLIER                                  \
    -volreg_align_e2a                                                     \
    -volreg_tlrc_warp                                                     \
    -volreg_warp_dxyz        3.0                                          \
    -volreg_compute_tsnr     yes                                          \
    -mask_epi_anat           yes                                          \
    -blur_size               6                                            \
    -blur_in_mask            yes                                          \
    -regress_stim_times      ${sdir_timing}/times.CONTROL.txt             \
                             ${sdir_timing}/times.TASK.txt                \
    -regress_stim_labels     CONTROL TASK                                 \
    -regress_stim_types      AM1                                          \
    -regress_basis_multi     'dmUBLOCK(-1)'                               \
    -regress_motion_per_run                                               \
    -regress_censor_motion   0.3                                          \
    -regress_censor_outliers 0.05                                         \
    -regress_compute_fitts                                                \
    -regress_fout            no                                           \
    -regress_opts_3dD        -jobs 8                                      \
                             -gltsym 'SYM: TASK -CONTROL'                 \
                             -glt_label 1 T-C                             \
                             -gltsym 'SYM: 0.5*TASK +0.5*CONTROL'         \
                             -glt_label 2 meanTC                          \
    -regress_3dD_stop                                                     \
    -regress_reml_exec                                                    \
    -regress_make_ideal_sum  sum_ideal.1D                                 \
    -regress_est_blur_errts                                               \
    -regress_run_clustsim    no                                           \
    -html_review_style       pythonic                                     \
    -bids_deriv              yes                                          \
------------------------------------------------------------------------------
```